\documentclass[a4paper,11pt]{article}
\pdfoutput=1
\usepackage{jheppub}
\usepackage[T1]{fontenc}
\usepackage{graphicx}
\usepackage{amsmath}
\usepackage{xspace}
\usepackage{subfig}
\usepackage{xcolor}
\usepackage{mathtools}
\usepackage{braket}
\usepackage{multirow}
\usepackage{colortbl}
\usepackage{setspace}
\usepackage{slashed}
\usepackage{appendix}
\usepackage{booktabs}
\usepackage{cancel}
\usepackage{array}
\usepackage{pst-node} 
\usepackage[off]{auto-pst-pdf} 
\usepackage{tikz}
\usepackage{wasysym}
\usepackage{amssymb}
\usepackage{cleveref}
\usepackage{makecell}
\usepackage{url}
\usetikzlibrary{tikzmark}
\usetikzlibrary{positioning}
\hypersetup{colorlinks=true, linkcolor=[rgb]{0,0.3,0.7}, urlcolor=[rgb]{0,0.3,0.7}, citecolor=[rgb]{0,0.3,0.7}}

\allowdisplaybreaks

\usepackage[detect-all]{siunitx}
\definecolor{newgray}{gray}{0.85}
\definecolor{newgray2}{gray}{0.95}

\newcommand{\ovbb}[0]{$0\nu\beta\beta$ }
\newcommand{\lr}[1]{\left( #1\right)}
\newcommand{\lrs}[1]{\big( #1\big)}

\Crefname{table}{}{}
\Crefname{figure}{}{}

\title{Probing Lepton Number Violation:\\A~Comprehensive Survey of Dimension-7 SMEFT
}
\author[a,b,c]{K\r{a}re Fridell}
\author[d,e]{Luk\'a\v{s} Gr\'af}
\author[f,c]{Julia Harz} 
\author[g,c]{Chandan Hati} 
\emailAdd{kare.fridell@kek.jp}
\emailAdd{lukas.graf@berkeley.edu}
\emailAdd{julia.harz@uni-mainz.de}
\emailAdd{chandan.hati@ulb.be}
\affiliation[a]{\AddrKEK}
\affiliation[b]{\AddrFSU}
\affiliation[c]{\AddrTUM}
\affiliation[d]{\AddrUCB}
\affiliation[e]{\AddrUCSD}
\affiliation[f]{\AddrJGU}
\affiliation[g]{\AddrULB}

\newcommand{\AddrKEK}{Theory Center, Institute of Particle and Nuclear Studies,\\ High Energy Accelerator Research Organization (KEK), Tsukuba 305-0801, Japan}
\newcommand{\AddrTUM}{Physik Department T70, Technische Universit\"at M\"unchen,\\ James-Franck-Stra{\ss}e 1, D-85748 Garching, Germany}
\newcommand{\AddrUCB}{Department of Physics, University of California, Berkeley, CA 94720, USA}
\newcommand{\AddrUCSD}{Department of Physics, University of California, San Diego, La Jolla, CA 92093-0319, USA}
\newcommand{\AddrJGU}{PRISMA+ Cluster of Excellence \& Mainz Institute for Theoretical Physics,\\ Johannes Gutenberg-Universit\"at Mainz, 55099 Mainz, Germany}
\newcommand{\AddrULB}{Service de Physique Th\'eorique, Universit\'e Libre de Bruxelles,\\ Boulevard du Triomphe, CP225,
1050 Brussels, Belgium}
\newcommand{\AddrFSU}{Department of Physics, Florida State University, Tallahassee, FL 32306, USA}

\abstract{Observation of lepton number violation would represent a groundbreaking discovery with profound consequences for fundamental physics and as such, it has motivated an extensive experimental program searching for neutrinoless double beta decay. However, the violation of lepton number can be also tested by a variety of other observables. We focus on the possibilities of probing this fundamental symmetry within the framework of the Standard Model Effective Field Theory (SMEFT) beyond the minimal dimension-5. Specifically, we study the bounds on $\Delta L = 2$ dimension-7 effective operators beyond the electron flavor imposed by all relevant low-energy observables and confront them with derived high-energy collider limits. We also discuss how the synergy of the analyzed multi-frontier observables can play a crucial role in distinguishing among different dimension-7 SMEFT operators.}
\begin{document}	
\preprint{KEK-TH-2486

\hfill N3AS-23-014

\hfill MITP-23-027

\hfill  ULB-TH/23-06}
\maketitle
\flushbottom

%%%%%%%%%%%%%%%%%%%%%%%%%%%%%%%%%%%%%%%%%%%%%%%%%%%%%%%%%%%%%%%%%%%%%%%%%%%%%%%%%
\section{Introduction}
%%%%%%%%%%%%%%%%%%%%%%%%%%%%%%%%%%%%%%%%%%%%%%%%%%%%%%%%%%%%%%%%%%%%%%%%%%%%%%%%%
The discovery of neutrino oscillations, implying small but non-vanishing neutrino masses, has been the first result of a laboratory experiment contradicting the predictions of the Standard Model (SM). While the non-vanishing masses of neutrinos strongly suggest the existence of new physics fields beyond the SM particle content, our current understanding of the nature of neutrino masses, including the underlying mechanism of neutrino mass generation, is far from complete. One of the central questions is whether neutrinos are Dirac or Majorana fermions. If the latter is the case then they are their own antiparticles, a consequence of which is that the lepton number ceases to be well-defined and can no longer be considered a symmetry of nature. New physics beyond the SM responsible for lepton number violation can potentially be probed at various low- and high-energy experiments. However, the existing literature offers hundreds of theoretically well-motivated models generating neutrino masses with an infinite set of possibilities to mix and match. That begs the question of how to probe and distinguish different scenarios using current and upcoming experiments. One of the avenues of approaching this problem systematically is to take a bottom-up approach and employ the techniques of effective field theory, which allow for a model-independent description of deviations from the SM due to new heavy states in an organized and comprehensive way.

In an effective field theory approach the heavy new physics degrees of freedom are integrated out to construct a low-energy theory consisting of operators involving only light degrees of freedom with the corresponding Wilson coefficient encoding all the effects of the heavy degrees of freedom. The resultant higher dimensional operators involving only light fields are suppressed by powers of the associated heavy new physics scale in the Wilson coefficient. If these new interactions are composed of the light SM fields and respect the full SM gauge symmetry $SU(3)_C\times SU(2)_L\times U(1)_Y$, then the corresponding set of operators is referred to as Standard Model Effective Field Theory (SMEFT). The accidental symmetries in the SM such as lepton and baryon number conservation generally need no longer be preserved by the higher-dimensional operators. When studying the physical effects of such operators at experiments operating at energy scales well below the electroweak symmetry breaking, it is appropriate to use an effective field theory composed of the lighter SM fields respecting the broken SM phase $SU(3)_C\times U(1)_{\text{EM}}$. Such an effective theory is often referred to as Low-Energy Effective Field Theory (LEFT). It is then necessary to run the SMEFT operators down to the electroweak symmetry breaking scale employing the renormalization group equations (RGEs) and match them onto the LEFT operator basis, which can then be in turn run down to the scale at which the matrix elements relevant for a given observable are evaluated. 

Lepton number violating (LNV) effective interactions have been a subject of extensive studies, mainly in connection to their potential role in understanding the nature and mass mechanism of neutrinos; see e.g.~\cite{Babu:2001ex,Deppisch:2017ecm,deGouvea:2007qla} for some surveys of effective LNV interactions up to dimension eleven. Typically, the literature discusses the lowest-dimensional $\Delta L=2$ effective interaction, which is the unique dimension-5 Weinberg operator~\cite{Weinberg:1979sa} of the form $LLHH$, where $L$ denotes the lepton doublet and $H$ is the Higgs doublet. The conventional type-I seesaw mechanism~\cite{Minkowski:1977sc,Gell-Mann:1979vob,Yanagida:1979as,Mohapatra:1979ia,Schechter:1980gr,Mohapatra:1980yp,Schechter:1981cv} is then a natural tree-level UV completion of this operator. Interestingly, SMEFT operators of higher dimensions ($d>5$) could provide the dominant source of Majorana neutrino masses independent of $\Delta L=2$ dimension-5 Weinberg operator or via the Weinberg operator generated at the loop level. Therefore, to probe also these possibilities for neutrino mass generation it is important to take the SMEFT operators of dimensions greater than five into account and study the potential sensitivities of high- and low-energy experiments to these exotic interactions.
The $\Delta L=2$ SMEFT operators are always of odd dimension, as shown in Ref.~\cite{Kobach:2016ami}. Therefore, the lowest dimension beyond the dimension-5 Weinberg operator at which LNV operators appear in SMEFT is dimension 7. A complete set of independent dimension-7 $\Delta L = 2$ SMEFT operators was first presented in Ref.~\cite{Lehman:2014jma}, which was reduced by identifying one redundant operator in Ref.~\cite{Liao:2016hru}. A number of works address some aspects related to $\Delta L=2$ dimension-7 SMEFT operators. For instance, in Ref.~\cite{Cirigliano:2017djv,Deppisch:2017ecm,Cirigliano:2018yza}, they were studied in the context of neutrinoless double beta decay. In Ref.~\cite{Li:2019fhz,Deppisch:2020oyx}, they were discussed in connection to rare Kaon decays, and in Ref.~\cite{Felkl:2021uxi} in the context of $B$-meson decays. The possibility of probing the dimension-5 SMEFT at high energy collider was explored in Ref.~\cite{Fuks:2020zbm}. To the best of our knowledge, no dedicated comprehensive collider study of dimension-7 $\Delta L = 2$ SMEFT operators exists. In light of the above, a comprehensive and dedicated analysis of the complete set of the independent dimension-7 $\Delta L = 2$ SMEFT operators in the context of various low- and high-energy experimental probes is desirable and timely and that is the subject of this work.

In the present manuscript, we provide a first dedicated global comparison of all the independent dimension-7 $\Delta L = 2$ SMEFT operators in the context of different relevant low- and high-energy experiments. We also single out the most interesting operators that can be probed in the near future using the synergy of low- and high-energy experimental tests. To make our approach consistent and comprehensive, we adopt the basis of independent dimension-7 $\Delta L = 2$ SMEFT operators from Ref.~\cite{Lehman:2014jma,Liao:2016hru}, and provide all the relevant matching relations of the Wilson coefficients of dimension-7 $\Delta L = 2$ SMEFT operators with the Wilson coefficients of dimension-6 $\Delta L = 2$ LEFT operators. We also include other relevant matching to dimension-7 and -9 $\Delta L = 2$ LEFT operators. The salient aspects covered in this work are the following: 
\begin{itemize}
\item The results from a systematic analysis of the complete set of dimension-7 $\Delta L = 2$ SMEFT operators in the context of the LHC and the upcoming collider experiments presented in this work are completely new. 
\item We extend upon the existing constraints from non-standard interaction-mediated long baseline neutrino oscillation on the dimension-7 $\Delta L = 2$ SMEFT operators. 
\item  We review and update (where applicable to the latest available experimental results) existing constraints on the dimension-7 $\Delta L = 2$ SMEFT operators from neutrinoless double beta decay, rare meson decays, charged lepton flavor violating lepton decays, neutrino magnetic moment, etc., to provide a comprehensive state of the art analysis of dimension-7 $\Delta L = 2$ SMEFT operators.
\item One of the goals of our study is also to provide guidance to both experimental searches and model building by discussing the possibilities of probing and distinguishing among various dimension-7 $\Delta L = 2$ SMEFT operators.
\end{itemize}
We note that a comprehensive analysis of various possible ultraviolet complete realizations of dimension-7 $\Delta L = 2$ SMEFT operators and their comparison in the context of various low- and high-energy experimental probes are beyond the scope of this work and will be the subject of a separate dedicated work\footnote{Some partial results of such an analysis can be found for example in Refs.~\cite{Cepedello:2017lyo,Cepedello:2017eqf,Herrero-Garcia:2019czj,Liao:2010rx,Krauss:2013gy,Gargalionis:2020xvt,Bonnet:2009ej,Angel:2012ug,Cai:2014kra,Fridell:2022wbz}}~\cite{Fridell:2023}. 

This paper is structured as follows. In Sec.~\ref{sec:framework} we describe the LEFT and SMEFT bases that we use to parametrize the new physics interactions as well as provide their matching. In Sec.~\ref{sec:LHC} we derive constraints on the LNV dimension-7 SMEFT operators from the current and future collider experiments. We also discuss the validity and caveats of the underlying EFT approach. In Sec.~\ref{sec:EFTlimits} we discuss the low-energy observables relevant for constraining the LNV dimension-7 SMEFT operators. In Sec.~\ref{sec:overview} we provide a global view of all the constraints on the LNV dimension-7 SMEFT operators. We also discuss how the interplay and synergy of various observables can lead to the possibilities for distinguishing between the different LNV dimension-7 SMEFT operators. Finally, we conclude in Sec.~\ref{sec:conclusion}.

For facile accessibility of the results presented in this work, at this point, we want to highlight that throughout the manuscript we employ the following notation for the effective operators: the SMEFT operators are denoted by ``$\mathcal{O}$'' and their corresponding Wilson coefficients by capital ``$C$'', while for LEFT operators we use ``O'' and small ``$c$''. The effective operators in $\chi$PT are denoted by small ``o'' and the corresponding Wilson coefficients by $\mathcal{C}$.
%%%%%%%%%%%%%%%%%%%%%%%%%%%%%%%%%%%%%%%%%%%%%%%%%%%%%%%%%%%%%%%%%%%%%%%%%%%%%%%%%
\section{The framework for dimension-7 $\Delta L = 2$ SMEFT operators}
\label{sec:framework}
%%%%%%%%%%%%%%%%%%%%%%%%%%%%%%%%%%%%%%%%%%%%%%%%%%%%%%%%%%%%%%%%%%%%%%%%%%%%%%%%%
The fact that neutrinos, the only electrically neutral fundamental fermions in the SM, acquire tiny masses can be very suggestively put in connection with the only non-anomalous global symmetry of the SM, $U(1)_{B-L}$. Breaking this symmetry at a certain high-energy scale $\Lambda$ allows neutrinos to acquire non-zero and naturally small Majorana masses.
All the renormalizable SM interactions respect $U(1)_{B-L}$ symmetry. Under the assumption that all the new physics responsible for a possible $U(1)_{B-L}$ symmetry breaking lies high above the electroweak scale, all the manifestations of this symmetry breaking at lower energies can then be parametrized in terms of operators of dimension $d\geq 4$ invariant under the SM gauge group. Consequently, the neutrino masses are suppressed by powers of the associated scale $\Lambda$.
Therefore, all high-energy models beyond the SM leading to small Majorana neutrino masses would induce SMEFT operators breaking the $B-L$ number. The operators with $\Delta L=2$ and $\Delta B=0$ are particularly relevant in this context.

\setlength{\extrarowheight}{4pt}
\begin{table}[b]
    \centering
    \begin{tabular}{|>{\raggedright}m{50pt}|>{\raggedright}m{50pt}|>{\centering}m{170pt}|c}
    \cline{1-3}
    Type & $\mathcal{O}$ & Operator & \\[2pt]
    \cline{1-3}
    $\Psi^2 H^4$ & $\mathcal{O}_{LH}^{pr}$ & $\epsilon_{ij}\epsilon_{mn}\lrs{\overline{L_p^c}{}^iL_r^m}H^jH^n\lrs{H^\dagger H}$ & \\[2pt]
     \cline{1-3}
    $\Psi^2H^3D$ & $\mathcal{O}_{LeHD}^{pr}$ & $\epsilon_{ij}\epsilon_{mn}\lrs{\overline{L_p^c}{}^i\gamma_\mu e_r}H^j\lrs{H^miD^\mu H^n}$ & \\[2pt]
     \cline{1-3}
    \multirow{2}{*}{$\Psi^2H^2D^2$} & $\mathcal{O}_{LHD1}^{pr} $ & $\epsilon_{ij}\epsilon_{mn}\lrs{\overline{L^c_p}{}^iD_\mu L^j_r}\lrs{H^mD^\mu H^n}$ & \\[2pt]
     \cline{2-3}
     & $\mathcal{O}_{LHD2}^{pr}$ & $\epsilon_{im}\epsilon_{jn}\lrs{\overline{L^c_p}{}^iD_\mu L_r^j}\lrs{H^mD^\mu H^n}$ & \\[2pt]
      \cline{1-3}
    \multirow{2}{*}{$\Psi^2H^2X$} & $\mathcal{O}_{LHB}^{pr} $ & $g\epsilon_{ij}\epsilon_{mn}\lrs{\overline{L_p^c}{}^i\sigma_{\mu\nu}L_r^m}H^jH^nB^{\mu\nu}$ & \\[2pt]
     \cline{2-3}
     & $\mathcal{O}_{LHW}^{pr} $ & $g'\epsilon_{ij}\lrs{\epsilon \tau^I}_{mn}\lrs{\overline{L_p^c}{}^i\sigma_{\mu\nu}L_r^m}H^jH^nW^{I\mu\nu}$ & \\[2pt]
     \cline{1-3}
    $\Psi^4D$ &  $\mathcal{O}_{\bar{d}uLLD}^{prst}$ & $\epsilon_{ij}\lrs{\overline{d_p}\gamma_\mu u_r}\lrs{\overline{L_s^c}{}^iiD^\mu L_t^j}$ & \\[2pt]
     \cline{1-3}
    \multirow{5}{*}{$\Psi^4H$} & $\mathcal{O}_{\bar{e}LLLH}^{prst}$ & $\epsilon_{ij}\epsilon_{mn}\lrs{\overline{e_p}L_r^i}\lrs{\overline{L_s^c}{}^j L_t^m}H^n$ & \\[2pt]   
     \cline{2-3}
    & $\mathcal{O}_{\bar{d}LueH}^{prst}$ & $\epsilon_{ij}\lrs{\overline{d_p}L_r^i}\lrs{\overline{u_s^c}e_t}H^j$ & \\[2pt]
     \cline{2-3}
     & $\mathcal{O}_{\bar{d}LQLH1}^{prst}$ & $\epsilon_{ij}\epsilon_{mn}\lrs{\overline{d_p}L_r^i}\lrs{\overline{Q_s^c}{}^jL_t^m}H^n$ & \\[2pt]
     \cline{2-3}
    & $\mathcal{O}_{\bar{d}LQLH2}^{prst}$ & $\epsilon_{im}\epsilon_{jn}\lrs{\overline{d_p}L_r^i}\lrs{\overline{Q_s^c}{}^jL_t^m}H^n$ & \\[2pt]
     \cline{2-3}
     & $\mathcal{O}_{\bar{Q}uLLH}^{prst}$ & $\epsilon_{ij}\lrs{\overline{Q_p}u_r}\lrs{\overline{L_s^c}L_t^i}H^j$ & \\[2pt]
     \cline{1-3}
    \end{tabular}
    \caption{List of independent dimension 7 $\Delta L= 2$ SMEFT operators following the basis of Ref.~\cite{Lehman:2014jma,Liao:2016hru}, where $D_\mu \psi^n$ denotes $(D_\mu\psi)^n$ for $\psi \in \{L_i,H\}$, and where $W^{I\mu\nu}$ and $B^{\mu\nu}$ are the $SU(2)_L$ and $U(1)_Y$ field strength tensors, respectively.}
    \label{tab:dim7op}
\end{table}

The $\Delta L = 2$ SMEFT operators are always of odd dimension, as was explicitly shown in Ref.~\cite{Kobach:2016ami}. The Lagrangian summarizing all the relevant interactions can be expressed in the form
\begin{align}
	\mathcal{L} = \mathcal{L}_\text{SM} + {{C}_5}{}\mathcal{O}_5 + \sum_{i}{C_7^i}{}\mathcal{O}_7^i 
	+ \sum_{i}{C_9^i}{}\mathcal{O}_9^i
	+ \dots .
 \end{align}
Here, $\mathcal{L}_\text{SM}$ is the renormalizable SM Lagrangian, $\mathcal{O}_5$ stands for the unique dimension-5 Weinberg operator~\cite{Weinberg:1979sa} and the SMEFT operators of higher dimensions are denoted as $\mathcal{O}_d^i$ with $d$ representing their canonical mass dimension. Each operator comes with its respective Wilson coefficient $C_d^i$ which scales as $\Lambda^{4-d}$ to keep the Lagrangian dimension four.

In this work, we will focus mainly on dimension-7 $\Delta L = 2$ SMEFT operators. The complete set of dimension-7 $\Delta L = 2$ SMEFT operators was first listed in Ref.~\cite{Lehman:2014jma}, which was further improved by reducing one redundant operator and the relevant one-loop anomalous dimension matrix was computed in Ref.~\cite{Liao:2016hru}. We note that in Ref.~\cite{Liao:2020zyx} another new basis for dimension-7 $\Delta L = 2$ SMEFT operators was proposed, which has some of the operators common with Ref.~\cite{Liao:2016hru}. In this work, we follow the basis of Refs.~\cite{Lehman:2014jma,Liao:2016hru}, and summarize the complete basis of 12 independent dimension-7 $\Delta L = 2$ SMEFT operators in Tab.~\ref{tab:dim7op}. Since we follow the basis of Ref.~\cite{Liao:2016hru}, we provide the complete matching for this basis to all independent dimension 6 $\Delta L= 2$ LEFT operators following the basis of Ref.~\cite{Jenkins:2017jig}. While some of the matching relations are the same as Ref.~\cite{Liao:2020zyx}, some of them are new and different due to the difference in basis between Ref.~\cite{Liao:2020zyx} and Ref.~\cite{Liao:2016hru}. To derive and compare the relevant phenomenological constraints from the plethora of low- and high-energy observables, we will often further assume only a single SMEFT operator to be present at a time on top of the SM Lagrangian. 

\begin{table}[t!]
    \centering
    \begin{tabular}{|>{\raggedright}m{40pt}|>{\centering}m{110pt}|>{\centering}m{240pt}|c}
    \cline{1-3}
    ${O}$ & Operator & Matching & \\[2pt]
    \cline{1-3}
    ${O}_{e\nu;LL}^{S,prst}$ & $\lrs{\overline{e_{Rp}}e_{Lr}}\lrs{\overline{\nu_s^c}\nu_t}$ & $\frac{4G_F}{\sqrt{2}} c_{e\nu;LL}^{S,prst}=-{\frac{\sqrt{2}v}{8}}\big(2C_{\bar{e}LLLH}^{prst}+C_{\bar{e}LLLH}^{psrt}+s\leftrightarrow t\big)$ & \\[2pt]
     \cline{1-3}
     ${O}_{e\nu;RL}^{S,prst}$ & $(\overline{e_{Lp}}e_{Rr})(\overline{\nu^c_s}\nu_t)$
 & $\frac{4G_F}{\sqrt{2}} c_{e\nu;RL}^{S,prst}=-{\frac{\sqrt{2}v}{2}}\big( C_{LeHD}^{sr}\delta^{tp}+ C_{LeHD}^{tr}\delta^{sp} \big)$ & \\[2pt]
     \cline{1-3}
     ${O}_{e\nu;LL}^{T,prst}$ & $(\overline{e_{Rp}}\sigma_{\mu\nu}e_{Lr})(\overline{\nu^c_s}\sigma^{\mu\nu}\nu_t)$ & $\frac{4G_F}{\sqrt{2}} c_{e\nu;LL}^{T,prst}=+{\frac{\sqrt{2}v}{32}}\big( C_{\bar{e}LLLH}^{psrt}-C_{\bar{e}LLLH}^{ptrs} \big)$ & \\[2pt]
     \cline{1-3}
 ${O}_{d\nu;LL}^{S,prst}$ & $(\overline{d_{Rp}}d_{Lr})(\overline{\nu^c_s}\nu_t)$
 & $ \frac{4G_F}{\sqrt{2}} c_{d\nu;LL}^{S,prst}=-{\frac{\sqrt{2}v}{8}}V_{xr}\lrs{C_{\bar dLQLH1}^{ptxs}+C_{\bar dLQLH1}^{psxt}}$ & \\[2pt]
     \cline{1-3}
 ${O}_{d\nu;LL}^{T,prst}$ & $(\overline{d_{Rp}}\sigma_{\mu\nu}d_{Lr})
 (\overline{\nu^c_s}\sigma^{\mu\nu}\nu_t)$
 & $\frac{4G_F}{\sqrt{2}} c_{d\nu;LL}^{T,prst}=-\frac{\sqrt{2}v}{32}V_{xr}\lrs{C_{\bar dLQLH1}^{ptxs}-C_{\bar dLQLH1}^{psxt}}$ & \\[2pt]
     \cline{1-3}
${O}_{u\nu;RL}^{S,prst}$ & $(\overline{u_{Lp}}u_{Rr})(\overline{\nu^c_s}\nu_t)$
 & $\frac{4G_F}{\sqrt{2}} c_{u\nu;RL}^{S,prst}=+{\frac{\sqrt{2}v}{4}}\big( C_{\bar QuLLH}^{prst}+C_{\bar QuLLH}^{prts}\big)$
 & \\[2pt]
     \cline{1-3}
 ${O}_{du\nu e;LL}^{S,prst}$ & $(\overline{d_{Rp}}u_{Lr})
 (\overline{\nu^c_s}e_{Lt})$
 & $\frac{4G_F}{\sqrt{2}} c_{du\nu e;LL}^{S,prst}=+{\frac{\sqrt{2}v}{8}}\lrs{2C_{\bar dLQLH1}^{ptrs}+C_{\bar dLQLH2}^{ptrs}-C_{\bar dLQLH2}^{psrt}}$
  & \\[2pt]
     \cline{1-3}
 ${O}_{du\nu e;RL}^{S,prst}$ & $(\overline{d_{Lp}}u_{Rr})
 (\overline{\nu^c_s}e_{Lt})$
 & $\frac{4G_F}{\sqrt{2}} c_{du\nu e;RL}^{S,prst}=+{\frac{\sqrt{2}v}{2}}V_{xp}^*C_{\bar QuLLH}^{xrts}$
  & \\[2pt]
     \cline{1-3}
     ${O}_{du\nu e;LL}^{T,prst}$ & $(\overline{d_{Rp}}\sigma_{\mu\nu}u_{Lr})
 (\overline{\nu^c_s}\sigma^{\mu\nu}e_{Lt})$
 & $\frac{4G_F}{\sqrt{2}} c_{du\nu e;LL}^{T,prst}=+\frac{\sqrt{2}v}{32}\lrs{2C_{\bar dLQLH1}^{ptrs}+C_{\bar dLQLH2}^{ptrs}+C_{\bar dLQLH2}^{psrt}}$
  & \\[2pt]
     \cline{1-3}
   ${O}_{du\nu e;LR}^{V,prst}$ & $(\overline{d_{Lp}}\gamma_\mu u_{Lr})(\overline{\nu^c_s}\gamma^\mu e_{Rt})$
 & $\frac{4G_F}{\sqrt{2}} c_{du\nu e;LR}^{V,prst}=+{ \frac{\sqrt{2}v}{2} }V_{rp}^*C_{LeHD}^{st}$
  & \\[2pt]
     \cline{1-3}
  ${O}_{du\nu e;RR}^{V,prst}$ & $(\overline{d_{Rp}}\gamma_\mu u_{Rr})(\overline{\nu^c_s}\gamma^\mu e_{Rt})$
 & $\frac{4G_F}{\sqrt{2}} c_{du\nu e;RR}^{V,prst}=+\frac{\sqrt{2}v }{4}C_{\bar dLueH}^{psrt}$
  & \\[2pt]
     \cline{1-3}
     ${O}_{d\nu;RL}^{S,prst}$ & $(\overline{d_{Lp}}d_{Rr})(\overline{\nu^c_s}\nu_t)$ &  & \\[2pt]
     \cline{1-2}
     $ {O}_{u\nu;LL}^{S,prst}$ & $(\overline{u_{Rp}}u_{Lr})(\overline{\nu^c_s}\nu_t)$ &  Not induced by $d=7$ $\Delta L=2$ SMEFT operators & \\[2pt]
     \cline{1-2}
     ${O}_{u\nu;LL}^{T,prst}$ & $(\overline{u_{Rp}}\sigma_{\mu\nu}u_{Lr})
 (\overline{\nu^c_s}\sigma^{\mu\nu}\nu_t)$ & & \\[2pt]
     \cline{1-3}
    \end{tabular}
    \caption{List of all independent dimension-6 $\Delta L= 2$ LEFT operators following the basis of Ref.~\cite{Jenkins:2017jig} and their matching relations with the Wilson coefficients of dimension-7 $\Delta L = 2$ SMEFT operators. Note that here we follow the convention that the CKM matrix needs to be multiplied with the left-handed down-type quark to diagonalize the SM charged current interactions. We also point out that in our convention the LEFT Wilson coefficients are dimensionless with the Lagrangian being normalized by an overall factor $4 G_F/\sqrt{2}$.}
\label{tab:left}
\end{table}
\setlength{\extrarowheight}{0pt}

Furthermore, in order to be able to identify and quantitatively assess manifestations of these operators at energies below the electroweak symmetry breaking it is often useful to employ the Low-Energy Effective Theory (LEFT), often also referred to also as Weak Effective Theory (WET). The LEFT does not respect the whole SM gauge group but is $SU(3)_c\times U(1)_{\text{EM}}$ invariant. In such cases, we use the bottom-up EFT approach for constraining the LNV SMEFT operators. Given the experimental limits on an observable from the non-observation of lepton number violation, first, we derive constraints on the LEFT operators corresponding to the low energy observable. Next assuming that a single $\Delta L = 2$ LEFT operator dominantly contributes to the observable we use the matching between LEFT and SMEFT operators to constrain the relevant SMEFT operators. If multiple SMEFT operators match into a single LEFT operator then to disentangle the different SMEFT operators we again use the assumption of single SMEFT operator dominance. Therefore, our limits in such scenarios represent the most stringent possible constraints on the LNV SMEFT operators. In the case where multiple SMEFT operators are generated simultaneously by a UV completion, these limits can be weaker in the presence of cancellations. For certain observables, e.g. the rare decay $K\rightarrow \pi\nu\nu$, additional matching onto Chiral EFT ($\chi$EFT) at the chiral symmetry breaking scale $\Lambda_\chi\sim 2$ GeV is also employed. The complete basis of dimension-6 $\Delta L = 2$ LEFT operators relevant for our discussion and the matching relations between the dimension-6 $\Delta L = 2$ LEFT Wilson coefficients to the Wilson coefficients of dimension-7 $\Delta L = 2$ SMEFT operators are presented in Tab.~\ref{tab:left}.

%%%%%%%%%%%%%%%%%%%%%%%%%%%%%%%%%%%%%%%%%%%%%%%%%%%%%%%%%%%%%%%%%%%%%%%%%%%%%%%%%
%%%%%%%%%%%%%%%%%%%%%%%%%%%%%%%%%%%%%%%%%%%%%%%%%%%%%%%%%%%%%%%%%%%%%%%%%%%%%%%%%
\section{Probing $\Delta L = 2$ dimension-7 SMEFT operators at colliders}~\label{sec:LHC}
%%%%%%%%%%%%%%%%%%%%%%%%%%%%%%%%%%%%%%%%%%%%%%%%%%%%%%%%%%%%%%%%%%%%%%%%%%%%%%%%%
At proton-proton colliders, LNV can potentially be probed by searching for processes with final state charged leptons and no missing energy, where the latter condition is needed in order to rule out final state neutrinos which also carry a non-zero lepton number. The number of positively charged leptons must also be different from the number of negatively charged ones. Here we focus on processes that violate the lepton number by two units, $\Delta L=2$, as these are most directly relevant for Majorana neutrino masses. In an EFT analysis using the dimension-5 operator, such processes~\cite{Fuks:2020zbm} have previously been shown to have the possibility of being constrained to $\Lambda < 8.3$~TeV, assuming an integrated luminosity of 300 fb$^{-1}$ at the LHC. Here we will focus on the dimension-7 $\Delta L = 2$ SMEFT operators.

\subsection{Same-sign dilepton plus dijet production at the LHC}

LNV can be studied at colliders through same-sign dilepton plus dijet final states with no missing energy. At the LHC, this corresponds to the process\footnote{ The LHC signal of like-sign dileptons in the context of one of the most minimal model realisation involving Majorana fields was studied in~\cite{Pilaftsis:1991ug}. On the other hand, in Left-Right Symmetric Models~\cite{Pati:1974yy,Mohapatra:1974gc,Senjanovic:1975rk,Hati:2018tge} with a massive Majorana fermion $N$ and heavy $W_R$ bosons, this interaction can be mediated via the Keung-Senjanović process~\cite{Keung:1983uu}, where a heavy $W_{R}$ produced in a proton-proton collision decays to two leptons and two jets via intermediate $N$ and $W_R$. For a review on the topic see for instance~\cite{Deppisch:2015qwa}. A similar topology can also arise in models with Majorana fermions and additional heavy scalars. In such scenarios, care must be taken not to let the same scalar have both quark-quark and quark-lepton couplings, as this could lead to proton decay at rates that have been excluded by Super-Kamiokande~\cite{Super-Kamiokande:2020wjk}. See also Refs.~\cite{Fornal:2017xcj,Hambye:2017qix,Hati:2018cqp,Fridell:2021gag} for baryon number violating phenomenology involving diquark and leptoquark couplings.}
\begin{equation}
    p p \to \ell^\pm \ell^\pm j j,
\end{equation}
where $\ell\in\{e,\mu,\tau\}$. Collider signatures for some UV-completions of the dimension-7 operators $\mathcal{O}_{LH}$ and $\mathcal{O}_{LeHD}$ are discussed in Ref.~\cite{Cepedello:2017lyo} and Ref.~\cite{delAguila:2012nu}, respectively. From a model-independent perspective, LNV effective operators can be constrained by LHC searches using the non-observation of same-sign dilepton events\footnote{This can also be used to falsify high-scale leptogenesis given a potential signal at colliders~\cite{Deppisch:2013jxa}. See also Refs.~\cite{Frere:2008ct,Deppisch:2013jxa,BhupalDev:2014hro,Dhuria:2015wwa,Dhuria:2015cfa,BhupalDev:2015khe} for some relevant discussions in the instance of a Left-Right Symmetric Model.}. In Ref.~\cite{Aoki:2020til,Fuks:2020att} collider signals of the dimension-5 operator $\ell^\pm\ell'^\pm W^\mp W^\mp$ were discussed in the context of LNV and neutrino masses. From a SMEFT perspective, such an operator can arise from the dimension-7 SMEFT operator of type $\Psi^2H^2D^2$, c.f. Table \ref{tab:dim7op}. In Refs.~\cite{Fuks:2020zbm} collider signatures of the dimension-5 Weinberg operator $\mathcal{O}_{LH}^{(5)}=\epsilon_{ij}\epsilon_{mn}\lrs{\overline{L^c}{}^iL^m}H^jH^n$, as well as the type-I seesaw UV-completion, were studied in the muon channel within a range of parameters for which the EFT description is valid. 

In what follows, we proceed to derive new limits on the LNV scale of all relevant dimension-7 $\Delta L = 2$ SMEFT operators based on collider searches. We obtain the constraints based on the non-observation of same-sign dilepton plus dijet signals at the ATLAS experiment for LHC Run 2~\cite{ATLAS:2023cjo}. We implement the LNV operators using \textsc{FeynRules}~\cite{Alloul:2013bka} and calculate the LO cross sections with \textsc{MadGraph5\_aMC@NLO}~\cite{Alwall:2014hca} using the basic generator level cuts provided by \textsc{MadGraph5\_aMC@NLO} (with the inclusion of the cuts $H_T> 100$ GeV, $m_{\mu^{1}\mu^{2}}>100$ GeV, and $m_{j^{1}j^{2}}>25.6$ GeV). The object selection cuts and the signal region cuts used in this work for event selection are presented in Tab.~\ref{tab:cuts}. We have also made sure that all of our selected events are consistent with the track-to-vertex association requirements quantified in terms of the longitudinal and transverse impact parameters $|z_0 \sin \theta < 5|$~mm and $|d_0 <1|$ {\textmu}m, respectively. In the latter requirement we take the conservative benchmark $\sigma(d_0)\sim 0.2$ {\textmu}m following Refs.~\cite{ATLAS:2017oro,Magliocca:2021bfg}.  The cuts for the $\sqrt{s}=13$~TeV case are identical to those used by the ATLAS collaboration in Ref.~\cite{ATLAS:2023cjo}\footnote{We note that we solely focus on the resolved channel of Ref.~\cite{ATLAS:2023cjo}, since the boosted channel case is particularly relevant in the particular scenario where a large hierarchy in the mass in the decay chain of the initial heavy state is present e.g. in the case of Keung-Senjanović process, a large mass difference between the right-handed $W$-boson $W_R$ and the right-handed neutrino $N_R$.}, and the PDF set \textsc{NNPDF30} provided by \textsc{LHAPDF6}~\cite{Buckley:2014ana} is used.  Hadronization is handled by \textsc{Pythia8}~\cite{Sjostrand:2014zea} and detector simulation is handled by \textsc{Delphes3}~\cite{deFavereau:2013fsa}. 

\begin{table}[]
    \centering
    \setlength{\extrarowheight}{8pt}
    \begin{tabular}{| l l |}
    \hline
    \specialrule{2pt}{2pt}{0pt}
         \multicolumn{2}{| c |}{\cellcolor{newgray2} \raisebox{2pt}{Cuts for $p p \to \mu^\pm\mu^\pm j j$ at $\sqrt{s}=13$ TeV} } \\
         \hline
         Object selection cuts & \\
         \hline
         $p_T^{\mu^{1(2)}}>25$ GeV & $p_T^{j^{1(2)}}>20$ GeV \\
         $|\eta^{\mu^{1(2)}}|<2.5$ & $|\eta^{j^{1(2)}}|<2.5$ \\
         \hline
         Track-to-vertex association cuts & \\
           \hline
         $|z_0 \sin \theta|< 5$ mm & $|d_0|< 1$ {\textmu}m\\
         \hline
         Signal region cuts & \\
         \hline
         $p_T^{\mu^{\text{leading}}}>40$ GeV & $p_T^{j^{1(2)}}>100$ GeV \\
         $H_T> 400$ GeV  & $\Delta R_{\mu\mu}< 3.9$ GeV \\
          \raisebox{4pt}{$m_{\mu^{1}\mu^{2}}>400$ GeV} & \raisebox{4pt}{$m_{j^{1}j^{2}}>110$ GeV}\\
    \specialrule{2pt}{0pt}{0pt}
          \multicolumn{2}{| c |}{\cellcolor{newgray2} \raisebox{2pt}{Cuts for $\sqrt{s}=100$ TeV}}\\
          \hline
            $p_T^{\mu^{1(2)}}>30$ GeV & $p_T^{j^{1(2)}}>100$ GeV \\
         $H_T> 400$ GeV  & $\sum_i p_T^{\mu_i}> 400$ GeV \\
         $|\eta^{j^{1(2)}}|<2.0$  & $|\eta^{\mu^{1(2)}}|<2.5$ \\
           $|\eta^{\mu^{1(2)}}|<4.5$ & $|\eta^{j^{1(2)}}|<4.0$ \\
           \raisebox{2pt}{$m_{\mu^1\mu^2}>700$ GeV} & \raisebox{2pt}{$m_{j^1j^2}>200$ GeV } \\
    \specialrule{2pt}{0pt}{2pt}
           \hline
    \end{tabular}
    \caption{Cuts used for the evaluation of $\sigma(p p \to j j \mu^\pm \mu^\pm)$. The cuts are taken from Ref.~\cite{ATLAS:2023cjo} in order to constrain the LNV scale using the data therein. The cuts for FCC-hh are based on the analysis of Ref.~\cite{ATLAS:2018dcj}, where the cuts on the rapidities and invariant masses are scaled up in order to account for the increased center-of-mass energy.}
    \label{tab:cuts}
\end{table}

\begin{figure}[h]
    \centering
    \includegraphics[width=0.69\textwidth]{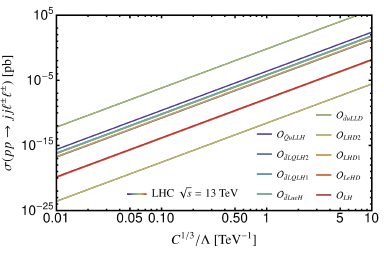}\\
    \includegraphics[width=0.69\textwidth]{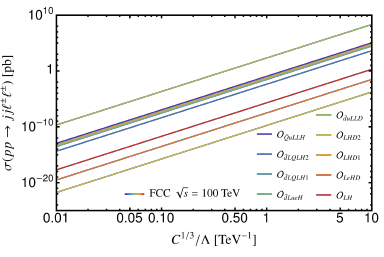}
    \caption{Cross sections for the process $p p \to j j \mu^\pm\mu^\pm$ at the LHC (top) and FCC-hh (bottom) using the cuts in Tab.~\ref{tab:cuts}.}
    \label{fig:LHCxs}
\end{figure}

The cross sections for all 7-dimensional $\Delta L = 2$ LNV operators that give rise to signal events for the process $p p \to \mu^\pm \mu^\pm j j$ are presented in Tab.~\ref{tab:op_lhc_HD} for both the LHC with $\sqrt{s}=13$ TeV and FCC-hh with $\sqrt{s}=100$~TeV. We focus on the muon channel for two reasons: the background is expected to be smaller than that of final state electrons, and the constraints on electron-flavour LNV operators are severely constrained by \ovbb decay experiments, while those of the muon flavor structure are not. The cross-sections in Tab.~\ref{tab:op_lhc_HD} are shown for $C^{1/3}=\Lambda^{-1}=1$~TeV$^{-1}$, where $C$ is the Wilson coefficient of a dimension-7 LNV operator, and $\Lambda$ is the corresponding LNV scale. The scaling with respect to increasing or decreasing $\Lambda$ is found to be described by
\begin{equation}
\label{eq:equation}
    \sigma(p p \to \mu^\pm \mu^\pm j j) = \sigma_0\times \left(\frac{100\text{ TeV}}{\Lambda}\right)^{1/6}
\end{equation}
for all operators, as can be seen in Fig.~\ref{fig:LHCxs}, where $\sigma_0$ is the cross section for $\Lambda = 100$~TeV.
We use the massless approximation for the first and second-generation SM fermions and consider jets to not include bottom quarks. Subsequently, we find exclusion limits corresponding to the different operators using the asymptotic formula
\begin{equation}
\label{eq:asmplikelihood}
    Z_0 = \sqrt{2\lr{s-b\ln\lr{\frac{s+b}{b}}}},
\end{equation}
where $b$ is the number of background events, and $s=\mathcal{L}\times \sigma(p p \to \mu^\pm \mu^\pm j j)$ is the number of expected signal events for a given luminosity $\mathcal{L}$, and $Z_0$ is the statistical significance. The ATLAS collaboration reports an expected number of background events $b=10$~\cite{ATLAS:2023cjo} for an integrated luminosity of $\mathcal{L}_0=139$ fb$^{-1}$ using the same cuts as presented in Tab.~\ref{tab:cuts}. Note that in Ref.~\cite{ATLAS:2023cjo} a $\sim$15\% relative uncertainty is reported for the signal region that is relevant for our study. We find that, when applied to the number of background events, this uncertainty has a $\sim$1\% level effect on the corresponding value of the new physics scale $\Lambda$. The reason for this small effect is the large power of $\Lambda$ in the expression for the signal cross section $\sigma(p p \to \mu^\pm \mu^\pm j j) \propto \Lambda^{-6}$. Only a small change in $\Lambda$ is needed to compensate for the change in the number of background events. Furthermore, we expect possible additional background channels generated by the operator we introduce to be insignificant due to suppression by the LNV scale as well as being largely rejected by the phase space cuts, and we therefore neglect them. Following Ref.~\cite{ATLAS:2023cjo} we therefore use $b=10\times \mathcal{L}/(139 \text{fb}^{-1})$ for a final state muon-flavour same-sign dilepton plus dijet. Solving Eq.~\eqref{eq:asmplikelihood} for $Z_0=1.96$ we then find the 95\% C.L.\  likelihood limits on $\Lambda_\text{LNV}$ for the different operators, which we show in Tab.~\ref{tab:op_lhc_HD}.

For the projection of the constraints from FCC-hh~\cite{FCC:2018vvp}, we base our analysis on the search from Ref.~\cite{ATLAS:2018dcj}. We assume a simple rescaling of the background events, use the increased lumminosity, and employ the cuts presented in Tab.~\ref{tab:cuts}. We show the projected future 95\% C.L.\ limits for $p p \to \mu^\pm \mu^\pm j j$ at an integrated luminosity 30 ab$^{-1}$ in FCC-hh.

\begin{figure}
    \centering
    \includegraphics[width=0.30\textwidth]{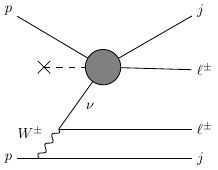}\hspace{5mm}
    \includegraphics[width=0.30\textwidth]{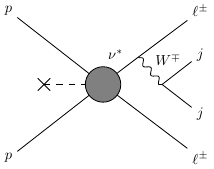}\hspace{5mm}
    \includegraphics[width=0.30\textwidth]{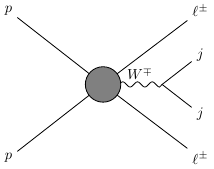}
    \caption{\textbf{Left:} Diagram for same-sign dilepton plus dijet production at colliders for LNV dimension-7 operators containing four fermions with an incoming neutrino. \textbf{Center:} Same but with an outgoing neutrino. \textbf{Right:} Same-sign dilepton plus dijet production at colliders for operator $\mathcal{O}_{\bar duLLD}$.}
    \label{fig:4fdiagrams}
\end{figure}

In Tab.~\ref{tab:op_lhc_HD}, the operators of type $\Psi^4H$ (i.e.\ those with four fermion legs) all have similar cross sections, as expected~\footnote{Note that in Fig.~\ref{fig:4fdiagrams} the vev insertion could in principle be replaced by a final state physical Higgs boson $h$ production. However, due to the large mass of $h$, this mode is severely suppressed.}. The difference between them comes from a number of effects. Given the $SU(2)_L$ structure of an operator there is a different number of $p p \to j j \mu^\pm\ell^\pm$ diagrams that can be drawn, leading to different total cross sections. For instance, the operator $\mathcal{O}_{\bar dLQLH2}$ has more diagrams than $\mathcal{O}_{\bar dLQLH1}$, since the $SU(2)_L$ indices of the lepton doublets are contracted with each other rather than the quark- and Higgs doublets. This leads to twice as many possibilities to generate diagrams of the topology shown in Fig.\ref{fig:4fdiagrams} (left). However, these $SU(2)_L$ contractions come with different signs due to the Levi-Civita tensor that contracts them, effectively reducing the cross section for $\mathcal{O}_{\bar dLQLH2}$ in diagrams where both leptons are in the final state, i.e.\ diagrams with the topology shown in Fig.~\ref{fig:4fdiagrams} (center). Such diagrams become relatively more dominant at  FCC-hh than at the LHC, due to the increased range of energy available to produce an off-shell neutrino~$\nu^*$, leading to the fact that the $\mathcal{O}_{\bar dLQLH1}$ and $\mathcal{O}_{\bar dLQLH2}$ cross sections compare differently in the two experiments. Another factor that affects the cross-section in four-fermion operators is the number of possible up-type quarks it contains. Since the prevalence of up-type quarks in a proton is greater than that of down-type quarks, generally speaking, more up-type quarks lead to a higher cross-section, as can be seen for operator $\mathcal{O}_{\bar QuLLH}$. 

\begin{table}[]
    \centering
    \setlength{\extrarowheight}{6pt}
    \begin{tabular}{|m{15mm} | m{20mm} | m{23mm} | m{12mm} | m{12mm} |}
    \hline
    \specialrule{2pt}{2pt}{0pt}
   \cellcolor{newgray2}  & \multicolumn{2}{c|}{\cellcolor{newgray2} $\sigma\lrs{p p \to \mu^\pm \mu^\pm j j}$ (pb) }  &  \cellcolor{newgray2} $\Lambda_{\text{LNV}}$ & \cellcolor{newgray2} $\Lambda_{\text{LNV}}^\text{future}$ \\
     %\cline{2-5}
     \cellcolor{newgray2} \multirow{-2}*{Operator}  & \cellcolor{newgray2}  \centering LHC &\cellcolor{newgray2}  \centering FCC  &\cellcolor{newgray2} \multirow{-1}*{  \centering [TeV]} &\cellcolor{newgray2} \multirow{-1}*{ \centering [TeV]}  \\
      \hline
       \raisebox{2pt}{$\mathcal{O}_{\bar QuLLH}$}  &  \centering $2.4\times 10^{-4}$ &  \centering $0.11$ & \centering $1.4$ & \hspace{5pt} $5.4$ \\
      \hline
      \raisebox{2pt}{$\mathcal{O}_{\bar dLQLH2}$} & \centering $1.5\times 10^{-5}$&  \centering $4.3\times 10^{-3}$ & \centering $0.90$ & \hspace{5pt} $3.1$ \\
      \hline
      \raisebox{2pt}{$\mathcal{O}_{\bar dLQLH1}$}  & \centering  $6.9\times 10^{-5}$ & \centering  $0.030$  & \centering $1.1$& \hspace{5pt} $4.3$\\
      \hline
      \raisebox{2pt}{$\mathcal{O}_{\bar dLueH}$}  & \centering $5.7\times 10^{-5}$ & \centering $0.035$ & \centering $1.1$ & \hspace{5pt} $4.5$ \\
      \hline
      \raisebox{2pt}{$\mathcal{O}_{\bar duLLD}$}  & \centering $0.64$ & \centering $210$ & \centering $5.0$ & \hspace{5pt} $19$ \\
      \hline
      \raisebox{2pt}{$\mathcal{O}_{LHD2}$} & \centering $2.7\times 10^{-12}$ & \centering $1.7\times 10^{-10}$ & \centering $0.075^*$ & \hspace{5pt} $0.18$ \\
      \hline
      \raisebox{2pt}{$\mathcal{O}_{LHD1}$}  & \centering $1.9\times 10^{-5}$ & \centering $0.061$ & \centering $1.1$ & \hspace{5pt} $4.9$ \\
      \hline
      \raisebox{2pt}{$\mathcal{O}_{LeHD}$}  & \centering $1.2\times 10^{-8}$ & \centering $3.1\times 10^{-8}$ & \centering $0.21^*$ & \hspace{5pt} $0.44$ \\
      \hline
      \raisebox{2pt}{$\mathcal{O}_{LH}$}  & \centering $1.5\times 10^{-8}$ & \centering $2.0\times 10^{-6}$ & \centering $0.35^*$ & \hspace{5pt} $0.87$ \\
      \specialrule{2pt}{0pt}{2pt}
      \hline
    \end{tabular}
    \caption{Cross sections and limits on the LNV scale for muon-flavour same-sign dilepton searches at the LHC and FCC-hh using the cuts in Tab.~\ref{tab:cuts}. In the second and third columns, the cross sections for the processes $p p \to \mu^\pm \mu^\pm j j$ are given. Operators $\mathcal{O}_{\bar eLLLH}$, $\mathcal{O}_{LHW}$ and $\mathcal{O}_{LHB}$ are not included since they do not lead to the given signals at leading order. The cross sections are calculated using Eq.~\eqref{eq:equation} assuming $1/\Lambda=1$ TeV$^{-1}$. The scales in the last and second-to-last columns are shown in TeV, where $\Lambda_\text{LNV}$ is the scale excluded at 95\%~C.L. by the ATLAS experiment at the LHC~\cite{ATLAS:2023cjo}, and $\Lambda_\text{LNV}^\text{future}$ is the scale that can potentially be excluded at 95\% C.L. at FCC-hh~\cite{FCC:2018vvp} assuming an integrated luminosity of 30 ab$^{-1}$. Note that some LNV limits fall below or very close to the limit imposed by the validity of the EFT, and should therefore be only taken as indicative rather than strict limits. These limits are marked by asterisks.}
    \label{tab:op_lhc_HD}
\end{table}

Note that some of the LNV limits given in Tab.~\ref{tab:op_lhc_HD} fall below or very close to the limit of the validity of the EFT for all perturbative values of the coupling constants, and should therefore be taken only as indicative rather than strict limits. We have marked these limits with an asterisk. The other LNV limits listed in Tab.~\ref{tab:op_lhc_HD} fall within the region of validity for the EFT for at least some values of the couplings, as can also be seen in Fig.~\ref{fig:EFTcontour}.

One operator that particularly distinguishes itself is $\mathcal{O}_{\bar duLLH}$, which leads to a cross-section that is significantly higher than in the case of the other operators. This is due to a unique topology where a $W$ boson is produced in the final state, as shown in Fig.~\ref{fig:4fdiagrams} (right).

Three out of the twelve $\Delta L = 2$ dimension-7 SMEFT operators, namely $\mathcal{O}_{\bar eLLLH}$, $\mathcal{O}_{LHB}$ and $\mathcal{O}_{LHW}$, do not give rise to the process $p p \to \mu^\pm \mu^\pm j j$ at tree-level. For $\mathcal{O}_{\bar eLLLH}$ this process can occur at 1-loop order by closing one lepton doublet $L$ and the singlet $e$ in a loop, leading to a suppression roughly by a factor $\sum_i y_\ell^i \times (1/16\pi^2)\approx 2\times 10^{-4}$ relative to the other four-fermion operators, assuming flavor-universal Wilson coefficients, where $y_\ell^i$ is the SM lepton Yukawa coupling for flavor $i$. Operators $\mathcal{O}_{LHB}$ and $\mathcal{O}_{LHW}$ are both flavors antisymmetric and therefore lead to charged lepton flavor violating processes in addition to lepton number violation. The final state in searches involving these operators must therefore involve a same-sign lepton pair $\ell^\pm\ell'^{\pm}$ with $\ell\neq\ell'$. 
%%%%%%%%%%%%%%%%%%%%%%%%%%%%%%%%%%%%%%%%%%%%%%%%%%%%%%%%%%%%%%%%%%%%%%%%%%%%%%%%%%%%%%%%%%%%%%%%%%%%%%%%%%%%
\subsection{Applicability and caveats of the EFT collider limits}\label{sec:EFTlimits}
It is well known that for EFT to be a valid description of a given process, the momentum exchange scales involved must be sufficiently small as compared to the heavy new physics scale involved. This can simply be seen from the expansion of a heavy new physics mediator field of mass $M_\text{med}$, which gets integrated out in the EFT description,
\begin{equation}
\label{eq:medexp}
    \frac{g^2}{Q^2-M_\text{med}^2}=-\frac{g^2}{M_\text{med}^2}\lr{1+\frac{Q^2}{M_\text{med}^2}+\mathcal{O}\lr{\frac{Q^4}{M_\text{med}^4}}},
\end{equation}
where $Q$ is the exchanged momentum. Note that Eq.~\ref{eq:medexp} describes an $s$-channel interaction. The $t$ or $u$-channel momentum exchange can always be Fierz rotated into a set of $s$-channel momentum exchanges. Therefore similar results should follow in these cases. If the value of $Q$ is large compared to the mediator mass $M_\text{med}$, this expansion series does not converge, and the EFT description becomes invalid. We, therefore, require the condition $M_\text{med}^2>Q^2$ to hold in order for the EFT description to be valid. For low-energy observables such as $0\nu\beta\beta$ decay, this does not pose a problem as the new physics scales being integrated out in the EFT are much greater than the energy of the process. However, for high-energy observables such as the LHC, where $Q$ can be quite large, the validity of an EFT description must be taken into account to assess the reliability of the relevant constraints~\cite{Buchmueller:2013dya,Busoni:2013lha}, as we proceed to discuss. 

Following the prescription in Ref.~\cite{Busoni:2013lha}, the momentum transfer for the processes of our interest at the LHC can be roughly estimated by the squared root of the averaged squared momentum transfer in proton collisions weighted with the parton distribution functions (PDFs). For the EFT treatment to be valid $M_\text{med}$ must be larger than the estimated momentum transfer. Since there are many different topologies entering into the LHC constraints there is no unique way to calculate an average momentum transfer that is applicable to all diagrams. Therefore, taking the case where momentum transfer should be the highest, we consider LNV scattering diagrams in which the partons directly enter into the LNV operator, such that all vertices external to the operator appear to its right, as illustrated in Fig.~\ref{fig:4fdiagrams} (right). The constraint on the momentum transfer from such diagrams should be the highest out of all types of topologies and therefore the limits we obtain therefrom can be viewed as the most conservative limit. Considering the collision of two partons $q_1$ and $\bar{q}_2$ with PDFs $f_{q_1}$ and $f_{\bar{q}_2}$, the transferred momentum $Q$ is given by
\begin{equation}
    Q=\sqrt{x_1x_2}\sqrt{s},
\end{equation}
where $x_1$ and $x_2$ are the fractions of the proton's momentum carried by each of the two colliding partons respectively, and $\sqrt{s}=13$ TeV is the center-of-mass energy. Assuming four quark flavors, the average squared momentum transfer $\langle Q^2\rangle$ in the type of diagram shown in Fig.~\ref{fig:4fdiagrams} (right) can then be expressed as\footnote{In addition to calculating the average value of $Q$ analytically it is possible to obtain it by generating events in Monte Carlo simulations using \textsc{MadGraph5\_aMC@NLO}. We have checked that doing so leads to the same conclusions as presented in the text.}
\begin{equation}
    \langle Q^2\rangle = \frac{\sum_{q_1=u,c}\sum_{q_2=d,s}\int dx_1 dx_2\lr{f_{q_1}(x_1)f_{\bar{q}_2}(x_2)+f_{q_1}(x_2)f_{\bar{q}_2}(x_1)}\Theta(Q-Q_0)Q^2}{ \sum_{q_1=u,c} \sum_{q_2=d,s}\int dx_1 dx_2\lr{f_{q_1}(x_1)f_{\bar{q}_2}(x_2)+f_{q_1}(x_2)f_{\bar{q}_2}(x_1)}\Theta(Q-Q_0)},
\end{equation}
where $\Theta$ is the Heaviside function and $Q_0$ is the minimum momentum transfer for the LNV process, which we take to be the sum of the minimum invariant masses of the lepton and jet pairs $Q_0=510$ GeV that we use in the cuts, c.f.\ Tab.~\ref{tab:cuts}. We also use $Q_0$ as the renormalization scale for the PDFs. Using the PDF set \textsc{NNPDF30}~\cite{Buckley:2014ana} we obtain $\langle Q^2\rangle^{1/2}\approx 900$ GeV. This is a result consistent with the literature, see e.g.~\cite{Busoni:2013lha}. However, we would like to point out that for increasing values of $Q_0$, the value of $\langle Q^2\rangle^{1/2}$ also increases monotonically. For instance, an invariant mass $Q_0\geq 1$ TeV leads to $\langle Q^2\rangle^{1/2}\sim \mathcal{O}(\text{few})$ TeV, which would require $M_\text{med}>\mathcal{O}(\text{few})$ TeV for the EFT approach to be valid. Therefore, a constraint derived using EFT from experimental data $\Lambda<1$ TeV is generally only valid provided that the couplings are relatively large to allow for a relatively large $M_\text{med}$. We now proceed to discuss this point in more detail below. 

In a decay or scattering with four external fermions the mediator in Eq.~\eqref{eq:medexp} is coupled to two vertices, for which we denote the coupling constants by $\lambda_1$ and $\lambda_2$. Furthermore, our effective operator being dimension-7, an additional Higgs vacuum expectation value must be involved to obtain the dilepton plus dijet signal, c.f. Fig.~\ref{fig:4fdiagrams} (left) and (center). UV completions with a single scalar mediator would then involve a trilinear coupling to the SM Higgs. Assuming for simplicity that this dimensionful coupling is of the order of the new mediator mass\footnote{Some dimension-7 $\Delta L=2$ SMEFT operators can only be UV-completed at tree level if at least two heavy SM-invariant mediator fields are present. However, the diagrams being induced by dimension-7 LNV operators that are most relevant for same-sign dilepton plus dijet searches at the LHC do involve a single mediator in the mass basis, even if two new physics fields are needed in the interaction basis. This is because a Higgs field that is part of the operator can provide mass insertion between the two new physics fields, leading to mixing. In the mass basis, this mediator is then written as a single field.}, this gives a relation for the LNV scale as
\begin{equation}
\label{eq:MmedL}
    \frac{\lambda_1\lambda_2}{M_\text{med}^3}=\frac{1}{(\Lambda_\text{LNV})^3}.
\end{equation}
 Taking the perturbative limit\footnote{A similar argument can be made using the unitarity of the $S$-matrix rather than the constraint that the interaction is perturbative, see e.g.\ Ref.~\cite{Shoemaker:2011vi}.} $\lambda_1, \lambda_2 < 4\pi$, the lower limit on the LNV scale for a given mediator mass scale is given by the constraint $M_\text{med}$ is $\Lambda_\text{LNV}>M_\text{med}/(4\pi)^{2/3}\approx 0.185\times M_\text{med}$. Combined with the momentum constraint $\langle Q^2\rangle^{1/2}\gtrsim 900$~GeV this leads to the limit on possible LNV scales for dimension-7 operators being probed at the LHC to be given by $\Lambda_\text{LNV}\gtrsim 170$~GeV. Note however that this limit only applies to couplings very close to the perturbative limit, for smaller couplings the limit on the LNV scale would be higher and scale according to Eq.~\eqref{eq:MmedL}.

\begin{figure}
    \centering
    \includegraphics[width=0.7\textwidth]{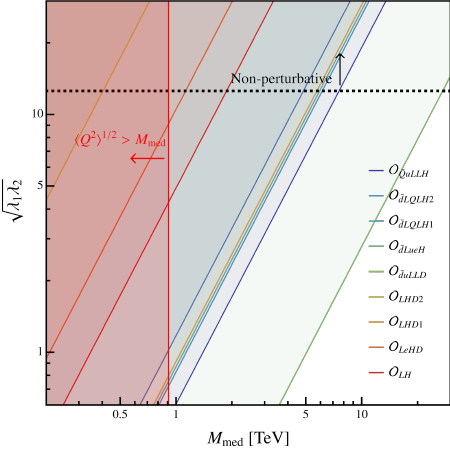}
    \caption{Constraints on the scale of dimension-7 $\Delta L=2$ LNV operators as a function of mediator mass $M_\text{med}$ and average coupling $\sqrt{\lambda_1\lambda_2}$. In the red-shaded region in the low-$M_\text{med}$ side, the EFT approach is invalid since the momentum scales involved in the process are typically larger than the mediator mass $M_\text{med}$. Therefore, the limits obtained in this section are not valid there. The other colored regions are excluded at 95\% C.L. by same-sign dilepton searches at the LHC. The black dashed line corresponds to $\sqrt{\lambda_1\lambda_2}=4\pi$, the limit at which the theory becomes non-perturbative.}
    \label{fig:EFTcontour}
\end{figure}

In Fig.~\ref{fig:EFTcontour}, the constraints on the different dimension-7 $\Delta L = 2$ LNV operators are shown as a function of mediator mass $M_\text{med}$ and average coupling $\sqrt{\lambda_1\lambda_2}$. Here it is assumed for simplicity that a single mediator mass scale is involved, such that the relation in Eq.~\eqref{eq:MmedL} holds. This assumption is valid for tree-level UV completions of the operators where only a single new physics field is required in the mass basis. Such UV-completions exist for all dimension-7 $\Delta L = 2$ LNV SMEFT operators. For $\mathcal{O}_{LH}$, $\mathcal{O}_{LeHD}$, and $\mathcal{O}_{LDH2}$ most of the region in which both the EFT description is valid, and the underlying UV-completion is perturbative, is excluded. For the other operators there are allowed regions, with varying upper limits on the mediator mass $M_\text{med}$, ranging all the way up to $M_\text{med} \lesssim \mathcal{O}(10)$~TeV for $\mathcal{O}_{\bar duLLD}$.
\begin{figure}
    \centering
    \includegraphics[width=0.7\textwidth]{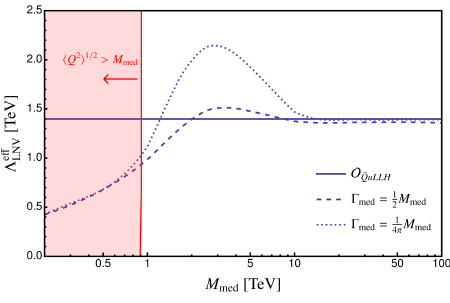}
    \caption{Effective 95\% C.L. limit on the effective scale of LNV $\Lambda_{\text{LNV}}^\text{eff}$ as a function of mediator mass $M_\text{med}$ for the operator $\mathcal{O}_{\bar{Q}uLLH}$ (solid), and a simplified model UV completion mediator with width $\Gamma_\text{med}=\frac{1}{2}M_\text{med}$ (dashed) and $\Gamma_\text{med}=\frac{1}{4\pi}M_\text{med}$ (dotted).}
    \label{fig:EFTline}
\end{figure}
To investigate the level of agreement between EFTs and simplified models for same-sign dilepton searches at the LHC, we now consider operator $\mathcal{O}_{\bar QuLLH}$ as an example to derive effective limits on the LNV scale using a generic UV-completion. In the mass basis, we take a scalar field $\chi$ with couplings to both quark-quark and lepton-lepton bilinears, according to the Lagrangian
\begin{equation}
\label{eq:UVlagQuLLH}
    \mathcal{L}\supset \lambda_1 \bar{d}_L u_R \chi^* + \lambda_2 \bar{e}_L^c\nu_L\chi + \text{h.c.},
\end{equation}
where $d_L$, $u_R$, $e_L$, and $\nu_L$ are the SM left-handed down-type quark, right-handed up-type quark, left-handed charged lepton, and neutrino, respectively. The field $\chi$ can arise as one component of the mixture between the fields $\varphi$ and $\Delta$ with representations $\varphi\in (1,2,\frac{1}{2})$ and $\Delta\in (1,3,1)$ under the SM gauge group. Now choosing 
\begin{equation}
\label{eq:UVscale}
    \frac{\lambda_1\lambda_2}{M_\text{med}^2}=\frac{v}{(\Lambda_\text{LNV}^\text{eff})^3},
\end{equation}
where $\Lambda_\text{LNV}^\text{eff}$ is an effective scale that would correspond to the LNV scale $\Lambda_\text{LNV}$ coming from operator $\mathcal{O}_{\bar QuLLH}$ if $\chi$ is integrated out, and where $M_\text{med}$ is the mass of $\chi$, we can compare the LHC signals of the simplified model described by Eq.~\eqref{eq:UVlagQuLLH} and the result on the scale of the operator $\mathcal{O}_{\bar QuLLH}$ given in Tab.~\ref{tab:op_lhc_HD}. In Fig.~\ref{fig:EFTline} we show the 95\% C.L. constraints on the effective scale $\Lambda_\text{LNV}^\text{eff}$ for both the simplified model and EFT approaches. Note that by effective LNV scale, we emphasize that the limit is dependent on the method used to derive it, EFT or simplified model, rather than corresponding to a fundamental scale. These constraints were obtained using \textsc{MadGraph5\_aMC@NLO}~\cite{Alwall:2014hca} in the same procedure and with the same cuts as described in Sec.~\ref{sec:LHC}. The red shaded area in the low-$M_\text{med}$ region is where we assume that the EFT breaks down due to the momentum scales involved in the process being greater than the mediator mass. The solid blue line corresponds to the constraint on the effective LNV scale $\Lambda^\text{eff}_\text{LNV}$ obtained from an EFT approach (c.f.\ Tab.~\ref{tab:op_lhc_HD}), and the dashed and dotted lines correspond to the effective limit on the LNV scale obtained from the model in Eq.~\eqref{eq:UVlagQuLLH} using the relation in Eq.~\eqref{eq:UVscale} using the width $\Gamma_\text{med}=\frac{1}{2}M_\text{med}$ and $\Gamma_\text{med}=\frac{1}{4\pi}M_\text{med}$, respectively. The former width is obtained by setting $\lambda_1=4\pi$ and assuming that $vM_\text{med}^2\ll (\Lambda_\text{LNV}^\text{eff})^3$ such that $\lambda_2\ll 1$, while the latter width is larger, allowing for the possibility of additional decay modes. Note that $\lambda_1=4\pi$ is at the limit of validity for perturbative couplings, and is chosen to reflect this limiting case. In Fig.~\ref{fig:EFTline} we see that a smaller width leads to an increasing $\Lambda_\text{LNV}^\text{eff}$ for a large range of mediator masses, while for a larger width, this effect is smaller. This increase is due to the resonant enhancement of an $s$-channel $\chi$ mediator, an effect that is not captured by the EFT approach. Therefore, for mediator masses, $\mathcal{O}(1)$ TeV $\lesssim M_\text{med} \lesssim$ $\mathcal{O}(10)$ TeV, the EFT approach underestimates the limit on $\Lambda_\text{LNV}$ for small widths, while there is practically no underestimation for large widths. For small mediator masses, we approach the limit where the EFT approach breaks down. The limits on $\Lambda_\text{LNV}^\text{eff}$ here obtained from the simplified model are lower due to the kinematic cuts removing a larger part of the phase space than for larger masses, which leads to the EFT approach overestimating the limit on the scale of LNV for $M_\text{med} \lesssim$ $\mathcal{O}(1)$ TeV. For large masses, $M_\text{med} \gtrsim$ $\mathcal{O}(1)$ TeV, the EFT and simplified model approaches agree quite well, as expected. The simplified model constraints lie slightly below the EFT line due to the nonzero decay width, which modifies the impact of $s$-channel mediators and slightly reduces the cross-section.
%%%%%%%%%%%%%%%%%%%%%%%%%%%%%%%%%%%%%%%%%%%%%%%%%%%%%%%%%%%%%%%%%%%%%%%%%%%%%%%%%
\section{Probing $d=7$ $\Delta L = 2$ SMEFT operators using low-energy observables}
%%%%%%%%%%%%%%%%%%%%%%%%%%%%%%%%%%%%%%%%%%%%%%%%%%%%%%%%%%%%%%%%%%%%%%%%%%%%%%%%%

\subsection{Neutrinoless double beta decay}
\begin{figure}[t]
    \centering
    \includegraphics[width=0.3\textwidth]{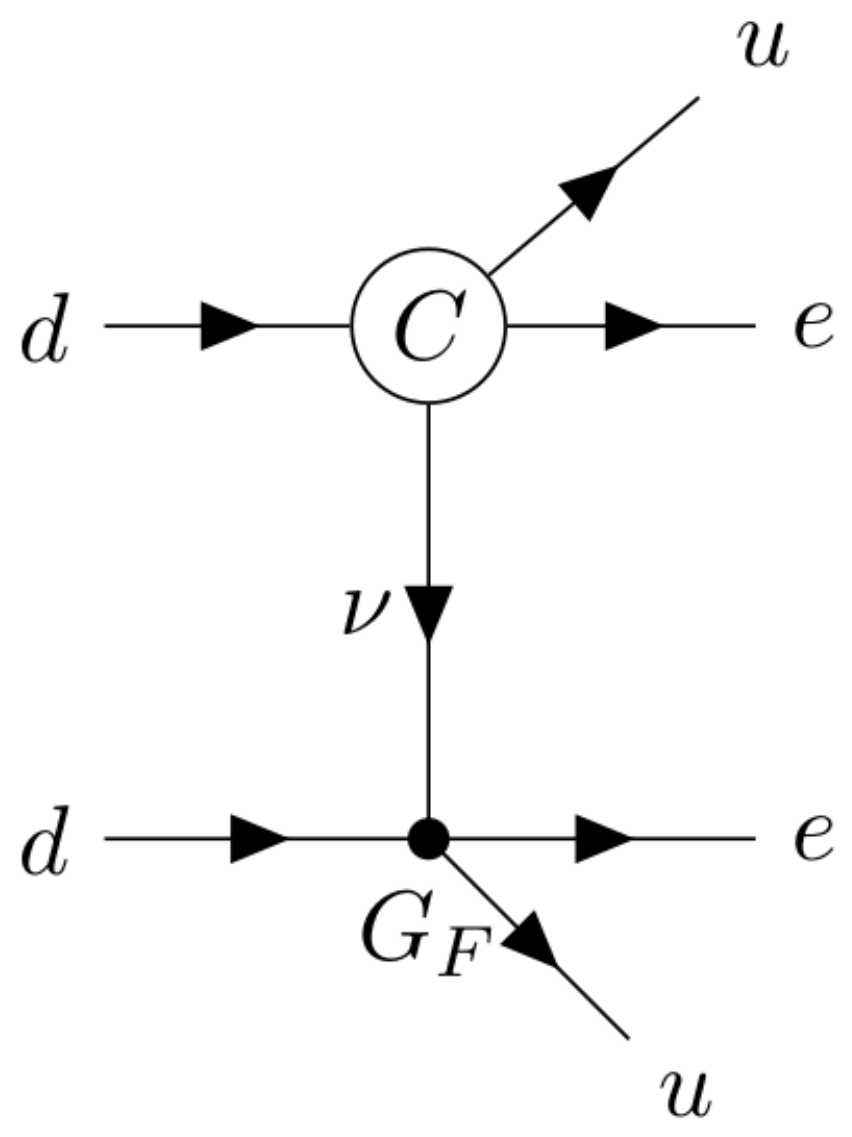}
    \caption{Schematic depiction of a long-range contribution to \ovbb decay.}
    \label{fig:0vbb-lr}
\end{figure}

A prominent way of probing lepton number violation in nature is the search for a neutrinoless mode of double beta decay. Although it is typically the light neutrino exchange, suppressed by the insertion of the light Majorana neutrino mass, that is assumed to trigger this hypothetical nuclear process, a variety of alternative mechanisms induced by higher-dimensional effective operators violating lepton number by two units can contribute~\cite{Pas:1999fc,Pas:2000vn,Deppisch:2012nb,deGouvea:2007xp,Ali:2007ec,Deppisch:2017ecm,Cirigliano:2017djv,Cirigliano:2018yza,Graf:2018ozy,Deppisch:2020ztt,Dekens:2020ttz}. The dimension-7 SMEFT operators studied in this work would trigger the so-called long-range \ovbb decay contributions, in which the lepton number is violated in one of the beta-decay vertices and the neutrino propagates between the two nucleons without the spin-flip. In this case, the effective operator replaces one of the SM beta-decay vertices with neutrino propagating between the two decaying nucleons, see Fig.~\ref{fig:0vbb-lr}.

In this work, we follow the formalism introduced in~Ref.~\cite{Cirigliano:2017djv}, which was recently automated in Ref.~\cite{Scholer:2023bnn}. In LEFT the $\Delta L = 2$ operators matching onto dimension 7 in SMEFT are of dimensions 6, 7, and 9. Therefore, the relevant Lagrangians are given by
\begin{align}
\begin{split}
    \mathcal{L}_{\Delta L=2}^{\text{LEFT} (d=6)} = \frac{4G_F}{\sqrt{2}}\bigg{[}
    &c_{du\nu e;LR}^{V} (\overline{d_{L}}\gamma_\mu u_{L})(\overline{\nu^c}\gamma^\mu e_{R})+
 c_{du\nu e;RR}^{V} (\overline{d_{R}}\gamma_\mu u_{R})(\overline{\nu^c}\gamma^\mu e_{R})\\
    +&c_{du\nu e;LL}^{S} (\overline{d_{R}}u_{L})
 (\overline{\nu^c}e_{L})+
 c_{du\nu e;RL}^{S} 
 (\overline{d_{L}}u_{R})
 (\overline{\nu^c}e_{L})\\
+&c_{du\nu e;LL}^{T} (\overline{d_{R}}\sigma_{\mu\nu}u_{L})
 (\overline{\nu^c}\sigma^{\mu\nu}e_{L})
\bigg{]} + \text{h.c.},
\end{split}
\end{align}
\begin{align}
\begin{split}
    \mathcal{L}_{\Delta L=2}^{\text{LEFT} (d=7)} = \frac{4G_F}{\sqrt{2}v}\bigg[&c_{du\nu e;LL}^{(7)V}\;\big(\overline{d_L}\gamma^\mu u_L\big)\; \big(\overline{\nu^c_L}\overset{\leftrightarrow}{D}_\mu e_L\big) 
    +c_{du\nu e;RL}^{(7)V}\;\big(\overline{d_R}\gamma^\mu u_R\big)\; \big(\overline{\nu^c_L}\overset{\leftrightarrow}{D}_\mu e_L\big)\bigg] + \text{h.c.} .
\end{split}
\end{align}
and
\begin{align}
\begin{split}
    \mathcal L^{\text{LEFT} (d=9)}_{\Delta L = 2}  =\ & \frac{8G_F^2}{v}\bar e_{L,i} C \bar e_{L,j}^T \bigg\{c^{(9);ij}_{V;LL}\,\bar u_L \gamma^\mu d_L\,  \bar u_L \gamma_\mu d_L +c^{(9);ij}_{V;LR}\, \bar u_L \gamma^\mu d_L\,  \bar u_R \gamma_\mu d_R \\ 
    &+c^{(9);ij}_{V';LR}\, \bar u^\alpha_L \gamma^\mu d^\beta_L\,  \bar u^\beta_R \gamma_\mu d^\alpha_R\bigg\} + {\rm h.c.}\;.
\end{split}
\end{align}
The matching of dimension-6 LEFT operators with dimension-7 SMEFT is already given in Table~\ref{tab:left}. The relevant dimension-7 and dimension-9 LEFT Wilson coefficients can be matched onto the dimension-7 SMEFT Wilson coefficients as follows~\cite{Liao:2020zyx,Cirigliano:2017djv}
\begin{eqnarray}
\frac{4 G_F}{\sqrt{2} v} c_{du\nu e;LL}^{(7)V; udst}&=& \frac{V^*_{du}}{2} (8 C^{ts}_{LHW}+C^{ts}_{LHD1}+C^{st}_{LHD1}+C^{ts}_{LHD2})\, ,\\
\frac{4 G_F}{\sqrt{2} v} c_{du\nu e;RL}^{(7)V;dust}&=&-\frac{1}{2} (C^{prts}_{\bar{d}uLLD}+C^{prst}_{\bar{d}uLLD})\, ,\\
\frac{8 G_F}{\sqrt{2}v} c^{(9);ij}_{V;LL} &=& -2V^2_{ud}(C_{LHD1}^{ij}+4C_{LHW}^{ij})^*\, , \\
\frac{8 G_F}{\sqrt{2}v} c^{(9);ij}_{V;LR} &=& -2 V_{ud} C_{\bar{d}uLLD}^{duij*}\, ,\quad \frac{8 G_F}{\sqrt{2}v} c^{(9);ss}_{V';LR} = 0.
\end{eqnarray}

Generally, the inverse half-life corresponding to a long-range \ovbb decay contribution induced by a single higher-dimensional operator is given by
\begin{align}
T^{-1}_{1/2} = |C|^2G_{0\nu}|M_{0\nu}|^2,
\end{align}
where $M_{0\nu}$ is the nuclear matrix element, $G_{0\nu}$ denotes the phase space factor arising from the lepton current, and $C$ stands for the SMEFT Wilson coefficient associated with the given operator. This coupling can be constrained employing the experimental bound on \ovbb decay half-life. The currently best limit on the \ovbb decay half-life $T_{1/2}^{\text{exp}} = 2.3\times10^{26}\,$y is provided by the KamLAND-Zen collaboration~\citep{KamLAND-Zen:2022tow} using the \textsuperscript{136}Xe isotope.

\begin{table}[t!]
    \centering
    \begin{tabular}{cc|cccc}\toprule
         %\multicolumn{5}{c}{}   \\
        \makecell{LEFT Wilson \\ [-0.25ex] Coefficient} & \makecell{Value \\ [-0.25ex] } & \makecell{SMEFT Wilson \\ [-0.25ex] Coefficient} & \makecell{Value \\ [-0.25ex] [$\text{TeV}^{-3}$]} & \makecell{$\Lambda_\text{NP}$ \\ [-0.25ex] [TeV]}   \\
        \midrule
       $c_{du\nu e;LL}^{S}$ & $1.86 \cdot 10^{-10}$ & $C_{\bar dLQLH1}$ & $7.06 \cdot 10^{-8}$ & $242$ \\
       $c_{du\nu e;RL}^{S}$ & $1.86 \cdot 10^{-10}$ & $C_{\bar QuLLH}$ & $3.62 \cdot 10^{-8} $ & $302$ \\
       $c_{du\nu e;LR}^{V}$ & $8.20 \cdot 10^{-10}$ & $C_{LeHD}$ & $1.55 \cdot 10^{-7}$ & $186$ \\
       $c_{du\nu e;RR}^{V}$ & $5.93 \cdot 10^{-8}$ & $C_{\bar dLueH}$ & $1.12 \cdot 10^{-5}$ & $44.7$ \\
       $c_{du\nu e;LL}^{T}$ & $4.51 \cdot 10^{-10}$ & $C_{\bar dLQLH1}$ & $6.83 \cdot 10^{-7}$ & $114$ \\
                                                 &  & $C_{\bar dLQLH2}$ & $3.41 \cdot 10^{-7}$ & $143$ \\
\midrule
       $c_{du\nu e;LL}^{(7)V}$ & $9.87 \cdot 10^{-6}$ 
                                                  & $C_{LHD1}$ & $1.36 \cdot 10^{-3}$ & $9.03$ \\
                                                  & & $C_{LHD2}$ & $2.71 \cdot 10^{-3}$ & $7.17$ \\
                                                  & & $C_{LHW}$ & $3.39 \cdot 10^{-4}$ & $14.3$ \\
       $c_{du\nu e;RL}^{(7)V}$ & $9.87 \cdot 10^{-6}$ & $C_{\bar{d}uLLD}$ & $1.32 \cdot 10^{-3}$ & $9.11$ \\
            \midrule
       $c^{(9);ij}_{V;LL}$ & $1.40 \cdot 10^{-5}$ & $C_{LHD1}$ & $9.91 \cdot 10^{-4}$ & $10.0$ \\
                                            & & $C_{LHW}$ & $2.48 \cdot 10^{-4}$ & $15.9$ \\
       $c^{(9);ij}_{V;LR}$ & $2.66 \cdot 10^{-8}$ & $C_{\bar{d}uLLD}$ & $1.83 \cdot 10^{-6}$ & $81.7$ \\
       \bottomrule
    \end{tabular}
    \caption{Current bounds and the corresponding new-physics scale for the relevant Wilson coefficients under the assumption of single operator dominance in contribution to \ovbb decay.}
    \label{tab:0vbb}
\end{table}

Employing the matching relations between our basis of $\Delta L =2$ SMEFT operators and the LEFT operators discussed above and using the experimental limit on \ovbb decay half-life one can readily derive the constraints on the SMEFT Wilson coefficients. Limits from \ovbb decay on the dimension-7 Wilson coefficients were presented in Ref.~\cite{Cirigliano:2017djv} and also analyzed in terms of the renormalization group running in Ref.~\cite{Liao:2019tep}. The limits corresponding to $^{136}\text{Xe}$ were based on the experimental constraints $T_{1/2}^{\text{old, exp}} = 1.07\times10^{26}\,$y~\cite{KamLAND-Zen:2016pfg}. Using the same framework we calculate the current limits $\Lambda_i$ for operators $\mathcal{O}_i$ considering the most recent experimental constraint $T_{1/2}^{\text{exp}} = 2.3\times10^{26}\,$y~\citep{KamLAND-Zen:2022tow}. For the different dimension-7 SMEFT operators, the obtained bounds are given in Tab.~\ref{tab:0vbb}. We discuss only tree-level contributions of the studied operators, omitting any loops, including higher-order contributions to the Weinberg operator, which were detailed e.g. in~\cite{Deppisch:2017ecm}.
Further, the presented limits are derived with the assumption of only a single operator added to the SM Lagrangian at a time. In our calculation, we employ the nuclear matrix elements computed using IBM-2 nuclear structure model~\cite{Deppisch:2020ztt}. As for the phase space factors, we use the values from Ref.~\cite{Graf:2022lhj}. From Tab.~\ref{tab:0vbb} we see that the largest contribution to the \ovbb decay rate is triggered by the scalar operators $O_{du\nu e;LL}^{S}$ and $O_{du\nu e;RL}^{S}$; therefore, the associated Wilson coefficients come with the most stringent constraints. Only slightly smaller rates are obtained for the tensor operator $O_{du\nu e;LL}^{T}$ and the vector operator $c_{du\nu e;LR}^{V}$, which are enhanced by large isovector magnetic moment~\cite{Cirigliano:2017djv}.

It is important to note that since the limits depend on the nuclear structure input, the numerical values of the Wilson coefficients may vary in dependence on the employed model and as such comes with an unspecified uncertainty, which however should not change the order of magnitude of the result. This is also because any potential deviation is further suppressed in the calculated scale of new physics $\Lambda_{\text{NP}}$, which is of the most interest. Somewhat higher uncertainty is associated with the limit on the dimension-9 operator $c^{(9);ij}_{V;LL}$, whose contribution depends on the unknown low-energy constants (LECs) and we set them to zero in our calculation.

%%%%%%%%%%%%%%%%%%%%%%%%%%%%%%%%%%%%%%%%%%%%%%%%%%%%%%%%%%%%
%%%%%%%%%%%%%%%%%%%%%%%%%%%%%%%%%%%%%%%%%%%%%%%%%%%%%%%%%%%%

\subsection{Coherent elastic neutrino-nucleus scattering and neutral current lepton number violation}
%%%%%%%%%%%%%%%%%%%%%%%%%%%%%
%%%%%%%%%%%%%%%%%%%%%%%%%%%%%%%%%%%%%%%%%
%% SM part
The SM neutral current interaction can lead to Neutrino-Nucleus Scattering and if the incoming neutrino is low energy as compared to the inverse size of the nucleus $E_\nu\lesssim 50$ MeV then the scattering cross section can be enhanced due to coherent scattering of the incoming neutrino with the protons and neutrons inside the nucleus. This leads to a large enhancement of the cross-section for a nucleus with a large number of protons ($Z$) and neutrons ($N$), with the differential coherent elastic neutrino-nucleus scattering cross section given by
\begin{equation}
\left.\frac{d\sigma}{dT}\right|_{ \text{ spin}=0}=\frac{\sigma_{0}^{{\text{ SM}}}}{M}\left(1-\frac{T}{T_{\text{max}}}\right),\label{eq:7s1}
 \end{equation}
 where $\sigma_{0}^{{\text{ SM}}}$ is given by~\cite{Freedman:1973yd}
 \begin{equation}
 \sigma_{0}^{{\text{ SM}}}\equiv\frac{G_{F}^{2}\left[N-(1-4s_{W}^{2})Z\right]^{2}F^{2}(q^{2})M^{2}}{4\pi}\,,\label{eq:7s2}
 \end{equation}
with $G_{F}$ denoting the Fermi constant, $s_{W}=\sin\theta_{W}$ denoting the weak-interaction Weinberg angle, and $M$ denoting the mass of the nucleus. $T$ denotes the recoil energy of the nucleus, which can have a maximal value of $T_{{\text{max}}}$, dictated by the incoming neutrino energy $E_{\nu}$ and $M$
 \begin{equation}
 T_\text{max}(E_{\nu})=\frac{2E_{\nu}^{2}}{M+2E_{\nu}}\,.\label{eq:7s3}
 \end{equation}
At the low-energies relevant for coherent elastic neutrino-nucleus scattering $s_{W}^{2}\approx0.238$~\cite{Erler:2004in}, giving $N-(1-4s_{W}^{2})Z\approx N-0.045Z$. This implies that a large number of neutrons should lead to an enhanced cross-section for coherent scattering. $F(q^{2})$ represents the form factor of the nucleus, with $F(q^{2})\rightarrow 1$ as $q^{2}\rightarrow 0$~\cite{Kerman:2016jqp}. We note that Eq.~\eqref{eq:7s1} is strictly valid if 
the nucleus is spin-0~\cite{Freedman:1973yd}. For a nucleus with spin-1/2, the relevant expression gets modified as
 \begin{equation}
 \left.\frac{d\sigma}{dT}\right|_{{{\text{ spin}}}=\frac{1}{2}}=\left.\frac{d\sigma}{dT}\right|_{{ {\text{ spin}}}=0}+\frac{\sigma_{0}^{{\text{ SM}}}}{M}\frac{T^{2}}{2E_{\nu}^{2}}\,.\label{eq:7s4}
 \end{equation}
However, the additional term  proportional to $T^{2}/E_{\nu}^{2}$ is highly suppressed and can therefore be neglected for a heavy nucleus. 
%%%%%%%%%%%%%%%%%%%%
%%%%%%%%%%%%%%%%%%%%%%%%%%%%%%%%%%%%%%%%%%%%%%%%%
In addition to the SM contributions, new interactions can also contribute to the scattering cross section modifying Eq. \eqref{eq:7s1}~\cite{Lindner:2016wff,Bolton:2021pey}. In the context of $\Delta L=2$ dimension-6 LEFT operators involving only left-handed neutrinos the relevant effective Lagrangian can be expressed as
\begin{align}
\begin{aligned}
\label{eq:7s5}
\mathcal{L}_\mathrm{eff}^{ \bar{\nu^c} \nu f f} = \frac{4G_F}{\sqrt{2}}\Bigg[
c_{d\nu;LL}^{S,prst} (\overline{d_{Rp}}d_{Lr})(\overline{\nu^c_s}\nu_t)
+c_{d\nu;RL}^{S,prst} (\overline{d_{Lp}}d_{Rr})(\overline{\nu^c_s}\nu_t)
+ c_{d\nu;LL}^{T,prst} (\overline{d_{Rp}}\sigma_{\mu\nu}d_{Lr})
 (\overline{\nu^c_s}\sigma^{\mu\nu}\nu_t)\\
+{c}_{u\nu;LL}^{S,prst} (\overline{u_{Rp}}u_{Lr})(\overline{\nu^c_s}\nu_t)
+{c}_{u\nu;RL}^{S,prst} (\overline{u_{Lp}}u_{Rr})(\overline{\nu^c_s}\nu_t)
+{c}_{u\nu;LL}^{T,prst} (\overline{u_{Rp}}\sigma_{\mu\nu}u_{Lr})
 (\overline{\nu^c_s}\sigma^{\mu\nu}\nu_t)\Bigg]\, .
\end{aligned}
\end{align}
Since we are interested in LNV interactions, we can impose the Majorana nature of the left-handed neutrinos. As a consequence, the above-mentioned operators satisfy the following relations under the exchange of $s$ and $t$
\begin{align}
\begin{aligned}
\label{eq:7s6}
c_{f\nu;LL}^{S,prst} &=c_{f\nu;LL}^{S,prts}\\
c_{f\nu;RL}^{S,prst}&= c_{f\nu;RL}^{S,prts}\\ c_{d\nu;LL}^{T,prst}&=-c_{d\nu;LL}^{T,prts} \, .
\end{aligned}
\end{align}
Using an alternative basis, frequently employed in the literature for coherent elastic neutrino-nucleus scattering (see e.g. Ref.~\cite{Bischer:2019ttk,Bolton:2020xsm}), the effective Lagrangian in Eq.~\eqref{eq:7s5} can be parameterized by
\begin{align} \label{eq:7s7}
\mathcal{L}_\mathrm{eff}^{ \bar{\nu^c} \nu f f}=-\frac{G_F}{\sqrt{2}}\sum_{\substack{a=S,P,T\\f=u,d}}\left(\overline{\nu^c}_{\alpha}\,\Gamma^a\nu_{\beta}\right)
\left(\overline{f_{\gamma}}\Gamma^a({c}_{\alpha\beta\gamma\delta}^a+{d}_{\alpha\beta\gamma\delta}^a i\gamma^5)f_{\delta}\right),
\end{align}
where the three combinations of Dirac matrices are defined as $\Gamma^a\in \left\{\mathbb{I},i\gamma^5,\sigma^{\mu\nu}\right\}$ corresponding to $a = S, P, T$ and the associated coefficients are denoted by $C^a$ and $D^{a}$.
The coefficients $c_{f\nu;XY}$ ($X,Y = L,R$) are simply linear combinations of the $c^{a}$ and $d^{a}$ coefficients
\begin{align} \label{eq:7s8}
\begin{aligned}
c_{f\nu;LL}^{S,prst} &= 
\frac12\left(c^S-id^S-c^P+id^P\right),\\
c_{f\nu;RL}^{S,prst} &=
\frac12\left(c^S+id^S+c^P+id^P\right),\\
c_{f\nu;LL}^{T,prst}&=
\frac14\left(c^T-id^T\right) \,,
\end{aligned}
\end{align}
where the relevant up- or down-type quarks are implied. For coherent elastic neutrino-nucleus scattering the nucleus can be treated as a spin-0 or spin-1/2 particle. Given that at energies relevant for coherent scattering the difference between the treatment of spin-0 or spin-1/2 nucleus is negligible, we will follow closely the approaches of~\cite{Lindner:2016wff,Bischer:2019ttk} which treat the nucleus as a spin-1/2 particle. In the presence of the dimension-6 LEFT LNV interactions presented in Eq.~\eqref{eq:7s7}, the differential cross section for coherent elastic neutrino-nucleus scattering in Eq.~\eqref{eq:7s1} is modified by~\cite{Lindner:2016wff}
 \begin{eqnarray}
\left. \frac{d\sigma}{dT}\right|_{\text{d=6 LEFT}} & = & \frac{G_{F}{}^{2}M}{4\pi}\left[\zeta_{S}^{2}\frac{MT}{2E_{\nu}{}^{2}} +\zeta_{T}^{2}\left(1-\frac{T}{T_{{\text {max}}}}+\frac{MT}{4E_{\nu}{}^{2}}\right) -\Xi\frac{T}{E_{\nu}}+{\cal O}\left(\frac{T^{2}}{E_{\nu}^{2}}\right)\right],\label{eq:7s9}
 \end{eqnarray}
 with $\zeta_S$, $\zeta_T$ and $\Xi$ defined as
 \begin{eqnarray}
 \zeta_{S}^{2}&=&\left({c'}_{S}^{2}+{d'}_{P}^{2}\right),\nonumber\\
 \zeta_{T}^{2}&=& {8}\left({c'}_{T}^{2}+{d'}_{T}^{2}\right) ,\nonumber\\
 \Xi &=& {2}(c'_{P}c'_{T}-c'_{S}c'_{T}+d'_{T}d'_{P}-d'_{T}d'_{S})\, .
 \label{eq:7s10}
 \end{eqnarray}
To simplify the analysis we will focus on the scenarios where either the scalar or tensor interactions are present at a time and will drop the $\Xi$ term hereafter (which is the interference term when both scalar and tensor interactions are present simultaneously). The reparameterized Wilson coefficients are related to the ones defined in Eq.~\eqref{eq:7s7} by
\begin{eqnarray}  \label{eq:7s11}
c'_S&=&\sum_{q=u,d}c^S_{(q)}\left[
N\frac{m_n}{m_q}f^n_{Tq}F_n(Q^2)+Z\frac{m_p}{m_q}f^p_{Tq}F_p(Q^2)
\right],\nonumber\\
d'_P&=&\sum_{q=u,d}d^P_{(q)}\left[
N\frac{m_n}{m_q}f^n_{Tq}F_n(Q^2)+Z\frac{m_p}{m_q}f^p_{Tq}F_p(Q^2)
\right],\nonumber\\
c'_T &=& N(\delta^n_u c_u^T + \delta^n_d c_d^T)F_n(Q^2)
  + Z(\delta^p_u c_u^T + \delta^p_d c_d^T ) F_p(Q^2)\,,
\end{eqnarray}
where $F_n$ and $F_p$ are functions dependent on energy transfer $Q$ and correspond to the form factors of neutron and proton, respectively. We will further neglect $d'_T$ since it is related to the spin-dependent part of the cross-section (which is usually suppressed because of the cancellation in the effective coupling while summing over spin-up and spin-down nucleons in heavy nuclei). The factors $f^n_{Tq}$ and $f^p_{Tq}$ are related to the fraction of nucleon mass due to a given quark type and calculated in chiral perturbation theory~\cite{Cheng:1988im}. The factors $\delta^n_q$ and $\delta^p_q$ correspond to the tensor charges corresponding to neutron and proton, respectively. Using following values for these parameters~\cite{Jungman:1995df,Anselmino:2008jk},
\begin{eqnarray}  \label{eq:7s12}
f^p_{Tu}&=&0.019\,,
f^n_{Tu}=0.023\,,\nonumber\\
f^p_{Td}&=&0.041\,,
f^n_{Td}=0.034\,,\nonumber\\
\delta^p_{u}&=&0.54\,,
\delta^p_{d}=-0.23\,,\nonumber\\
\delta^n_{u}&=&-0.23\,,
\delta^n_{d}=0.54. 
\end{eqnarray}
the relevant limits derived using the data from the COHERENT experiment (using a CsI detector) as derived in~\cite{AristizabalSierra:2018eqm} are given by\footnote{The limits derived in~\cite{AristizabalSierra:2018eqm} are for lepton number conserving modes. However, the underlying amplitude for the lepton number violating and conserving operators are the same and therefore these limits hold identically.}
\begin{eqnarray}\label{eq:7s13}
|\zeta_S|/NF(Q^2)&\leq & 0.620 \quad\text{at 90\% CL,}\,\nonumber\\
|\zeta_S|/NF(Q^2)&\leq & 1.065\quad\text{at 99\% CL,}\,\nonumber\\
|\zeta_T|/NF(Q^2) &\leq & 0.591\quad\text{at 90\% CL,}\,\nonumber\\ 
|\zeta_T|/NF(Q^2) &\leq & 1.072 \quad\text{at 99\% CL}\, .
\end{eqnarray}
The future projections for upcoming reactor neutrino-based coherent elastic neutrino-nucleus scattering experiments and Germanium target can be found for example in~\cite{Lindner:2016wff}
\begin{eqnarray}\label{eq:7s13f}
|\zeta_S|/NF(Q^2)&\leq & 0.21 \quad \text{at 3$\sigma$,}\,\nonumber\\
|\zeta_T|/NF(Q^2) &\leq & 0.25\quad \text{at 3$\sigma$}\, .
\end{eqnarray}

These limits have been used for example in~\cite{Bischer:2019ttk} to constrain dimension-6 SMEFT operators. Here we will use these limits to constrain the LNV dimension-7 SMEFT operators. Note that the dipole portal and  neutral current operators involving up quarks at LEFT level give constraints on $\mathcal{O}_{LHW}, \mathcal{O}_{LHB} $ and $\mathcal{O}_{\bar{Q}uLLH}$ , respectively. Given that these operators are more tightly constrained by other observables, we will not discuss them at length here. A comprehensive study in the context of dimension-6 and dimension-7 SMNEFT can be found for example in \cite{Li:2020lba}, which includes the dipole portal, neutral current operators involving up quarks and sterile neutrino. The relevant dimension-6 LNV LEFT operators of interest are the ones corresponding to $c_{d\nu;LL}^{S,prst}$, $c_{d\nu;RL}^{S,prst}$ and $c_{d\nu;LL}^{T,prst}$. We can use the constraints presented in Eq.~\eqref{eq:7s13} for the CsI target-based COHERENT experiment by assuming the Helm form factors for protons and neutron to be equal ($F_p(Q^2)\simeq F_n(Q^2)=F(Q^2)$). Combining the relations given in Eq.~\eqref{eq:7s8} and Eq.~\eqref{eq:7s10} with Eq.~\eqref{eq:7s11} leads to
\begin{eqnarray}\label{eq:7s14}
\frac{\zeta_S^2}{N^2F^2(Q^2)}
&=& \sum_{i=\text{Cs},\text{I}}\frac{1}{N_i^2}\Big((c^S_{d})^2+(d^P_{d})^2\Big)\left[
N_i\frac{m_n}{m_d}f^n_{Td}+Z_i\frac{m_p}{m_d}f^p_{Td}
\right]^2\nonumber\\
&=& 4\left|c_{d\nu;LL(RL)}^{S,11st}\right|^2\,\sum_{i=\text{Cs},\text{I}}\frac{1}{N_i^2}\left[
N_i\frac{m_n}{m_d}f^n_{Td}+Z_i\frac{m_p}{m_d}f^p_{Td}
\right]^2,
\end{eqnarray}
and 
\begin{eqnarray}\label{eq:7s15}
\frac{\zeta_T^2} {N^2F^2(Q^2)} &=&  \sum_{i=\text{Cs},\text{I}}
8 (c_T^d)^2 \left(\delta^n_d + \frac{Z_i}{N_i} \delta^p_d\right)^2\nonumber\\
&=&  
128 \left|c_{d\nu;LL}^{T,11st}\right|^2 \sum_{i=\text{Cs},\text{I}}\left(\delta^n_d + \frac{Z_i}{N_i} \delta^p_d\right)^2.
\end{eqnarray}

Using $N=77.9$, $Z=55$ for Caesium nuclei, and $N=73.9$, $Z=53$ for Iodine nuclei, the current constraints from the COHERENT experiment listed in Eq.~\eqref{eq:7s13} combined with Eq.~\eqref{eq:7s14} and Eq.~\eqref{eq:7s15} lead to the constraints 
\begin{eqnarray}\label{eq:7s16}
\left| c_{d\nu;LL(LR)}^{S,11st} \right| \leq 0.017 \quad\text{at 90\% CL,} \nonumber\\
\left| c_{d\nu;LL(LR)}^{S,11st} \right| \leq 0.030 \quad\text{at 99\% CL,} \nonumber\\
\left| c_{d\nu;LL}^{T,11st} \right| \leq 0.098 \quad\text{at 90\% CL,} \nonumber\\
\left| c_{d\nu;LL}^{T,11st} \right| \leq 0.178 \quad\text{at 99\% CL}\, .
\end{eqnarray}
Using the future projections listed in Eq.~\eqref{eq:7s13f} and taking $N=40.6$, $Z=32$ for Germanium (and replacing CsI by Ge in Eq.~\eqref{eq:7s14} and Eq.~\eqref{eq:7s15}) we find the projected constraints to be
\begin{eqnarray}\label{eq:7s16}
\left| c_{d\nu;LL(LR)}^{S,11st} \right| \leq 0.008 \quad\text{at 3$\sigma$,} \nonumber\\
\left| c_{d\nu;LL}^{T,11st} \right| \leq 0.062 \quad\text{at 3$\sigma$}\, .
\end{eqnarray}
In Table~\ref{tab:so_cevns} we summarise the current best limits and future projections for the sensitivity of the coherent elastic neutrino-nucleus scattering to  $\Delta L=2$ dimension-7 SMEFT operators and the relevant new physics scales\footnote{In passing we note that long-range potential induced by light neutrino exchange can also constrain the neutral current LNV interactions, see e.g. ~\cite{Bolton:2020xsm}. However, since the relevant constraints are much weaker, we do not discuss them here.}. 
\begin{table}[htb!]
    \centering
    \begin{tabular}{ccccc}\toprule
         \multicolumn{5}{c}{Current Bound}   \\
        LEFT Wilson \\ [-0.25ex] Coefficient & \makecell{Value \\ [-0.25ex] } & \makecell{ $C_{\bar{d}LQLH1}$ \\ [-0.25ex] [$\text{TeV}^{-3}$]} & \makecell{$\Lambda_{\text{NP}}$ \\ [-0.25ex] [TeV]} & Experiment  \\
        \midrule
        $c_{d\nu;LL(LR)}^{S,11\mu\mu}$ & $0.030$ & 11.3 & 0.4 & COHERENT \\
        $c_{d\nu;LL}^{T,11st} $ & $0.178$ & 540.2 & 0.1 & COHERENT  \\
        \bottomrule
           \multicolumn{5}{c}{Future Sensitivity} \\
        \midrule
        $c_{d\nu;LL(LR)}^{S,11\alpha\alpha} $ &  $0.008$ & 3.0 & 0.7 & Ge \\
        $c_{d\nu;LL}^{T,11st}$ &  $0.062$ & 186.9 & 0.2 & Ge \\
        \bottomrule
    \end{tabular}
    \caption{Most restrictive current bounds, future sensitivity, and the corresponding new-physics scale for the relevant Wilson coefficients under the assumption of single LEFT operator dominance to coherent elastic neutrino-nucleus scattering. Note that "Value" referes to the dimensionless LEFT Wilson coefficients.}
    \label{tab:so_cevns}
\end{table}
%
%%%%%%%%%%%%%%%%%%%%%%%%%%%%%%%%%%%%%%%%%%%%%%%%%%%%%%%%%%%%%%%%%%%%
\subsection{Neutrino oscillations and charged current lepton number violation}
%%%%%%%%%%%%%%%%%%%%%%%%%%%%%%%%%%%%%%%%%%%%%%%%%%%%%%%%%%%%%%%%%%%%

Although most of the literature studying BSM physics in neutrino oscillation experiments assumes only lepton number conserving non-standard neutrino interactions (NSI), using oscillations to probe lepton number violation has been also proposed, see e.g. Ref.~\cite{Bolton:2019wta}. The idea suggested therein is to consider a charge-current LNV NSI rather than the heavily suppressed standard Majorana neutrino helicity flip. Indeed, the LNV transition ``$\nu_\alpha \rightarrow \nu_\beta$" would introduce the usual helicity suppression factor $(m_\nu/E_\nu)^2$, which for typical neutrino energies at large oscillation experiments ($E_\nu \sim 5 \text{MeV} - 2 \text{GeV}$) and $m_\nu \sim 0.1$ eV leads to suppression of the magnitude in the range $\sim 10^{-21} - 10^{-16}$. If, however, the interaction vertex at production or detection violates lepton number, the chiralities of the respective lepton currents are opposite and the helicity reversal is not required (similarly as in the case of the non-standard long-range \ovbb decay mechanisms). 

Oscillation experiments usually are not sensitive to the charge of the outgoing lepton produced at the far detector that would allow distinguishing between incoming neutrino and antineutrino. However, as pointed out in Ref.~\cite{Bolton:2019wta} there exist such data from the long-baseline (LBL) oscillation experiment MINOS and the LBL reactor (and solar) oscillation experiment KamLAND. These measurements can be therefore used for constraining the LNV NSIs. Ref.~\cite{Bolton:2019wta} focuses on LNV vector currents; specifically, they employ the experimental data to derive bounds on Wilson coefficients of dimension-6 LEFT operators with ($V+A$) Lorentz structure in the leptonic current. Here we generalize their treatment to scalar and tensor currents also for dimension-6 LEFT operators. Finally, based on these limits on LNV dimension-6 LEFT operators and assuming a single LEFT operator dominance at a time we derive new constraints on the LNV dimension-7 SMEFT operators and the relevant new physics scale in a comprehensive manner.

Considering an LBL oscillation experiment blind to the charge of the lepton $\ell_\alpha^{\pm}$ at the production process, but sensitive to the charge of the outgoing lepton $\ell_\beta^{\pm}$, the ratio
\begin{align*}
    R_{\alpha\beta} \equiv \frac{N_{\ell_\beta^+}}{N_{\ell_\beta^-}}
    =\frac{\Gamma_{\nu_{\alpha}\rightarrow\bar{\nu}_{\beta}}+\Gamma_{\bar{\nu}_{\alpha}\rightarrow\bar{\nu}_{\beta}}}{\Gamma_{\nu_{\alpha}\rightarrow\nu_{\beta}}+\Gamma_{\bar{\nu}_{\alpha}\rightarrow\nu_{\beta}}}
\end{align*}
can be used to constrain new LNV physics. Following the MINOS analysis~\cite{Danko:2009qw}, the above ratio consists of a signal part $S_{\mu\mu}$ corresponding to the $\nu_\mu \rightarrow \bar{\nu_\mu}$ process and a background part $B_{\mu\mu}$ describing the standard oscillation process $\nu_\mu \rightarrow \nu_\mu$. After removing the background from the measured ratio $R_{\mu\mu}$, Ref.~\cite{Danko:2009qw} constrains the signal part. Hence, assuming the new physics induced oscillation rate is factorizable, one can write~\cite{Bolton:2019wta}
\begin{align}
    S_{\mu\mu}\approx\frac{\int dE_{\mathbf{q}}~\sum\limits_{\rho,\sigma}\frac{d\Gamma_{\nu_{\mu}}}{dE_{\mathbf{q}}}\cdot P^{(\rho,\sigma)}_{\nu_{\mu}\rightarrow\bar{\nu}_{\mu}}\cdot\sigma^{}_{\bar{\nu}_{\mu}}}{\int dE_{\mathbf{q}}~\sum\limits_{\rho,\sigma}\frac{d\Gamma_{\nu_{\mu}}}{dE_{\mathbf{q}}}\cdot P^{(\rho, \sigma)}_{\nu_{\mu}\rightarrow\nu_{\mu}}\cdot\sigma^{}_{\nu_{\mu}}}\lesssim 0.026,
    \label{eq:signalpart}
\end{align}
where the $(\rho,\sigma)$ run over all the possible Dirac structures.
Since, the oscillation probability $P^{(\rho,\sigma)}_{\nu_{\mu}\rightarrow\bar{\nu}_{\mu}}$ entering the above expression depends on the new-physics coupling $\varepsilon_{\mu\lambda}$ with $\lambda \in {e,\mu,\tau}$, the corresponding bound on these effective couplings can be derived, for detail see Ref.~\cite{Bolton:2019wta}. Therein, the focus is on operators with vector currents, specifically,
\begin{align}
\begin{aligned}
\label{eq:osc1}
\mathcal{L}_\mathrm{LEFT}^{ d=6} &\supset \frac{4G_F}{\sqrt{2}}
\Big[c_{du\nu e;LR}^{V,prst} (\overline{d_{Lp}}\gamma_\mu u_{Lr})(\overline{\nu^c_s}\gamma^\mu e_{Rt})
+
c_{du\nu e;RR}^{V,prst} (\overline{d_{Rp}}\gamma_\mu u_{Rr})(\overline{\nu^c_s}\gamma^\mu e_{Rt})\Big]\; .\\
\end{aligned}
\end{align}
%

% %
Following Ref.~\cite{Formaggio:2012cpf} the cross sections entering Eq.~\eqref{eq:signalpart} can be assumed to be equal in the quasi-elastic scattering limit, i.e.,
\begin{align}
    \sigma_{\bar{\nu}_{\mu}p\rightarrow\ell_{\beta}^{+}n}(E_{\mathbf{q}})\backsimeq \sigma_{\nu_{\mu}n\rightarrow\ell_{\beta}^{-}p}(E_{\mathbf{q}})\backsimeq\frac{G_{F}^{2}|V_{ud}|^{2}}{\pi}\big(g_{\mathrm{V}}^{2}+3g_{\mathrm{A}}^{2}\big)E_{\mathbf{q}}^{2}.
    \label{eq:crosssection}
\end{align}

The strategy is similar in the case of the KamLAND experiment, which can constrain the LNV NSI process $\nu_{\mu,\tau}p\rightarrow e^+ n$ at detection using the measured limit on the signal ratio $S_{ee}$ given by
\begin{align}\label{eq:osc2}
    S_{ee}\approx\frac{\int dE_{\mathbf{q}}~\sum\limits_{\beta}\frac{d\Gamma^{}_{\nu_{e}}}{dE_{\mathbf{q}}}\cdot P^{\mathrm{eff}}_{\nu_{e}\rightarrow\nu_{\beta}}\cdot\varepsilon_{e\beta}^2\cdot \sigma^{}_{\bar{\nu}_{\beta}} }{\int dE_{\mathbf{q}}~\frac{d\Gamma^{}_{\nu_{e}}}{dE_{\mathbf{q}}}\cdot P^{\mathrm{eff}}_{\nu_{e}\rightarrow\nu_{e}}\cdot\sigma^{}_{\nu_{e}}
}\lesssim 2.8\times 10^{-4},
\end{align}
where the non-standard oscillation probability was approximated by the effective standard probability, taking into account the fact that solar neutrinos make up an incoherent mixture of flavour eigenstates by the time they reach Earth, which effectively smears out any dependence on the Majorana phases. Employing the above equation and experimental limit Ref.~\cite{Bolton:2019wta} sets bounds on the effective couplings $\varepsilon_{e\lambda}$ with $\lambda \in {e,\mu,\tau}$ employing the ${}^8\text{B}$ flux from the solar model in Ref.~\cite{Bahcall:1996qv}.

To make the constraints of Ref.~\cite{Bolton:2019wta} compatible with our analysis we need to relate the effective couplings $\varepsilon_{\alpha\beta}^{L/R}$ to the Wilson coefficients of operators in our LEFT basis, which can be subsequently used to constrain the SMEFT Wilson coefficients employing the matching conditions given in Table~\ref{tab:left}.
\begin{table}[htb!]
    \centering
    \begin{tabular}{ccccccc}\toprule
         \multicolumn{6}{c}{}   \\
        \makecell{LEFT Wilson \\ [-0.25ex] Coefficient} & \makecell{Value \\ [-0.25ex] } & \makecell{SMEFT Wilson \\ [-0.25ex] Coefficient} & \makecell{Value \\ [-0.25ex] [$\text{TeV}^{-3}$]} & \makecell{$\Lambda_{\text{NP}}$ \\ [-0.25ex] [TeV]} & Experiment  \\
        \midrule
       $c_{du\nu e;LR}^{V,11ee (e\mu)}$   & $0.017$   & $C_{LeHD}^{ee(e\mu)}$           & 3.2     & 0.7 & KamLAND\\
       $c_{du\nu e;RR}^{V,11ee(e\mu)}$    & $0.017$   & $C_{\bar dLueH}^{1e1e(1e1\mu)}$ & 6.4     & 0.5  & KamLAND\\
       $c_{du\nu e;LR}^{V,11e\tau}$   & $0.015$       & $C_{LeHD}^{ee(e\tau)}$          & 2.8    &  0.7 & KamLAND \\
       $c_{du\nu e;RR}^{V,11e\tau}$   & $0.015$       & $C_{\bar dLueH}^{1e1\tau}$      & 5.7    & 0.6 & KamLAND \\
       $c_{du\nu e;LR}^{V,11\mu e}$   & $0.22-3.47$   & $C_{LeHD}^{\mu e}$              &  41.7-658.1 & 0.1-0.3 & MINOS\\
       $c_{du\nu e;RR}^{V,11\mu e}$   & $0.22-3.47$   & $C_{\bar dLueH}^{1\mu1e}$       & 83.4-1316.2 & 0.1-0.2  & MINOS\\
       $c_{du\nu e;LR}^{V,11\mu\mu}$  & $0.16-0.63$   & $C_{LeHD}^{\mu\mu}$             & 30.3-119.5 & 0.2-0.3 & MINOS \\
       $c_{du\nu e;RR}^{V,11\mu\mu}$  & $0.16-0.63$   & $C_{\bar dLueH}^{1\mu1\mu}$     & 60.7-239.0 & 0.2-0.3 & MINOS \\
       $c_{du\nu e;LR}^{V,11\mu\tau}$ & $0.16-0.71$   & $C_{LeHD}^{\mu\tau}$            & 30.3-134.7 & 0.2-0.3 & MINOS \\
       $c_{du\nu e;RR}^{V,11\mu\tau}$ & $0.16-0.71$   & $C_{\bar dLueH}^{1\mu1\tau}$    & 60.7-269.31 & 0.2-0.3  & MINOS \\
        \bottomrule
    \end{tabular}
    \caption{Current bounds, future sensitivity and the corresponding new-physics scale for the LNV dimension-7 SMEFT Wilson coefficients under the assumption of single vector LEFT operator dominance to LNV LBL oscillations. The limits obtained from the MINOS experiment have a form of a range, which reflects their dependence on Majorana phases.}
    \label{tab:so_cevns1}
\end{table}
\begin{table}[htb!]
    \centering
    \begin{tabular}{ccccccc}\toprule
         \multicolumn{6}{c}{}   \\
        \makecell{LEFT Wilson \\ [-0.25ex] Coefficient} & \makecell{Value \\ [-0.25ex] } & \makecell{SMEFT Wilson \\ [-0.25ex] Coefficient} & \makecell{Value \\ [-0.25ex] [$\text{TeV}^{-3}$]} & \makecell{$\Lambda_{\text{NP}}$ \\ [-0.25ex] [TeV]} & Experiment  \\
        \midrule
       $c_{du\nu e;LL}^{S,11ee}$      & $0.040$      & $C_{\bar dLQLH1}^{1e1e}$ & 15.2 & 0.4  & KamLAND\\
       $c_{du\nu e;RL}^{S,11ee}$      & $0.040$      & $C_{\bar QuLLH}^{11ee}$ & 7.5 & 0.5 & KamLAND\\
       $c_{du\nu e;LL}^{S,11e\mu}$    & $0.040$      & $C_{\bar dLQLH1}^{1\mu 1e}$ & 15.2 & 0.4  & KamLAND \\
                                      &              & $C_{\bar dLQLH2}^{1\mu 1e(1e1\mu)}$ & 30.3 & 0.3 & KamLAND \\
       $c_{du\nu e;RL}^{S,11e\mu}$    & $0.040$      & $C_{\bar QuLLH}^{11\mu e}$ & 7.6 & 0.5  & KamLAND \\
       $c_{du\nu e;LL}^{S,11e\tau}$   & $0.036$      & $C_{\bar dLQLH1}^{1\tau 1e}$ & 13.7 & 0.4  & KamLAND \\
                                      &              & $C_{\bar dLQLH2}^{1\tau 1e(1e1\tau)}$ & 27.3 & 0.3  & KamLAND \\
       $c_{du\nu e;RL}^{S,11e\tau}$   & $0.036$      & $C_{\bar QuLLH}^{11\tau e}$ & 6.8 & 0.5  & KamLAND \\
       $c_{du\nu e;LL}^{S,11\mu e}$   & $0.52-3.47$  & $C_{\bar dLQLH1}^{1e1\mu}$ & 197.2-1316.2 & 0.1-0.2  & MINOS\\
                                      &              & $C_{\bar dLQLH2}^{1e1\mu (1\mu 1e)}$ & 394.5-2631.4 & 0.1 & MINOS \\
       $c_{du\nu e;RL}^{S,11\mu e}$   & $0.52-3.47$  & $C_{\bar QuLLH}^{11e\mu}$ &  98.6-658.1 & 0.1-0.2  & MINOS\\
       $c_{du\nu e;LL}^{S,11\mu\mu}$  & $0.38-1.49$  & $C_{\bar dLQLH1}^{1\mu1\mu}$ & 144.1-565.2 & 0.1-0.2   & MINOS \\
       $c_{du\nu e;RL}^{S,11\mu\mu}$  & $0.38-1.49$  & $C_{\bar QuLLH}^{11\mu\mu}$  &  72.1-282.6 & 0.1-0.2  & MINOS \\
       $c_{du\nu e;LL}^{S,11\mu\tau}$ & $0.38-1.68$  & $C_{\bar dLQLH1}^{1\tau 1\mu}$ & 144.1-637.2 & 0.1-0.2 & MINOS \\
                                  &              & $C_{\bar dLQLH2}^{1\tau 1\mu (1\mu 1\tau)}$ & 288.3-1274.5 & 0.1-0.2  & MINOS \\
       $c_{du\nu e;RL}^{S,11\mu\tau}$ & $0.38-1.68$  & $C_{\bar QuLLH}^{11\tau\mu}$ & 72.1-318.6 & 0.1-0.2  & MINOS \\
       \midrule
       $c_{du\nu e;LL}^{T,11ee}$          & $0.041$      & $C_{\bar dLQLH1(2)}^{1e1e}$ & 62.2 & 0.3  & KamLAND\\
       $c_{du\nu e;LL}^{T,11e\mu}$        & $0.041$      & $C_{\bar dLQLH1}^{1\mu 1e}$ & 62.2 & 0.3  & KamLAND \\
                                          &              & $C_{\bar dLQLH2}^{1\mu 1e(1e1\mu)}$ & 124.4 & 0.2 & KamLAND\\
       $c_{du\nu e;LL}^{T,11e\tau}$       & $0.036$      & $C_{\bar dLQLH1}^{1\tau 1e}$ & 54.6 & 0.3  & KamLAND \\
                                              &          & $C_{\bar dLQLH2}^{1\tau 1e(1e1\tau)}$ & 109.2 & 0.3 & KamLAND\\
       $c_{du\nu e;LL}^{T,11\mu e}$       & $0.53-8.38$  & $C_{\bar dLQLH1}^{1e1\mu}$ & 804.1-12714.5 & 0.0-0.1  & MINOS\\
                                          &              & $C_{\bar dLQLH2}^{1e1\mu(1\mu1e)}$ & 1608.3-25429.0 & 0.0-0.1  & MINOS\\
       $c_{du\nu e;LL}^{T,11\mu\mu}$      & $0.39-1.52$  & $C_{\bar dLQLH1(2)}^{1\mu 1\mu}$ & 591.7-2306.2 & 0.1 & MINOS \\
       $c_{du\nu e;LL}^{T,11\mu\tau}$     & $0.39-1.71$  & $C_{\bar dLQLH1}^{1\tau 1\mu}$ & 591.7-2594.5 & 0.1  & MINOS \\
                                          &              & $C_{\bar dLQLH2}^{1\tau1\mu(1\mu1\tau)}$ & 1183.5-5189.0 & 0.1 & MINOS\\
        \bottomrule
    \end{tabular}
    \caption{Current bounds, future sensitivity and the corresponding new-physics scale for the LNV dimension-7 SMEFT Wilson coefficients under the assumption of single scalar and tensor LEFT operator dominance to LNV LBL oscillations.}
    \label{tab:so_cevns2}
\end{table}
In analogy with the above-discussed vector currents, one can put constraints also on other effective operators with different Lorentz structures, namely scalar and tensor currents
\begin{align}
\begin{aligned}
\mathcal{L}_\mathrm{LEFT}^{ d=6} &= \frac{4G_F}{\sqrt{2}}
\Big[
{c}_{du\nu e;LL}^{S,prst}(\overline{d_{Rp}}u_{Lr})
 (\overline{\nu^c_s}e_{Lt})
 +
 c_{du\nu e;RL}^{S,prst} (\overline{d_{Lp}}u_{Rr})
 (\overline{\nu^c_s}e_{Lt}) \nonumber\\
& +
 c_{du\nu e;LL}^{T,prst} (\overline{d_{Rp}}\sigma_{\mu\nu}u_{Lr})
 (\overline{\nu^c_s}\sigma^{\mu\nu}e_{Lt})\Big]\; .\\
\end{aligned}
\end{align}

The corresponding cross sections in the quasi-elastic limit are similar to Eq.~\eqref{eq:crosssection}. For both right-handed and left-handed scalar currents, the cross-section reads
\begin{align}
    \sigma_{\bar{\nu}_{\mu}p\rightarrow\ell_{\beta}^{+}n}(E_{\mathbf{q}})\backsimeq \sigma_{\nu_{\mu}n\rightarrow\ell_{\beta}^{-}p}(E_{\mathbf{q}})\backsimeq\frac{G_{F}^{2}|V_{ud}|^{2}}{\pi}\big(g_{\mathrm{S}}^{2}\big)E_{\mathbf{q}}^{2},
    \label{eq:crosssection1}
\end{align}
with $g_S = 1.02$~\cite{Gonzalez-Alonso:2018omy} denoting the scalar form factor. For tensor currents, one gets the following expression
\begin{align}
    \sigma_{\bar{\nu}_{\mu}p\rightarrow\ell_{\beta}^{+}n}(E_{\mathbf{q}})\backsimeq \sigma_{\nu_{\mu}n\rightarrow\ell_{\beta}^{-}p}(E_{\mathbf{q}})\backsimeq\frac{G_{F}^{2}|V_{ud}|^{2}}{\pi}\big(12 g_{\mathrm{T}}^{2}\big)E_{\mathbf{q}}^{2},
    \label{eq:crosssection2}
\end{align}
where $g_T = 1$~\cite{Adler:1975he} stands for the tensor form factor.

Employing the above cross sections we can constrain the Wilson coefficients of dimension-6 LEFT operators with the corresponding Lorentz structures and consequently constraint also the respective dimension-7 SMEFT operators. 
We collect the relevant current bounds on Wilson coefficients of dimension-6 LEFT, dimension-7 SMEFT operators, and the new physics scale in Tables~\ref{tab:so_cevns1} and~\ref{tab:so_cevns2}.

%%%%%%%%%%%%%%%%%%%%%%%%%%%%%%%%%%%%%%%%%%%%%%%%%%%%%%%%%%%%%%%

%%%%%%%%%%%%%%%%%%%%%%%%%%%%%%%%%%%%%%%%%%%%%%%%%%%%%%%%%%%%%%%%%%%%%%%%%%%%%%%%%%%%%%%%%%%%%%%%%%%%%%%%%%%%%%%%%%%%%%%%%%%%%%%%%%%%%%%%%%%%%%%%%%%%%%%%%%%%%%%%%%%%%%%%%%%%%%%%%%%%%%%%%%%%%%%%%%%%%%%%%%%%%%%%%%%%%%%%%%%%%%%%%

\subsection{Rare meson decays}
Lepton number violating rare meson decays are among some of the best probes for probing lepton number violation complementary to neutrinoless double beta decay. In this section, we discuss various lepton number violating rare meson and charged lepton decays which can probe the LNV dimension-7 SMEFT operators and provide the relevant constraints where they are relevant. 

%%%%%%%%%%%%%%%%%%%%%%%%%%%%%%%%%%%%%%%%%
%%%%%%%%%%%%%%%%%%%%%%%%%%%%%%%%%%%%%%%%%
%%%%%%%%%%%%%%%%%%%%%%%%%%%%%%%%%%%%%%%%%%%%%%%%%%
\subsubsection{Lepton number violating $K\rightarrow(\pi)\nu\nu$ decays}\label{sec:rarekaon}
$\Delta L=2$ dimension-7 SMEFT operators can induce long-distance contributions\footnote{The interaction between free quarks and leptons producing an effective local four-fermion coupling corresponds to the short-distance contribution, while the interaction involving hadronic (instead of free quark) degrees of freedom
in intermediate states are referred to as long-distance contributions.} to LNV $K\rightarrow(\pi)\nu\nu$ at tree level. Such a contribution from SMEFT corresponds to the sole dimension-5 LEFT operator that contributes to the neutrino magnetic moment and is given by
\begin{eqnarray}\label{numm0}
\mathcal{L}^{LEFT(d=5)}_{\Delta L=2}\supset c^{5\gamma}_{\nu\nu F} {O}^{5\gamma}_{\nu\nu F}&\equiv& \mu \,e \,{O}^{5\gamma}_{\nu\nu F}\, ,
\end{eqnarray}
where $\mu$ represents the magnetic dipole moment of neutrino and $e$ corresponds to the electric charge. The operator ${O}^{5\gamma}_{\nu\nu F}$ is defined as
\begin{eqnarray}\label{numm1}
 {O}^{5\gamma}_{\nu\nu F}&=&(\overline{\nu^C}i\sigma_{\mu\nu}\nu)F^{\mu\nu}+h.c. \, .
\end{eqnarray}
Here $F_{\mu\nu}$ is the electromagnetic field strength tensor
and $\nu =P_L \nu$ denotes left-handed active SM neutrinos. The Wilson coefficients of $\Delta L=2$ dimension-7 SMEFT operators can be matched onto the Wilson coefficient for the dimension-5 LEFT operator as follows~\cite{Cirigliano:2017djv}
\begin{eqnarray}\label{numm2}
c^{5\gamma}_{\nu\nu F}/e \equiv\mu_{i j} =  \frac{1}{2 v} \, \left(v^3  C^{ij}_{LHB} -v^3 \frac{ C^{ij}_{LHW}- C^{ji}_{LHW}}{2}\right)\,,
\end{eqnarray}
where $\mu$ and $\mathcal C_{LHB}$ are anti-symmetric in flavor indices. We note that the magnetic moment is subject to constraints from neutrino-electron scattering (in solar and reactor experiments~\cite{Bell:2006wi,Giunti:2014ixa,Canas:2015yoa}), from astrophysical limits due to globular clusters~\cite{Raffelt:1998xu} and from Coherent Elastic Neutrino Nucleus Scattering (CE$\nu$NS)~\cite{Bolton:2021pey,Miranda:2021kre}, etc. In fact the relevant constraints are rather strong  $|c_{\nu\nu F}|\leq 4\times10^{-9}~\text{GeV}^{-1}$~\cite{Canas:2015yoa}. The long-distance contribution to the $K\to \pi \nu \nu$ process proceeds through the $s\to d\gamma$ transition operator $\bar s\sigma_{\mu\nu}P_{L/R}dF^{\mu\nu}$, with the corresponding coefficient estimated to be $c_{sdF}\sim 10^{-9}~\text{GeV}^{-1}$ in the SM~\cite{Tandean:1999mg}. However, given the strong constraints on $c_{\nu\nu F}$ such long-distance contributions to $K\to\pi\nu\nu$ decay are negligible.

The LEFT LNV $|\Delta L|=2$ dimension-6 operators relevant for $K\to\pi\nu\nu$ decay are given by
\begin{align}
{O}_{d\nu;LL}^S&= (\overline{d_R} d_L)(\overline{\nu^C} \nu)+h.c.\; , \\
{O}_{d\nu;RL }^S&= (\overline{d_L} d_R)(\overline{\nu^C} \nu)+h.c.\; ,\\
{O}_{d\nu;LL}^T&= (\overline{d_R} \sigma^{\mu\nu}d_L)(\overline{\nu^C}\sigma_{\mu\nu} \nu)+h.c.\; .
\end{align}
We note that for the tensor operator $\overline{\nu_\alpha^C} \sigma^{\mu\nu} \nu_\beta$ to be non-vanishing one must have $\alpha\neq\beta$. The $\Delta L=2$ dimension-7 SMEFT operators can induce the dimension-6 LEFT operator ${O}^{S(T)}_{d\nu;LL}$. We also note that the $\mathcal{O}^S_{d\nu;RL}$ operator cannot be induced at tree-level from the LNV dimension-7 SMEFT operators as can be easily verified from Tab.~\ref{tab:left}; see also the discussion in Ref.~\cite{Deppisch:2020oyx}. The leading contributions to both ${O}^{S(T)}_{d\nu;LL}$ are induced by the SMEFT operator $\mathcal{O}_{\bar{d}LQLH1}$, with the relevant matching between the Wilson coefficient of LEFT and SMEFT given in Table~\ref{tab:left}. The LNV operator ${O}_{d\nu;LL}^S$ can contribute to the leading order chiral Lagrangian, while the tensor operator ${O}_{d\nu;LL}^T$ can only contribute at the next-to-leading order in the expansion momentum $\mathcal{O}(p^4)$ in the chiral Lagrangian and is therefore suppressed by a factor $p^2/\Lambda_\chi^2$. Therefore, in the rest of this subsection, we will focus on the scalar operator ${O}^S_{d\nu;LL}$ to constrain the SMEFT operator $\mathcal{O}_{\bar{d}LQLH1}$. To simplify the analysis we will further impose the assumption that contributions from the lepton number conserving dimension-6 LEFT operators are negligible, i.e.\ the SM contribution dominantly dictates the lepton number conserving decay mode branching fractions. It is then possible to use the experimental measurements of (semi)-invisible Kaon decays to constrain the LNV contribution due to the ${O}_{d\nu;LL}^S$ arising from dimension-7 SMEFT operator $\mathcal{O}_{\bar{d}LQLH1}$.

Firstly, let us consider the invisible Kaon decays $K_{L(S)}\to \nu \nu$. The effective Lagrangian at the leading order for $K_L\to \nu \nu$ decay can be expressed in terms of the Wilson coefficient associated with ${O}_{d\nu;LL}^S$ as~\cite{Li:2019fhz}  
\begin{eqnarray}\label{klinv}
\mathcal{L}_{K_L\to \nu\nu}&=& {i 4 G_F BF_0\over 2\sqrt{2}} \Big[ (c_{d\nu;LL}^{S, sd\alpha\beta}+c_{d\nu;LL}^{S, ds\alpha\beta})(\overline{\nu^c_\alpha}\nu_\beta)-(c_{d\nu;LL}^{S, ds\alpha\beta\ast}+c_{d\nu;LL}^{S, sd\alpha\beta\ast})(\overline{\nu_\alpha}\nu^c_\beta) \Big]K_L \; ,\nonumber\\
\end{eqnarray}
and that for $K_S\to \nu\nu$ decay as
\begin{eqnarray}\label{ksinv}
\mathcal{L}_{K_S\to \nu\nu}&=& {i 4 G_FBF_0\over 2\sqrt{2}} \Big[ (c_{d\nu;LL}^{S, sd\alpha\beta}-c_{d\nu;LL}^{S, ds\alpha\beta})(\overline{\nu^c_\alpha}\nu_\beta)-(c_{d\nu;LL}^{S, ds\alpha\beta\ast}-c_{d\nu;LL}^{S, sd\alpha\beta\ast})(\overline{\nu_\alpha}\nu^c_\beta) \Big]K_S \; ,\nonumber\\
\end{eqnarray}
where $B$ and $F_0$ are constants related by the quark condensate $B=-\langle \bar{q}q\rangle_0/(3F_0^2)$, with their numerical values given by $F_0=87$ MeV~\cite{Colangelo:2003hf} and $B\simeq 2.8$ GeV~\cite{Cirigliano:2017djv}. We note that for the pseudoscalar Kaon to decay invisibly a helicity flip is necessary for one of the neutrinos, which makes the potential contribution from lepton number conserving dimension-6 LEFT LNC dim-6 operators to these modes helicity-suppressed by light neutrino masses. The one-loop QCD running effects from the electroweak scale ($\Lambda_{\text{EW}}\approx m_W$) to the chiral symmetry breaking scale ($\Lambda_\chi\approx 2~\text{GeV}$) leads to an order one effect~\cite{Li:2019fhz}
\begin{eqnarray}
c_{d\nu;LL}^{S}(\Lambda_\chi)=1.656 \ c_{d\nu;LL}^{S}(\Lambda_{\text{EW}}).
\end{eqnarray}
%

%%%%%%%%%%%%%%%%%%%%%%%%%%%%%
%%%%%%%%%%%%%%%%%%%%%%%%%%%%%

Now at the electroweak scale under the simplifying assumption
\begin{eqnarray}\label{sim:cdLQLH1}
C_{\bar{d}LQLH1}^{1\alpha 2\beta}(\Lambda_{\text{EW}})=C_{\bar{d}LQLH1}^{1\beta 2\alpha}(\Lambda_{\text{EW}})
=C_{\bar{d}LQLH1}^{2\alpha 1\beta}(\Lambda_{\text{EW}})=C_{\bar{d}LQLH1}^{2\beta 1\alpha}(\Lambda_{\text{EW}})
\equiv C_{\bar{d}LQLH1},\nonumber\\
\end{eqnarray}
and further using $\alpha=\beta$,
 the $K_L$ branching ratio of invisible decay simplifies to be given by~\cite{Li:2019fhz}
\begin{eqnarray}
\mathcal{B}_{K_L\rightarrow \nu\nu}= 0.014\, {m_{K_L}\over \Gamma_{K_L}^{\text{Exp}}} \left| \frac{ BF_0}{\sqrt{G_F}}\, C_{\bar{d}LQLH1} \right|^2\; .
\end{eqnarray}
The relevant bound for the $K_L$ invisible decay (by subtracting all the known visible modes from Particle
Data Group~\cite{ParticleDataGroup:2012pjm}) is given by~\cite{Gninenko:2014sxa}
\begin{eqnarray}
\mathcal{B}_{K_L\rightarrow \text{invisible}}&<&6.3\times 10^{-4} \ (95\%~ \text{C.L.})\, .
\end{eqnarray}
%

%%%%%%%%%%%%%%%%%%%%%%%%%%%
%%%%%%%%%%%%%%%%%%%%%%%%%%%
The effective Lagrangians for $K^0\to \pi^0\nu\nu$ and $K^+\to \pi^+\nu\nu$ at the leading order in $\chi$PT relevant for our analysis are given by~\cite{Li:2019fhz}
\begin{eqnarray}
\mathcal{L}_{K^0\to \pi^0\nu{\nu}}&=&
{G_F B}\Big[
c_{d\nu;LL}^{S,sd\alpha\beta}(\overline{\nu^c_\alpha}\nu_\beta)
+c_{d\nu;LL}^{S,ds\alpha\beta*}(\overline{\nu_\alpha}\nu_\beta^c)\Big]K^0\pi^0\; ,
\\
\mathcal{L}_{K^+\to \pi^+\nu{\nu}}&=&
-{\sqrt{2} G_F B}\Big[
c_{d\nu;LL}^{S,sd\alpha\beta}(\overline{\nu^c_\alpha}\nu_\beta)
+c_{d\nu;LL}^{S,ds\alpha\beta*}(\overline{\nu_\alpha}\nu_\beta^c)\Big]K^+\pi^-
\; ,
\end{eqnarray}
which follow the isospin relation
\begin{eqnarray}
{\langle\pi^0| \mathcal{L}_{K^0\to \pi^0\nu{\nu}}|K^0\rangle \over \langle\pi^-|\mathcal{L}_{K^+\to \pi^+\nu{\nu}}|K^+\rangle} =-\frac{1}{\sqrt{2}} \; .
\label{eq:isospin1over2}
\end{eqnarray}
By neglecting the small CP violation in $K^0-\bar K^0$ system, and given the absence of interference between the SM contribution and the LNV contribution, the branching ratios for the decays $K_L\to \pi^0\nu{\nu}$ and $K^+\to \pi^+\nu{\nu}$ can be expressed as
\begin{eqnarray}
\mathcal{B}_{K_L\rightarrow\pi^0\nu\nu}&=&J_1^{K_L} \sum_{\alpha\leq \beta} \left(1-{1\over2}\delta_{\alpha\beta}\right) \left( \frac{4G_F}{\sqrt{2}}\right)^2\left|c_{d\nu;LL}^{S,ds\alpha\beta}+c_{d\nu;LL}^{S,sd\alpha\beta}\right|^2
+\mathcal{B}_{K_L\rightarrow\pi^0\nu\bar{\nu}}^{\text{SM}}\; ,
\\
\mathcal{B}_{K^+\rightarrow\pi^+\nu\nu}&=&J_1^{K^+}\sum_{\alpha\leq \beta} \left(1-{1\over2}\delta_{\alpha\beta}\right) \left( \frac{4G_F}{\sqrt{2}}\right)^2
\left(\left|c_{d\nu;LL}^{S,ds\alpha\beta}\right|^2+\left|c_{d\nu;LL}^{S,sd\alpha\beta}\right|^2\right)
+\mathcal{B}_{K^+\rightarrow\pi^+\nu\bar{\nu}}^{\text{SM}} \; ,\nonumber\\
\end{eqnarray}
where the three-body decay kinematics are parameterized by  
\begin{eqnarray}
	J_1^{K_L}&=&{1 \over \Gamma_{K_L}^{\text{Exp}}}{B^2\over 2^9\pi^3m_{K_L}^3}\int ds\, s\left((m_{K_L}^2+m_{\pi^0}^2-s)^2-4m_{K_L}^2m_{\pi^0}^2\right)^{1/2},
\\
J_1^{K^+}&=&{1 \over \Gamma_{K^+}^{\text{Exp}}}{ B^2\over 2^8\pi^3m_{K^+}^3}\int ds\, s\left((m_{K^+}^2+m_{\pi^+}^2-s)^2-4m_{K^+}^2m_{\pi^+}^2\right)^{1/2},
\end{eqnarray}
with $m_{K_L}(m_{K^+})$ denoting the physical mass of $K_L(K^+)$, and $\Gamma_{K_L}^{\text{Exp}}(\Gamma_{K^+}^{\text{Exp}})$ denoting the decay width of $K_L(K^+)$, respectively. Here $m_{\pi^0}(m_{\pi^+})$ denotes the mass of $\pi^0(\pi^+)$, and $s$ denotes the invariant mass squared for the neutrino pair. Again using the simplification as Eq.~\eqref{sim:cdLQLH1} and $\alpha=\beta$, the branching ratios for the decay modes $K_L\to \pi^0\nu{\nu}$ and $K^+\to \pi^+\nu{\nu}$ can be simplified to
\begin{eqnarray}
\mathcal{B}_{K_L\rightarrow\pi^0\nu\nu}&=&
{58.8 \over G_F^3} \left| C_{\bar{d}LQLH1} \right|^2+\mathcal{B}_{K_L\rightarrow\pi^0\nu\bar{\nu}}^{\text{SM}} \; ,
\\
\mathcal{B}_{K^+\rightarrow\pi^+\nu\nu}&=&
{13.0\over G_F^3}\left| C_{\bar{d}LQLH1} \right|^2+\mathcal{B}_{K^+\rightarrow\pi^+\nu\bar{\nu}}^{\text{SM}} \; .
\end{eqnarray}
The SM GIM suppressed rates for these flavor changing neutral current (FCNC) processes are $\mathcal{B}_{K_L\to \pi^0\nu \bar{\nu}}^{\text{SM}}=(3.4\pm 0.6)\times 10^{-11}$ and $\mathcal{B}_{K^+\to \pi^+ \nu \bar{\nu}}^{\text{SM}}=(8.4\pm 1.0)\times 10^{-11}$, respectively~\cite{Buras:2006gb,Brod:2010hi,Buras:2015qea}. The current best experimental limits on these processes are due to the KOTO experiment at J-PARC~\cite{Kitahara:2019lws} and the NA62 experiment at CERN~\cite{NA62:2020fhy}, with the relevant limits being
\begin{eqnarray}\label{koto}
\mathcal{B}_{K_L\to \pi^0\nu \bar{\nu}}^{\text{KOTO}}&=&2.1^{+4.1}_{-1.7}\times 10^{-9},
\end{eqnarray}
at 95\% confidence level (CL) and 
\begin{eqnarray}\label{na62}
 \mathcal{B}_{K^+\to \pi^+ \nu \bar{\nu}}^{\text{NA62}}&<&1.7\times 10^{-10} ,
\end{eqnarray}
at 90\% confidence level (CL). The NA62 experiment is expected to reach a sensitivity of $\pm10\%$ in the branching ratio for $K^+\to \pi^+ \nu \bar{\nu}$~\cite{Newson:2014ahc,Romano:2014xda}, while the KOTO experiment should improve over the final results from E391a experiment by a factor 100, providing a sensitivity $\sim 10^{-10}$ to explore the interesting region below the Grossman-Nir upper limit~\cite{Grossman:1997sk} for $K_L\to \pi^0\nu \bar{\nu}$ branching fraction~\cite{Komatsubara:2012pn,Shiomi:2014sfa,KOTOWG}.

\begin{table}[htb!]
    \centering
    \begin{tabular}{ccccc}\toprule
         \multicolumn{5}{c}{Current Bound}   \\
        LEFT Wilson \\ [-0.25ex] Coefficient & \makecell{Value \\ [-0.25ex] } & \makecell{ $C_{\bar{d}LQLH1}$ \\ [-0.25ex] [$\text{TeV}^{-3}$]} & \makecell{$\Lambda_{\text{NP}}$ \\ [-0.25ex] [TeV]} & Observable  \\
        \midrule
        ${c}^{{S,ds\gamma\gamma}}_{d\nu ;LL}$ & $1.3\times 10^{-6}$ & $4.8\times 10^{-4}$ & 12.8 & $K_L \to \nu\nu$ \\
        ${c}^{{S,ds\gamma\gamma}}_{d\nu ;LL}$ & $2.5\times 10^{-7}$ & $9.6\times 10^{-5}$  & 21.8 & $K^+\to \pi^+\nu\nu$  
        \\
        ${c}^{{S,ds\gamma\gamma}}_{d\nu ;LL}$ & $2.6\times 10^{-7}$ & $9.9\times 10^{-5}$  &  21.6  & $K^0\to \pi^0\nu\nu$ \\
        \bottomrule
           \multicolumn{5}{c}{Future Sensitivity } \\
        \midrule
        ${c}^{{S,ds\gamma\gamma}}_{d\nu ;LL}$ & $8.4\times 10^{-8}$  & $3.2\times 10^{-5}$ & 31.5 & $K^+\to \pi^+\nu\nu$ \\
        ${c}^{{S,ds\gamma\gamma}}_{d\nu ;LL}$ & $1.4\times 10^{-7}$   & $5.2\times 10^{-5}$& 26.8 & $K^0\to \pi^0\nu\nu$ \\
        \bottomrule
    \end{tabular}
    \caption{Most restrictive current bounds, future sensitivity, and the corresponding new-physics scale for the relevant Wilson coefficients under the assumption of single LEFT operator dominance to $K\rightarrow (\pi) \nu\nu$ decays.}
    \label{tab:kpnn1}
\end{table}
%%%%%%%%%%%%%%%%%%%%%%%%%%%%%%%%%%%%%%%%%%%%%%%%%%%%%%%%%%%%%%%%%%%%%%%%%%%%
%%%%%%%%%%%%%%%%%%%%%%%%%%%%%%%%%%%%%%%%%%%%%%%%%%%%%%%%%%%%%%%%%%%%%%%%%%%%%
%%%%%%%%%%%%%%%%%%%%%%%%%%%%%%%%%%%%%%%%%%%%%%%%%%%%%%%%%%%%%%%%%%%%%%%%%%%%
\subsubsection{Lepton number violating $K^+\rightarrow\pi^-\ell^+\ell^+$ decays}
The short-distance contributions to LNV $K^+\rightarrow\pi^-\ell^+\ell^+$ decays can be induced at the leading order from $\Delta L=2$ dimension-7 SMEFT operators. The EFT approach for the short-distance contributions to $K^+\rightarrow\pi^-\ell^+\ell^+$ in the context of dimension 7 SMEFT operators has been discussed in detail for instance in Ref.~\cite{Liao:2019gex}, where the $\Delta L=2$ dimension-7 SMEFT operators are matched onto dimension-9 LEFT operators, which are suppressed by $\frac{v^3}{\Lambda^3}$. As a consequence, the currently available measurements can only constrain the new physics energy scale rather loosely $\Lambda > \mathcal O(10)$ GeV, which is in fact far too low for the validity of the SMEFT approach.

On the other hand, the long-distance contributions to $K^+\rightarrow\pi^-\ell^+\ell^+$ can arise from the LNV dimension-5 and dimension-7 SMEFT operators. Ref.~\cite{Liao:2020roy} finds that the long-range contributions can be overwhelmingly dominant over the short-distance ones by orders of magnitude in the branching ratios for the decay. Relating the branching ratios for $K^+\rightarrow\pi^-\ell^+\ell^+$ to the leading LNV SMEFT dimension-7 operators using a LEFT and subsequently $\chi$PT, in Ref.~\cite{Liao:2020roy} it is found that the current experimental upper bounds on the branching ratios are too weak to set a useful constraint on the new physics scale of LNV SMEFT dimension 7 operators. Adopting their results, we find that for $\Lambda\sim 1$ TeV the branching fractions for $K^+\rightarrow\pi^-\ell^+\ell^+$ lie seven orders of magnitude below the current experimental upper bounds. It is interesting to note, however, that in the presence of sterile neutrinos in addition to the SM neutrinos (e.g. in the framework of $\nu$SMEFT) it is possible to have resonant enhancements which can lead to sizeable contributions to $K^+\rightarrow\pi^-\ell^+\ell^+$~\cite{Zhou:2021lnl}. However, such scenarios are beyond the scope of this work and will be addressed elsewhere. 
%%%%%%%%%%%%%%%%%%%%%%%%%%%%%%%%%%%%%%%%%%%%%%%%%%
%%%%%%%%%%%%%%%%%%%%%%%%%%%%%%%%%%%%%%%%%%%%%%%%%%%%%%%%%%%%%%%%%%%%%%%%%%%%%
%%%%%%%%%%%%%%%%%%%%%%%%%%%%%%%%%%%%%%%%%%%%%%%%%%%%%%%%%
\subsubsection{Lepton number violating $B\rightarrow K^{(*)} \nu\nu$ decays}
The $\Delta L=2$ dimension-7 SMEFT operators can induce LNV $b\to s \nu \nu$ transition leading to $B\rightarrow K^{(*)} \nu\nu$ decays. Such contributions correspond to the dimension-6 LEFT Lagrangian~\cite{Aebischer:2017gaw,Jenkins:2017jig,Felkl:2021uxi}
\begin{align}{\label{eq:bsnn1}}
    \mathcal{L} &= \frac{4G_F}{\sqrt{2}} \left(  c_{d\nu;LL}^{{S}} {O}_{d\nu;LL}^{{S}}
    +c_{d\nu;LL}^{{T}} {O}_{d\nu;LL}^{{T}}
    +\mathrm{h.c.}\, ,
    \right)
\end{align}
the relevant effective operators and the matching of the Wilson coefficients corresponding to dimension-7 SMEFT operators and dimension-6 LEFT operators are defined in Table~\ref{tab:left}. Note that we neglect the dimension-5 neutrino dipole interaction of the form  $\overline{\nu_L^c}\sigma^{\mu\nu}\nu_L F_{\mu\nu}$, since it can only appear together with the down quark dipole operator $\overline{d_L}\sigma^{\mu\nu}d_R F_{\mu\nu}$. The contribution of such operators is subject to double loop suppression as well as stringent constraints from searches for magnetic dipole moments of neutrinos~\cite{Beda:2012zz,Borexino:2017fbd}. Since experimentally the $B\rightarrow K^{(*)} \nu\nu$ mesonic decay cannot be distinguished for lepton number violating vs. conserving modes based on the neutrino nature, the relevant lepton number conserving SM contribution can be written as
\begin{align}{\label{eq:bsnn2}}
    \mathcal{L}^{\text{LNC}} &= \frac{4G_F}{\sqrt{2}}
    c^{V;\text{SM}}_{d\nu ;XL} (\overline{d_{L}}\gamma^\mu d_{L})(\overline{\nu}\gamma_\mu\nu)\; ,
\end{align}
where the relevant SM Wilson coefficient corresponding to the left-chiral vector interaction is given by ~\cite{Brod:2010hi}
\begin{equation}\label{eq:SM}
    c_{d\nu;LL}^{{V,sb\alpha\alpha},\text{SM}} = - \frac{\alpha}{2\pi} V_{ts}^* V_{tb} \left(\frac{X}{\sin^2\theta_W}\right) \;.
\end{equation}
Here the function $X$ parametrises the electroweak corrections induced by the top quark given by~\cite{Straub:2018kue,Felkl:2021uxi}
$X = 6.402 \sin^2\theta_W$. We note that the LNV scalar operator ${O}_{d\nu;LL}^{{S}}$ is symmetric in the neutrino flavors, while the tensor operator $\mathcal{O}_{d\nu;LL}^{{T}}$ is antisymmetric in the neutrino flavors. The dominant RG running effects for the relevant scalar and tensor operators are due to QCD. Numerically, the Wilson coefficients at the hadronic scale $\Lambda_H=4.8$ GeV can be expressed in terms of the Wilson coefficients at the electroweak scale $\Lambda_{\text{EW}}$~\cite{Felkl:2021uxi}
 \begin{equation}
     \begin{aligned}
     C_{d\nu;LL}^{{S}}(4.8\, \mathrm{GeV}) & = 1.4\, C_{d\nu;LL}^{{S}}(\Lambda_{\text{EW}})\;, 
     &
     C_{d\nu;LL}^{{T}}(4.8\, \mathrm{GeV}) & = 0.9\, C_{d\nu;LL}^{{T}}(\Lambda_{\text{EW}})\;.
 \end{aligned}
 \end{equation}
 Searches for $B\rightarrow K^{(*)} \nu\nu$ have been performed by BaBar and Belle experiments using hadronic and semileptonic tagging with the relevant upper limits being a factor two to five above the SM predictions. Belle II is expected to improve the predictions to about $10\%$ of the SM predictions with 50$\text{ab}^{-1}$ data~\cite{Belle-II:2018jsg}. In Table~\ref{tab:btosnn} we have collected the SM predictions, current upper limits, and future sensitivities of Belle II. 
 \begin{table}[h]
    \centering
    \begin{tabular}{lcc cc}\toprule
        Observable & SM prediction & current constraint & {Belle II \cite{Belle-II:2018jsg}}  \\ 
         & LQCD+LCSR  & & 50 ab$^{-1}$  \\ \midrule
         Br($B^0\to K^0 \nu\nu$) & $(4.1\pm0.5) \times 10^{-6}$ &
         $<2.6\times 10^{-5}$~\cite{Belle:2017oht}
         &   \\
         Br($B^+\to K^+ \nu\nu$) & $(4.4\pm0.7)\times 10^{-6}$ & $<1.6\times 10^{-5}$~\cite{BaBar:2013npw} &  11\% \\
         Br($B^0\to {K^*}^{0} \nu\nu$) & $(11.6\pm 1.1)\times 10^{-6}$ &$<1.8\times 10^{-5}$~\cite{Belle:2017oht} &  9.6\% \\
         Br($B^+\to K^{*+} \nu\nu$) & $(12.4\pm 1.2)\times 10^{-6}$ &$<4.0\times 10^{-5}$~\cite{Belle:2013tnz} &  9.3\% \\\bottomrule
    \end{tabular}
    \caption{The SM predictions for exclusive $b\to s \nu\nu$ decays (based on light-cone sum rules (LCSR) and lattice QCD~\cite{Gubernari:2018wyi,Bharucha:2015bzk,Felkl:2021uxi}), current constraints, and future Belle-II sensitivities (with respect to the SM predictions)~\cite{Belle-II:2018jsg}.} 
    \label{tab:btosnn}
\end{table}

A detailed discussion on the formalism to compute the exclusive decay $B\rightarrow K^{(*)} \nu\nu$ can be found for instance in Ref.~\cite{Felkl:2021uxi}. We note that the scalar and tensor operators do not interfere with the vector operators due to the different chiralities of the neutrinos. Therefore, in the absence of interference with the SM contributions the scalar and tensor operators are subject to strong constraints from $B\rightarrow K^{(*)} \nu\nu$ decays\footnote{After the submission of this paper to arXiv, the Belle~II collaboration reported an excess in the measurement of $B\rightarrow K \nu\nu$ decays~\cite{Belle-II:2023esi}. For a scalar current dimension-6 LEFT operator with massless neutrinos this excess corresponds to a new physics scale of about $\Lambda_\text{dim-6}\sim 4.1$~TeV~\cite{Fridell:2023ssf}. Matching onto the dimension-7 SMEFT basis operator $\mathcal{O}_{\bar dLQLH1}$ this corresponds to a SMEFT new physics scale $\Lambda_{\bar dLQLH1}\sim 0.8$~TeV. Note however that $\mathcal{O}_{\bar dLQLH1}$ is constrained by kaon decays at a much greater scale, as shown in Sec.~\ref{sec:rarekaon}, in the absence of any special quark flavor dependent coupling structure of new physics. As we wait for the confirmation of the anomaly, we use the limit for $B\rightarrow K \nu\nu$ based on Ref.~\cite{BaBar:2013npw}.}. In Table~\ref{tab:so_dom1} we summarise the limits on the LNV dimension-7 SMEFT operator $C_{\bar{d}LQLH1}$ using the limits on the LEFT operators presented in Ref.~\cite{Felkl:2021uxi} and then using our matching relations between SMEFT and LEFT operators in Tab.\ref{tab:left}. We note that these limits are valid under the assumption that only one of the LEFT operators contributes dominantly at a time and the neutrinos are massless.
\begin{table}[htb!]
    \centering
    \begin{tabular}{ccccc}\toprule
         \multicolumn{5}{c}{Current Bound}   \\
        LEFT Wilson \\ [-0.25ex] Coefficient & \makecell{Value \\ [-0.25ex] } & \makecell{ $C_{\bar{d}LQLH1}$ \\ [-0.25ex] [$\text{TeV}^{-3}$]} & \makecell{$\Lambda_{\text{NP}}$ \\ [-0.25ex] [TeV]} & Observable  \\
        \midrule
        ${c}^{{S,sb\gamma\gamma}}_{d\nu ;LL}$ & $3.6\times 10^{-4}$ & 0.14 & 1.9 & $B\to K^{(*)}\nu\nu$  \\
        ${c}^{{S,sb\gamma\delta}}_{d\nu ;LL }$ & $2.7\times 10^{-4}$ & 0.21 & 1.7 & $B\to K^{(*)}\nu\nu$  \\
        ${c}^{{T,sb\gamma\delta}}_{d\nu;LL }$ & $0.6\times 10^{-4}$ & 0.18 &  1.75  & $B\to K^*\nu\nu$ \\
        \bottomrule
           \multicolumn{5}{c}{Future Sensitivity (50 ab$^{-1}$)} \\
        \midrule
        ${c}^{{S,sb\gamma\gamma}}_{d\nu ;LL}$ &  $0.6\times 10^{-4}$ & 0.02 & 3.5 & $B\to K\nu\nu$ \\
        ${c}^{{S,sb\gamma\delta}}_{d\nu ;LL }$ &  $0.6\times 10^{-4}$ & 0.05 & 2.8 & $B\to K\nu\nu$ \\
        ${c}^{{T,sb\gamma\delta}}_{d\nu;LL }$ &  $0.3\times 10^{-4}$ & 0.08 & 2.3 & $B\to K^*\nu\nu$ \\
        \bottomrule
    \end{tabular}
    \caption{Most restrictive current bounds, future sensitivity of the LNV dimension-7 SMEFT operator $\mathcal{O}_{\bar{d}LQLH1}$ and the corresponding new-physics scale for the relevant Wilson coefficients under the assumption of single LEFT operator dominance to exclusive $B\rightarrow K^{(*)} \nu\nu$ decays and for massless neutrinos.}
    \label{tab:so_dom1}
\end{table}
%

%%%%%%%%%%%%%%%%%%%%%%%%%%%%%%%%%%%%%%%%%%%%%%%%%%%%%%%%%%%%%%%%%%%%%%%%%%%%%
%%%%%%%%%%%%%%%%%%%%%%%%%%%%%%%%%%%%%%%%%%%%%%%%%%%%%%%%%%%%%%%%%%%%%%%%%%%%%
\subsection{Charged Lepton Flavor violating lepton decays}
Charged lepton flavor violating lepton decays provides yet another alternative for probing lepton number violation. In particular, these observables are very exciting because they give access to lepton number violation in the second and third generation of charged leptons in contrast to the neutrinoless double beta decay which is sensitive to the first-generation leptons. In this subsection, we discuss the various relevant charged lepton flavor violating decays which can be relevant for probing $\Delta L=2$ dimension-7 SMEFT operators. 
%%%%%%%%%%%%%%%%%%%%%%%%%%%%%%%%%%%%%%%%%%%%%%%%%%%%%%%%%%%%%%%%%%%%%%%
\subsubsection{Lepton number violating $\tau$ decays}
The three-body semileptonic $\tau$ lepton decays such as $\tau^\pm\rightarrow \ell^\mp_\alpha P_i^\pm P_j^\pm$, with $P_{i,j}^\pm=\pi^\pm,~K^\pm$ can potentially constrain LNV dimension-7 SMEFT operators. These processes can be of particular interest since the relevant SMEFT Wilson coefficients involve first and second-generation quarks in combination with different generations of leptons. However, the bottleneck of using these modes to derive useful constraints is the suppression of amplitude due to the light left-handed neutrino mass. The current best upper limits on the branching ratios of these decays are due to the Belle experiment~\cite{Belle:2012unr}
\begin{align}
	\label{ul1}
	&\mathcal{B}(\tau^-\rightarrow e^+\pi^-\pi^-)<2.0\times10^{-8},
	&&\mathcal{B}(\tau^-\rightarrow \mu^+ \pi^-\pi^-)<3.9\times10^{-8},\\
	\label{ul2}
	&\mathcal{B}(\tau^-\rightarrow e^+K^-K^-)<3.3\times10^{-8},
	&&\mathcal{B}(\tau^-\rightarrow \mu^+ K^-K^-)<4.7\times10^{-8},\\
	\label{ul3}
	&\mathcal{B}(\tau^-\rightarrow e^+K^-\pi^-)<3.2\times10^{-8},
	&&\mathcal{B}(\tau^-\rightarrow \mu^+ K^-\pi^-)<4.8\times10^{-8}.
\end{align}
The above limits are expected to be improved by orders of magnitude by the Belle II experiment~\cite{Belle-II:2018jsg,RodriguezPerez:2019nhw}, with the projected sensitivities at $50\, {\text{ab}}^{-1}$ being
\begin{align}
	\label{ul1}
	&\mathcal{B}(\tau^-\rightarrow e^+\pi^-\pi^-)<3\times10^{-10},
	&&\mathcal{B}(\tau^-\rightarrow \mu^+ \pi^-\pi^-)<7\times10^{-10},\\
	\label{ul2}
	&\mathcal{B}(\tau^-\rightarrow e^+K^-K^-)<6 \times10^{-10},
	&&\mathcal{B}(\tau^-\rightarrow \mu^+ K^-K^-)<9\times10^{-10},\\
	\label{ul3}
	&\mathcal{B}(\tau^-\rightarrow e^+K^-\pi^-)<6\times10^{-10},
	&&\mathcal{B}(\tau^-\rightarrow \mu^+ K^-\pi^-)<9\times10^{-10}.
\end{align}
In spite of the expected significant improvements in the experimental sensitivities, even if the new physics scale is as low as 100 GeV in the SMEFT Wilson coefficients the relevant branching fractions induced by the $\Delta L=2$ dimension-7 SMEFT operators are well below the BELLE II sensitivities (by at least three orders of magnitude)~\cite{Liao:2021qfj}. Therefore, we do not include the constraints from these processes in our analysis. However, it is worth noting that in the presence of additional heavy neutral leptons in addition to the SM neutrinos (e.g. in the context of $\nu$SMEFT) these processes can potentially give useful constraints complimentary to $0\nu\beta\beta$ and LNV kaon decays~\cite{Atre:2005eb,Atre:2009rg,Abada:2017jjx,Abada:2019bac}.
%%%%%%%%%%%%%%%%%%%%%%%%%%%%%%%%%%%%%%%%%%%
%%%%%%%%%%%%%%%%%%%%%%%%%%%%%%%%%%%%%%%%%%%
\subsubsection{Non-standard muon decay}
\label{ss:NSMD}
Non-standard muon decays can be induced by $\mathcal O_{\bar eLLL H}$ at the tree level. It is interesting to note that this particular operator does not contribute to $0\nu\beta\beta$ decay at the tree level. Therefore, non-standard muon decays can provide crucial constraints on $\mathcal C_{\bar eLLL H}$. After electroweak symmetry breaking, the $\Delta L = 2$ Lagrangian relevant for the non-standard muon decay can be written as~\cite{Cirigliano:2017djv}
\begin{eqnarray}
\mathcal L &=& - \frac{4 G_F}{\sqrt{2}} \Bigg\{  c_{e\nu;LL}^{{S},\mu e e\mu}\,  (\overline{\mu_R} e_L) \, (\bar{\nu^c_e}  \nu_{\mu})  + c_{e\nu;LL}^{{S},e \mu  e\mu}\,  (\overline{e_R} \mu_L) \, (\bar{\nu^c_e}  \nu_{\mu})  \nonumber \\
& &  +  c_{e\nu;LL}^{{T},\mu e e\mu}\,  (\overline{\mu_R} \sigma_{\mu \nu} e_L) \, (\bar{\nu^c_e}  \sigma^{\mu \nu} \nu_{\mu})  
     +  c_{e\nu;LL}^{{T},e \mu  e\mu}\,  (\overline{e_R} \sigma_{\mu \nu}  \mu_L) \, (\bar{\nu^c_e} \sigma^{\mu \nu}  \nu_{ \mu})
\Bigg\} + \text{h.c.}\,\,,
\end{eqnarray}
where only $c_{e\nu;LL}^{ S (T),\mu e e\mu}$ can mediate the $\Delta L = 2$ decays $\mu^+ \rightarrow e^+ \bar{\nu}_e \bar{\nu}_\mu$, which was searched for experimentally. The experiment~\cite{Armbruster:2003pq} used the charged current processes 
$p \, \bar{\nu}_e \rightarrow e^+ n$ and $^{12}{\rm C} \, \bar{\nu}_e \rightarrow e^+\, n\, ^{11}$B following the decay of the muon to identify $\bar{\nu}_e$ in the decay products of a $\mu^+$ at rest. The muonic neutrino on the other hand is not identified. The hermitian conjugate operators $c_{e\nu;LL}^{ S (T),\mu e e\mu *}$ mediate the $\Delta L = 2$ decays  $\mu^- \rightarrow e^-  {\nu}_e {\nu}_\mu$ and $c_{e\nu;LL}^{ S (T),e\mu  e\mu}$ (and $c_{e\nu;LL}^{ S (T),e\mu  e\mu *}$) can induce  $\mu^- \rightarrow e^- \bar{\nu}_e \bar{\nu}_\mu$  (and $\mu^+ \rightarrow e^+  {\nu}_e {\nu}_\mu$). The relevant coefficients $c_{e\nu;LL}^{{S},\mu e e\mu}$ and $c_{e\nu;LL}^{{T},\mu e e\mu}$ are related to the SMEFT dimension 7 Wilson coefficient $\mathcal C_{\bar eLLL H}$ by
\begin{eqnarray}
c_{e\nu;LL}^{{S},\mu e e\mu}  &=&-\frac{3 v^3}{8\sqrt{2}} \left(  \mathcal   C^{\mu e\, \mu e}_{\bar eLLL H} + \mathcal C^{\mu e\, e \mu}_{\bar eLLL H}  \right)\;, \nonumber\\
c_{e\nu;LL}^{{T},\mu e e\mu}  &=&-\frac{v^3}{32\sqrt{2}} \left(  \mathcal C^{\mu\mu\, e e}_{\bar eLLL H} -  \mathcal C^{\mu e\, e \mu}_{\bar eLLL H}  \right)\;. 
\end{eqnarray}
 as can be verified from Table~\ref{tab:left}. Under the assumption that there are no $\Delta L =0$ lepton-flavor violating operators contributing to $\mu^+ \rightarrow e^+ \bar{\nu}_e  {\nu}_\mu$, the experimental limits on the branching ratio can be used to constrain $c^{\mu\, e}_{\rm S,\, T}$. The relevant branching ratio can then be expressed in terms of $c^{\mu e}_{\rm S,\, T}$ to be~\cite{Cirigliano:2017djv}
\begin{eqnarray}
\textrm{BR} \left(\mu^+ \rightarrow e^+ \bar{\nu}_e \bar{\nu}_\mu\right) = \frac{\Gamma\left(\mu^+ \rightarrow e^+ \bar{\nu}_e \bar{\nu}_\mu\right)}{\Gamma\left(\mu^+ \rightarrow e^+ {\nu}_e \bar{\nu}_\mu\right)}
= \frac{1}{4} \left|c_{e\nu;LL}^{{S},\mu e e\mu}\right|^2 + \frac{3}{4} \left| c_{e\nu;LL}^{{T},\mu e e\mu} \right|^2.
\end{eqnarray}
 The 90\% C.L. limits on the branching ratio using the experimental limit from~\cite{Armbruster:2003pq}
\begin{equation}
\textrm{BR} \left(\mu^+ \rightarrow e^+ \bar{\nu}_e \bar{\nu}_\mu, \tilde{\rho} = 0.75 \right) < 0.9 \cdot 10^{-3}, \qquad \
\textrm{BR} \left(\mu^+ \rightarrow e^+ \bar{\nu}_e \bar{\nu}_\mu, \tilde{\rho} = 0.25 \right) < 1.3 \cdot 10^{-3},
\end{equation}
implies~\cite{Cirigliano:2017djv} $|c_{e\nu;LL}^{{S},\mu e e\mu}| < 0.06$ and $|c_{e\nu;LL}^{{T},\mu e e\mu}| < 0.04$. Here $\tilde{\rho}$ is the Michel parameter determining the dependence of the decay rate on the $\bar{\nu}_e$ energy. The relevant limits on the SMEFT Wilson coefficient and the new physics scales are tabulated in Table~\ref{tab:mudecay}.
\begin{table}[htb!]
    \centering
    \begin{tabular}{ccccc}\toprule
         \multicolumn{5}{c}{Current Bound}   \\
        LEFT Wilson \\ [-0.25ex] Coefficient & \makecell{Value \\ [-0.25ex] } & \makecell{ $C_{\bar eLLL H}$ \\ [-0.25ex] [$\text{TeV}^{-3}$]} & \makecell{$\Lambda_{\text{NP}}$ \\ [-0.25ex] [TeV]} & Observable  \\
        \midrule
        $c_{e\nu;LL}^{{S},\mu e e\mu}$ & $0.06$ & 15.2 & 0.4 & $\mu^+ \rightarrow e^+ \bar{\nu}_e \bar{\nu}_\mu, \tilde{\rho} = 0.75$  \\
        $c_{e\nu;LL}^{{T},\mu e e\mu}$ & $0.04$ & 121.6 & 0.2 & $\mu^+ \rightarrow e^+ \bar{\nu}_e \bar{\nu}_\mu, \tilde{\rho} = 0.25$  \\
        \bottomrule
    \end{tabular}
    \caption{Current bounds  and the corresponding new-physics scale for the relevant Wilson coefficients under the assumption of single LEFT operator dominance to $\mu^+ \rightarrow e^+ \bar{\nu}_e \bar{\nu}_\mu$.}
    \label{tab:mudecay}
\end{table}
%
%%%%%%%%%%%%%%%%%%%%%%%%%%%%%%%%%%%%%%%%%%%%%%%%%%%%%%%%%%%%
%%%%%%%%%%%%%%%%%%%%%%%%%%%%%%%%%%%%%%%%%%%%%%%%%%%%%%%%%%%%
\subsubsection{$\mu^-$ to $e^+$ conversion}
The experiments searching for the lepton flavor violating neutrinoless $\mu^-\to e^-$ conversion in nuclei can potentially also be sensitive to the LNV mode $\mu^-\to e^+$ conversion in nuclei~\cite{Mu2e:2014fns,Kuno:2015tya}. The current best limit for the LNV mode is due to the SINDRUM II collaboration which used titanium~\cite{SINDRUMII:1998mwd}
\begin{figure}[h!]
    \centering
    \includegraphics[width=0.39\textwidth]{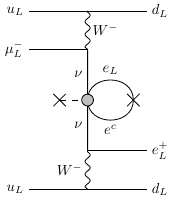}
    \caption{Leading order contribution to LNV $\mu^-\to e^+$ conversion due to $\mathcal{O}_{\bar{e}LLLH}$.}
    \label{fig:lnvmue}
\end{figure}
\begin{equation}
\label{sinII}
R_{\mu^- e^+}^\text{Ti} \equiv \frac{\Gamma(\mu^- + \text{Ti} \to e^+ + \text{Ca})}{\Gamma(\mu^- + \text{Ti} \to \nu_\mu + \text{Sc})} < \left\{
\begin{array}{l}
1.7 \times 10^{-12} \text{ ( 90\% CL)}\; , \\ 
3.6 \times 10^{-11} \text{ ( 90\% CL)}\; ,  \end{array}  \right.
\end{equation}
where the top and bottom limits correspond to the coherent scattering to ground state calcium and transition to giant dipole resonance states, respectively. There are several upcoming experiments planned for $\mu^- \to e^-$ conversion which promises a significant enhancement in the future sensitivities e.g., Mu2e~\cite{Mu2e:2014fns}, DeeMe~\cite{Natori:2014yba}, COMET~\cite{COMET:2009qeh} and its upgrade PRISM \cite{Kuno:2012pt})
\begin{align*}
\text{DeeMe:} & \quad R^{\text{SiC}}_{\mu^- e^-} > 5 \times 10^{-14} \text{ (90\% CL)}, \\
\text{Mu2e:} & \quad R^{\text{Al}}_{\mu^-e^-} >  6.6 \times 10^{-17} \text{ (90\% CL)}, \\
\text{COMET Phase-I:} & \quad R^{\text{Al}}_{\mu^-e^-} > 7.2 \times 10^{-15} \text{ (90\% CL)}, \\
\text{COMET Phase-II:} & \quad R^{\text{Al}}_{\mu^-e^-} > 6 \times 10^{-17} \text{ (90\% CL)}, \\
\text{PRISM:} & \quad R^{\text{Al}}_{\mu^-e^-} > 5 \times 10^{-19} \text{ (90\% CL)}.
\end{align*}
Among these experiments, Mu2e and COMET Phase-I can determine if the final state lepton is an electron or a positron. Therefore these are sensitive to the LNV mode $\mu^-\to e^+$ conversion in nuclei. A detailed discussion regarding a simplified effective operator approach in the context of LNV dimension-5, -7, and -9 SMEFT operators (in the absence of any covariant derivatives and field strength tensors) can be found in~\cite{Berryman:2016slh}. Unfortunately, the current constraints from SINDRUM II are too weak to provide any useful constraint on the LNV dimension-7 SMEFT operators. In spite of the future sensitivity increasing by 5-6 orders of magnitude, for an $\mathcal{O}(\text{TeV})$ new physics mass scale and $\mathcal{O}$(1) couplings the projected conversion ratios are still beyond the reach of the upcoming experiments\footnote{Furthermore, ab-initio computations for the relevant nuclear matrix elements are also not abundant in literature and need more attention.}. To discuss the sensitivity of neutrinoless $\mu^-\to e^-$ conversion to LNV dimension-7 SMEFT operators let us consider the operator $\mathcal{O}_{\bar{e}LLLH}$ and use the prescription of ~\cite{Berryman:2016slh}. The leading contribution from this operator to LNV $\mu^-\to e^+$ conversion occurs at the one-loop level as shown in Fig.~\ref{fig:lnvmue}. The amplitude for such an operator should scale with $\frac{1}{\Lambda^3}$ and the loop should contribute as $\int \frac{d^4p}{(2\pi)^4 p^2}\sim \frac{\Lambda^2}{16 \pi^2}$, where $p$ is the loop momentum. Each $W$ boson propagator along with gauge interaction vertices should contribute $G_F/\sqrt{2}$ and the nearly massless neutrino propagators should scale with the inverse of the momentum $q$. Therefore the conversion rate should scale as
\begin{eqnarray}
R_{\mu^-e^+} &\equiv& \frac{\Gamma(\mu^- + \text{N} \to e^+ + \text{N}')}{\Gamma(\mu^- + \text{N} \to \nu_\mu + \text{N}'')}\nonumber\\
&=&\frac{|g_{e\mu}|^2 \left(\frac{G_F}{\sqrt{2}}\right)^4 \left( \frac{1}{q^2}\right)^2 \left(\frac{y_\tau v \Lambda^2}{16\pi^2}\right)^2 v^2 \left(\frac{1}{\Lambda^3}\right)^2 Q^8 \, |\psi_{100}(0)|^2 }
{\left(\frac{G_F}{\sqrt{2}}\right)^2 Q^2 \, |\psi_{100}(0)|^2 }\nonumber\\
&=&|g_{e\mu}|^2 \left(\frac{G_F}{\sqrt{2}}\right)^2 \left( \frac{1}{q^2}\right)^2 \left(\frac{y_\tau v^2}{16\pi^2}\right)^2 \left(\frac{Q^6}{\Lambda^2}\right).
\end{eqnarray}
where $\psi_{100}(0)$ is the wave function of the incoming muon in the $1s$ ground state of the atom and from a mass dimension scaling $|\psi_{100}(0)|^2$ should scale as $[M]^3$. $Q$ is a mass dimension one quantity that parametrizes the dependence on phase space, nuclear matrix elements, etc. We note that the above estimation mainly tries to capture the new physics scaling behavior and ignores most of the technicalities involved in a careful estimation of the actual rates. Nevertheless, for our purpose, the above-mentioned estimate shows that for a new physics scale $\Lambda \sim 1$ TeV the conversion rate is $\mathcal{O}(10^{-24})$, which is well beyond the expected sensitivity of Mu2e. Given the order of magnitude estimate as well as the largely unsettled uncertainties due to the lack of nuclear matrix elements, we do not include LNV $\mu^-\to e^+$ conversion in nuclei in our final parameter space analysis\footnote{Note that a recent study in Ref.~\cite{Conlin:2020veq} discuss the charged lepton flavour violating muonium-antimuonium oscillations in the context of LEFT operators. However, as such a process does not violate total lepton number, it does not provide constraints for the LNV dimension-7 SMEFT operators.}. 

%%%%%%%%%%%%%%%%%%%%%%%%%%%%%%%%%%%%%%%%%%%%%%%%%%%%%%
%%%%%%%%%%%%%%%%%%%%%%%%%%%%%%%%%%%%%%%%%%%%%%%%%%%%%%%
\subsection{Constraints from neutrino magnetic moment}
\label{ref:mag_mom}
In addition to the low-energy observables discussed above, the LNV dimension-7 LNV operators $\mathcal{O}_{LHB}$ and $\mathcal{O}_{LHW}$ can induce a neutrino magnetic moment, which is subject to very stringent constraints from neutrino-electron scattering in solar and reactor experiments~\cite{Bell:2006wi,Giunti:2014ixa,Canas:2015yoa,AristizabalSierra:2021fuc,deGouvea:2022znk}\footnote{In the presence of additional sterile states the coherent elastic nucleus-neutrino scattering also becomes a very relevant probe for neutrino magnetic moment~\cite{Magill:2018jla,Brdar:2020quo,Schwetz:2020xra} and can even lead to the possibility to distinguish the Dirac vs.\ Majorana nature of neutrinos~\cite{Bolton:2021pey}.}, as well as from other astrophysical probes~\cite{Raffelt:1998xu}. We remind the readers that the LNV dimension-7 LNV operators $\mathcal{O}_{LHB}$ and $\mathcal{O}_{LHW}$ are related to the neutrino magnetic moment via the relation (c.f.\ Eq.~\eqref{numm2})
\begin{eqnarray}\label{mm1}
c^{5\gamma}_{\nu\nu F}/e \equiv\mu_{i j} =  \frac{1}{2 v} \, \left(v^3  C^{ij}_{LHB} -v^3 \frac{ C^{ij}_{LHW}- C^{ji}_{LHW}}{2}\right)\,,
\end{eqnarray}
where for Majorana neutrinos the magnetic moment has the form 
\begin{equation}
{\cal L}_M\supset\frac{1}{2}\left(
\begin{array}{ccc}
\nu_1 & \nu_2 &  \nu_3 
\end{array}
\right) \sigma_{\mu\nu}
\left(
\begin{array}{ccc}
0 & \mu_{12} &  \mu_{13} \\
-\mu_{12} & 0 &   \mu_{23} \\
-\mu_{13} & -\mu_{23} &  0 
\end{array}
\right)
\left(
\begin{array}{c}
\nu_1 \\ \nu_2 \\  \nu_3 
\end{array}
\right) F^{\mu\nu} + \text{h.c.}, 
\end{equation}
with $\nu_{1,2,3}$ denoting the three mass eigenstates of neutrinos. Solar neutrino experiments such as Borexino use the incoherent admixture of solar neutrino mass eigenstates scattering off of electrons to derive limits on neutrino magnetic moments. Such an experiment is sensitive to the combination
\begin{equation}
|\mu|^2_{\rm solar}=P_1|\mu_1^{\rm eff}|^2+P_2|\mu_2^{\rm eff}|^2+P_3|\mu_3^{\rm eff}|^2\; ,
\end{equation}
where the incoming flux of solar neutrino is a mixture of the neutrino mass eigenstate $\nu_i$ with probability $P_i$, with $i=1,2,3$. On the other hand, reactor-based experiments such as GEMMA, TEXONO, CONUS, etc. employ a reactor-based electron antineutrino flux, while accelerator-based experiments such as LSND, DUNE, etc. have access to a mixture of electron and muon neutrinos produced from pion and muon decays. A recent detailed analysis of the same can be found for example in~\cite{deGouvea:2022znk}, which provides the combined limits $|\mu_{12}|< 0.64 \times 10^{-11} \mu_B$, $|\mu_{13}|< 0.75 \times 10^{-11} \mu_B$ and  $|\mu_{13}|< 1.1 \times 10^{-11} \mu_B$. The relevant limits on the flavor eigenstate basis are subject to the PMNS mixing matrix. For our purposes, we will follow the approach of~\cite{Cirigliano:2017djv} and use the following conservative estimate in the flavor basis
\begin{equation}
|\mathcal C_{LHB}^{ij}-\mathcal C_{LHW}^{ij}|_{i\neq j}\lesssim \frac{10^{-11}}{4 m_e v^2}
\end{equation}
which under the assumption of single SMEFT operator dominance leads to an order of magnitude limit $\Lambda>11$ TeV. 
%%%%%%%%%%%%%%%%%%%%%%%%%%%%%%%%%%%%%%%%%%%%%%%%%%%%%%%%%%%%%%%%%%%%%%%%%%%%%%%%%

\section{A global view and implications of the current and projected future constraints}
\label{sec:overview}
%%%%%%%%%%%%%%%%%%%%%%%%%%%%%%%%%%%%%%%%%%%%%%%%%%%%%%%%%%%%%%%%%%%%%%%%%%%%%%%%%

\begin{figure}
    \centering
    \includegraphics[width=\textwidth]{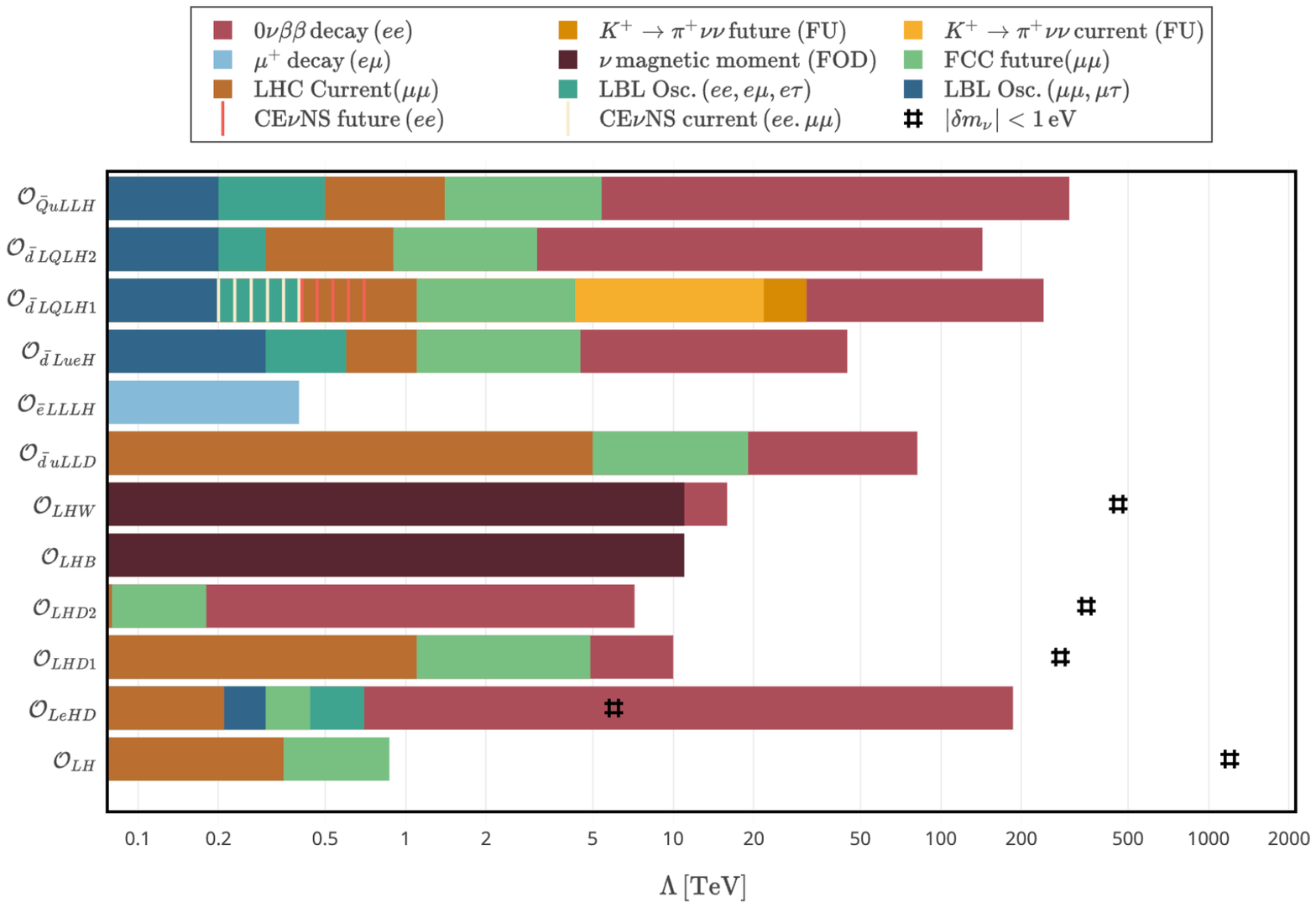}
    \caption{A summary of the current and projected future constraints on the new physics scales corresponding to the 12 independent LNV dimension-7 SMEFT operators from various current and future experiments considered in this work. Different observables that we consider are represented by the given colors indicated in the plot legend. The new physics reach of different observables increases from left to right. The relevant lepton flavor structure for the operators constrained by a given observable is also indicated in the legend in the parenthesis after the observable. Some of the abbreviations used are flavor off-diagonal (FOD) and flavor universal (FU). We indicate in the plot also the limits on some of the operators from a naive neutrino mass correction argument using $\#$ symbols. The operator $\mathcal{O}_{LH}$ is constrained from $0\nu\beta\beta$ decay $\Lambda> 1.9\times 10^{12}$ TeV (not shown in the plot explicitly to keep the other constraints clearly visible). }
    \label{fig:barplotflavor}
\end{figure}

As we have discussed in previous sections, the new physics scale of LNV dimension-7 SMEFT operators is constrained by a number of different experiments at various energy scales with several of those constraints expected to be improved significantly in the near future. In this section, we will discuss a global view of all relevant constraints and discuss their implications. Fig.~\ref{fig:barplotflavor} summarises all the current and future sensitivities from various ongoing and future experiments considered in this work for the new physics scale of different LNV dimension-7 SMEFT operators. Each observable that we consider is represented by a given color indicated in the plot legend, with the new physics reach increasing from left to right. However, most of the constraints are particularly applicable to the given flavor structure for the operators as indicated in the plot legend. If a color is missing from a bar, the operator does not trigger the corresponding observable at tree level with the exception of $\mathcal{O}_{LH}$, with the constraint from $0\nu\beta\beta$ decay being $\Lambda> 1.9\times 10^{12}$ TeV (induced trivially via the neutrino mass term with two additional electroweak vacuum expectation value insertions as compared to dimension-5 Weinberg operator). We have omitted this bound from Fig.~\ref{fig:barplotflavor} in favor of making the other constraints more clearly visible. The lepton flavor structures for the SMEFT operator constrained by a given observable are indicated next to the observable in the legend.

In addition to various experimental constraints, we also indicate in the plot the limits on some of the operators from a naive neutrino mass naturalness argument using floating $\#$ symbols, which of course is indicative and applicable only in the absence of fine-tuning or some additional symmetry. These naturalness limits are an artifact of the fact that the operator $\mathcal{O}_{LH}$ can induce a correction to the neutrino mass at tree level through VEV insertion $\delta m_\nu^{ij}= -v/2 (v^3 {C_{LH}^{ij}})$ and the operators $\mathcal{O}_{LHD1}$, $\mathcal{O}_{LHD2}$, $\mathcal{O}_{LeHD}$ and $\mathcal{O}_{LHW}$ can induce $C_{LH}$ through RG running-induced operator mixing effects between the new physics scale and the electroweak symmetry breaking scale~\cite{Cirigliano:2017djv}. The limits indicated in our plot are derived under the assumption that $|\delta m_\nu^{ij}|<1$ eV.
\begin{table}[]
    \centering
    \begin{tabular}{l c r r r r r r r}
    \toprule
    Operator && Collider & $0\nu\beta\beta$ & LBL Osc. & $\mu_\nu$ & $\mu^+$-decay & CE$\nu$NS & Meson decay  \\
    \cmidrule{1-1} \cmidrule{3-9}
$\mathcal{O}_{LH}$ && \textcolor{red}{$\checkmark$} & \textcolor{red}{$\checkmark$} & \text{\--} & \text{\--} & \text{\--} & \text{\--} & \text{\--}\\
$\mathcal{O}_{LeHD}$ && \textcolor{red}{$\checkmark$} & \textcolor{red}{$\checkmark$} & \textcolor{red}{$\checkmark$} & \text{\--} & \text{\--} & \text{\--} & \text{\--} \\
$\mathcal{O}_{LHD1}$ && \textcolor{red}{$\checkmark$} & \textcolor{red}{$\checkmark$} & \text{\--} & \text{\--} & \text{\--} & \text{\--} & \text{\--}\\
$\mathcal{O}_{LHD2}$ && \textcolor{red}{$\checkmark$} & \textcolor{red}{$\checkmark$} & \text{\--} & \text{\--} & \text{\--} & \text{\--} & \text{\--}\\
$\mathcal{O}_{LHB}$ && \text{\--} & \text{\--} & \text{\--} & \textcolor{red}{$\checkmark$} & \text{\--} & \textcolor{red}{$\checkmark$} & \text{\--}\\
$\mathcal{O}_{LHW}$ && \text{\--} & \textcolor{red}{$\checkmark$} & \text{\--} & \textcolor{red}{$\checkmark$} & \text{\--} & \textcolor{red}{$\checkmark$} & \text{\--}\\
$\mathcal{O}_{\bar{d}uLLD}$ && \textcolor{red}{$\checkmark$} & \textcolor{red}{$\checkmark$} & \text{\--} & \text{\--} & \text{\--} & \text{\--} & \text{\--}\\
$\mathcal{O}_{\bar{e}LLLH}$ && \text{\--} & \text{\--} & \text{\--} & \text{\--} & \textcolor{red}{$\checkmark$} & \text{\--} & \text{\--}\\
$\mathcal{O}_{\bar{d}LueH}$ && \textcolor{red}{$\checkmark$} & \textcolor{red}{$\checkmark$} & \textcolor{red}{$\checkmark$} & \text{\--} & \text{\--} & \text{\--} & \text{\--} \\
$\mathcal{O}_{\bar{d}LQLH1}$ && \textcolor{red}{$\checkmark$} & \textcolor{red}{$\checkmark$} & \textcolor{red}{$\checkmark$} & \text{\--} & \text{\--} & \textcolor{red}{$\checkmark$} & \textcolor{red}{$\checkmark$} \\
$\mathcal{O}_{\bar{d}LQLH2}$ && \textcolor{red}{$\checkmark$} & \textcolor{red}{$\checkmark$} & \textcolor{red}{$\checkmark$} & \text{\--} & \text{\--} & \text{\--} & \text{\--} \\
$\mathcal{O}_{\bar{Q}uLLH}$ && \textcolor{red}{$\checkmark$} & \textcolor{red}{$\checkmark$} & \textcolor{red}{$\checkmark$} & \text{\--} & \text{\--} & \textcolor{red}{$\checkmark$} & \text{\--}\\
\bottomrule
    \end{tabular}
    \caption{Table showing complementarity of various observables in probing the twelve independent LNV dimension-7 SMEFT operators. C.f. Fig.~\ref{fig:barplotflavor} for the quantitative constraints and specification of flavor sensitivities of the various observables.}  
    \label{tab:observables}
\end{table}

As expected, $0\nu\beta\beta$ decay provides the most stringent experimental constraints for the electron-flavored charged leptons ($ee$) for most of the LNV dimension-7 SMEFT operators, with the exception of $\mathcal{O}_{LHB}$ and $\mathcal{O}_{\bar{e}LLLH}$, which do not induce $0\nu\beta\beta$ decay at the leading order. However, if the LNV interactions dominantly occur in heavier (mixed) charged lepton flavors then we can identify several other observables which can constrain the LNV dimension-7 SMEFT operators. Among them is the measurement of $K^+\rightarrow \pi^+ \nu\nu$ by the NA62 experiment which currently provides an impressive limit of $\Lambda> 21.8$ TeV for the LNV dimension-7 SMEFT operator $\mathcal{O}_{\bar{d}LQLH1}$ under the assumption of flavor universality and same flavor of the two leptons. In the future, NA62 will improve the sensitivity to about $31.5$ TeV. Given that the measured mode involves light neutrinos in the final states of $K^+\rightarrow \pi^+ \nu\nu$, this is one of the unique probes sensitive to all three flavors of leptons. A similar line of argument is also true for $B \rightarrow K^{(*)} \nu\nu$ decay, which has already been searched for by the BaBar and Belle experiments. The sensitivity for this observable will further be improved by the Belle II experiment in the near future. However, the limits from $B \rightarrow K^{(*)} \nu\nu$ decay are about an order of magnitude weaker than those from the $K^+\rightarrow \pi^+ \nu\nu$ measurement. The collider searches can also provide very competitive limits for LNV dimension-7 SMEFT operators with the possibility of improving sensitivities by an order of magnitude in future generation experiments. The current limit from LHC applicable for the $\mu\mu$ flavor of leptons is most stringent for the SMEFT operator $\mathcal{O}_{\bar{d}uLLD}$, providing a bound of 4 TeV. This limit is expected to be improved to 19 TeV at FCC-hh. On the other hand, the LHC currently provides  $O(\text{TeV})$ bounds for $\mathcal{O}_{\bar{d}LueH}$, $\mathcal{O}_{\bar{d}LQLH1}$, and  $\mathcal{O}_{\bar{Q}uLLH}$. While for $\mathcal{O}_{\bar{d}LQLH2}$ and $\mathcal{O}_{LHD1}$ the current limits are order half a TeV. With the exception of $\mathcal{O}_{LH}$, $\mathcal{O}_{LeHD}$, $\mathcal{O}_{LHD1}$, $\mathcal{O}_{LHB}$, $\mathcal{O}_{LHW}$ and $\mathcal{O}_{\bar{e}LLLH}$, all the other LNV dimension-7 SMEFT operators can be probed at FCC-hh to $\mathcal{O}(\text{several TeV})$ new physics scales. 

Another observable which can constrain five of the LNV dimension-7 SMEFT operators is the search for charged current LNV non-standard interactions at LBL neutrino oscillation experiments. Currently they provide constraints of $O(\text{sub TeV})$ in $ee$, $e\mu$, $e\tau$ flavors for $\mathcal{O}_{LeHD}$, $\mathcal{O}_{\bar{d}LueH}$, $\mathcal{O}_{\bar{d}LQLH1}$, $\mathcal{O}_{\bar{d}LQLH2}$ and $\mathcal{O}_{\bar{Q}uLLH}$. While for the $\mu\mu$ and $\mu\tau$ flavors, the constraints from LBL neutrino oscillation experiments are a factor half weaker than the electron flavor. On the other hand, the scattering of solar, reactor, and accelerator neutrinos with electrons can provide some of the most stringent constraints on the neutrino magnetic moment which can be induced by the LNV dimension-7 SMEFT operators $\mathcal{O}_{LHB}$ and $\mathcal{O}_{LHW}$. In fact, the operator $\mathcal{O}_{LHB}$ is dominantly constrained by the constraint on neutrino magnetic moment. It is worth noting that due to the Majorana nature of the neutrinos, the constraints from the neutrino magnetic moment are only applicable when the flavors of the lepton pair in the relevant LNV dimension-7 SMEFT operators are different from each other. A probe sensitive to the operator $\mathcal{O}_{\bar{d}LQLH1}$ is also provided by the coherent elastic neutrino-nucleus scattering (CE$\nu$NS) which currently provides a bound $\Lambda>0.4$ TeV from the COHERENT experiment (where the neutrino flux is dominantly of muon flavor). CE$\nu$NS can also provide subdominant constraints on $\mathcal{O}_{LHW}, \mathcal{O}_{LHB} $ and $\mathcal{O}_{\bar{Q}uLLH}$~\cite{Li:2020lba}. This limit is expected to be improved by a factor of two in future generation experiments using a Germanium detector and using electron anti-neutrino flux from nuclear reactors.  Finally, the operator $\mathcal{O}_{\bar{e}LLLH}$, which does not involve any quarks, is mainly constrained by the muon decay mode $\mu^+ \rightarrow e^+ \bar{\nu}_e \nu_\mu$ to $\Lambda>0.4$ TeV.

Given the discussion above it is apparent that the synergy of different observables can potentially distinguish between the different LNV dimension-7 SMEFT operators and further can shed light on the flavor structure of the LNV interactions. If the underlying LNV new physics is flavor universal then for most operators we would expect to see the first sign of lepton number violation at \ovbb decay experiments. However, for instance, if an LNV signal is instead seen at $\mu^+$-decay experiments, without a corresponding signal of \ovbb decay, then, assuming that a dimension-7 $\Delta L = 2$ operator is responsible for such signal, one can deduce that the signal is caused by $\mathcal{O}_{\bar eLLLH}$. Similarly, if neutrinos are found to have a magnetic moment, we could deduce that $\mathcal{O}_{LHB}$ is realized in case we do not see a \ovbb signal, while $\mathcal{O}_{LHW}$ could be realized in nature if we do. Indeed, for LNV new physics violating lepton flavor non-universality, it may be the case that an LNV signal is first observed in a different observable than the \ovbb decay. For example, if LNV is dominantly mediated in the muon flavor, an LNV signal may first appear at the LHC, meson decay experiments, or neutrino oscillation experiments, rather than via \ovbb decay. On the other hand, positive signals at multiple experiments can be crucial in disentangling the flavor information. For instance, if both a signal at \ovbb decay and neutrino magnetic moment is detected then that (assuming a dimension-7 $\Delta L = 2$ operator is responsible for such a signal) would imply that LNV is not only prevalent only for $ee$ flavors but also for $e\mu$, $e\tau$ or $\mu\tau$ flavors. Tab.~\ref{tab:observables} provides a simplified overview of the complementarity of various observables in probing the LNV dimension-7 SMEFT operators. Given positive signal(s) for LNV at a single experiment or multiple experiments, the complementarity of various observables discussed in this work can give valuable hints towards understanding the flavor and effective operator structure of the underlying LNV new physics, which can then provide guidance for identifying potential UV completions for LNV interactions realized by nature.

%%%%%%%%%%%%%%%%%%%%%%%%%%%%%%%%%%%%%%%%%%%%%%%%%%%%%%%%%%%%%%%%%%%%%%%%%%%%%%%%%
\section{Conclusions}
\label{sec:conclusion}
%%%%%%%%%%%%%%%%%%%%%%%%%%%%%%%%%%%%%%%%%%%%%%%%%%%%%%%%%%%%%%%%%%%%%%%%%%%%%%%%%
In this work, we have performed a state-of-the-art global comparison of all the independent dimension-7 $\Delta L = 2$ SMEFT operators to identify the prospects of probing them at various low- and high-energy experiments. Unsurprisingly, $0\nu\beta\beta$ decay provides some of the best current experimental constraints for most of the LNV dimension-7 SMEFT operators in the electron flavor. However, we have explored in detail also other relevant observables that can test the conservation of lepton number in other flavors, such as rare meson decays, charged lepton flavor violating leptonic processes, coherent elastic neutrino-nucleus scattering, non-standard interaction mediated long baseline neutrino oscillation, neutrino-electron scattering, etc. Studying these probes in the context of dimension-7 $\Delta L = 2$ SMEFT operators, we identified the most promising ways of testing lepton number non-conservation triggered by operators not constrained by $0\nu\beta\beta$ decay or providing valuable information about lepton number violation in flavors other than the electron one. In particular, the muon decay mode $\mu^+ \rightarrow e^+ \bar{\nu}_e \nu_\mu$ and constraints on neutrino magnetic moment from neutrino-electron scattering provide bounds on two of the operators not constrained by $0\nu\beta\beta$ decay. Further, observables like rare meson decay, non-standard interactions at LBL neutrino oscillation, or coherent elastic neutrino-nucleus scattering, give complementary limits on dimension-7 $\Delta L = 2$ SMEFT operators for different combinations of charged lepton flavors. 

Besides that, we have also presented a comprehensive and systematic analysis of the complete set of dimension-7 $\Delta L = 2$ SMEFT operators in the context of the LHC and the proposed collider accelerator FCC, putting them in comparison with the low-energy probes. The current limit from the LHC applicable for the muon flavor of leptons is most stringent for the SMEFT operator $\mathcal{O}_{\bar{d}uLLD}$ with the associated new-physics scale reaching about 5 TeV. For the other studied operators, $\mathcal{O}_{\bar{d}LueH}$, $\mathcal{O}_{\bar{d}LQLH1}$, $\mathcal{O}_{\bar{d}LQLH2}$, $\mathcal{O}_{\bar{Q}uLLH}$, and $\mathcal{O}_{LHD1}$ the LHC currently sets order TeV bounds. All of these limits are expected to be improved by an order of magnitude at FCC-hh in the future. 

If a signal or even multiple signals of lepton number violation are revealed by future experiments, we have also discussed how the synergy of the analyzed multi-frontier observables can play a crucial role in distinguishing among different LNV dimension-7 SMEFT operators, and therefore, further, shed light on the flavor structure of the new interactions. This can in turn provide valuable guidance for identifying the potential UV completions triggering the observed manifestation(s) of LNV. We conclude that the future generation of multi-frontier experimental efforts seems to offer a promising and exciting pathway toward probing lepton number violation in different flavors beyond dimension 5.
%%%%%%%%%%%%%%%%%%%%%%%%%%%%%%%%%%%%%%%%%%%%%%%%%%%%%%%%%%%%%%%%%%%%%%%%%%%%%%%%%

%%%%%%%%%%%%%%%%%%%%%%%%%%%%%%%%%%%%%%%%%%%%%%%%%%%%%%%%%%%%%%%%%%%%%%%%%%%%%%%%%
\section*{Acknowledgements}
%%%%%%%%%%%%%%%%%%%%%%%%%%%%%%%%%%%%%%%%%%%%%%%%%%%%%%%%%%%%%%%%%%%%%%%%%%%%%%%%%
The authors are grateful to Arvind Bhaskar, Patrick Bolton, Olivier Mattelaer, Gang Li, Michael Schmidt, and Oliver Scholer for helpful discussions and clarifications. CH would also like to thank Benjamin Grinstein for very useful discussions. KF, JH, and CH acknowledge support from the Deutsche Forschungsgemeinschaft (DFG, German Research Foundation) Emmy Noether grant HA 8555/1-1. JH acknowledges additional support from the Cluster of Excellence “Precision Physics, Fundamental Interactions, and
Structure of Matter” (PRISMA$^+$ EXC 2118/1) funded by the Deutsche Forschungsgemeinschaft (DFG, German Research
Foundation) within the German Excellence Strategy (Project No. 390831469). LG acknowledges support from the National Science Foundation, Grant PHY-1630782, and the Heising-Simons Foundation, Grant 2017-228. KF acknowledges support from the Japan Society for the Promotion of Science (JSPS) Grant-in-Aid for Scientific Research B No.~21H01086 and No.~22K21350.

\bibliographystyle{JHEP}
\bibliography{References}

\providecommand{\href}[2]{#2}\begingroup\raggedright\begin{thebibliography}{100}

\bibitem{Babu:2001ex}
K.~S. Babu and C.~N. Leung, \emph{{Classification of effective neutrino mass
  operators}},
  \href{http://dx.doi.org/10.1016/S0550-3213(01)00504-1}{\emph{Nucl. Phys. B}
  {\bf 619} (2001) 667--689}, [\href{http://arxiv.org/abs/hep-ph/0106054}{{\tt
  hep-ph/0106054}}].

\bibitem{Deppisch:2017ecm}
F.~F. Deppisch, L.~Graf, J.~Harz and W.-C. Huang, \emph{{Neutrinoless Double
  Beta Decay and the Baryon Asymmetry of the Universe}},
  \href{http://dx.doi.org/10.1103/PhysRevD.98.055029}{\emph{Phys. Rev. D} {\bf
  98} (2018) 055029}, [\href{http://arxiv.org/abs/1711.10432}{{\tt
  1711.10432}}].

\bibitem{deGouvea:2007qla}
A.~de~Gouvea and J.~Jenkins, \emph{{A Survey of Lepton Number Violation Via
  Effective Operators}},
  \href{http://dx.doi.org/10.1103/PhysRevD.77.013008}{\emph{Phys. Rev. D} {\bf
  77} (2008) 013008}, [\href{http://arxiv.org/abs/0708.1344}{{\tt 0708.1344}}].

\bibitem{Weinberg:1979sa}
S.~Weinberg, \emph{{Baryon and Lepton Nonconserving Processes}},
  \href{http://dx.doi.org/10.1103/PhysRevLett.43.1566}{\emph{Phys.Rev.Lett.}
  {\bf 43} (1979) 1566}.

\bibitem{Minkowski:1977sc}
P.~Minkowski, \emph{{$\mu \to e\gamma$ at a Rate of One Out of $10^{9}$ Muon
  Decays?}}, \href{http://dx.doi.org/10.1016/0370-2693(77)90435-X}{\emph{Phys.
  Lett. B} {\bf 67} (1977) 421--428}.

\bibitem{Gell-Mann:1979vob}
M.~Gell-Mann, P.~Ramond and R.~Slansky, \emph{{Complex Spinors and Unified
  Theories}}, {\emph{Conf. Proc. C} {\bf 790927} (1979) 315--321},
  [\href{http://arxiv.org/abs/1306.4669}{{\tt 1306.4669}}].

\bibitem{Yanagida:1979as}
T.~Yanagida, \emph{{Horizontal gauge symmetry and masses of neutrinos}},
  {\emph{Conf. Proc. C} {\bf 7902131} (1979) 95--99}.

\bibitem{Mohapatra:1979ia}
R.~N. Mohapatra and G.~Senjanovic, \emph{{Neutrino Mass and Spontaneous Parity
  Nonconservation}},
  \href{http://dx.doi.org/10.1103/PhysRevLett.44.912}{\emph{Phys. Rev. Lett.}
  {\bf 44} (1980) 912}.

\bibitem{Schechter:1980gr}
J.~Schechter and J.~W.~F. Valle, \emph{{Neutrino Masses in SU(2) x U(1)
  Theories}}, \href{http://dx.doi.org/10.1103/PhysRevD.22.2227}{\emph{Phys.
  Rev. D} {\bf 22} (1980) 2227}.

\bibitem{Mohapatra:1980yp}
R.~N. Mohapatra and G.~Senjanovic, \emph{{Neutrino Masses and Mixings in Gauge
  Models with Spontaneous Parity Violation}},
  \href{http://dx.doi.org/10.1103/PhysRevD.23.165}{\emph{Phys. Rev. D} {\bf 23}
  (1981) 165}.

\bibitem{Schechter:1981cv}
J.~Schechter and J.~W.~F. Valle, \emph{{Neutrino Decay and Spontaneous
  Violation of Lepton Number}},
  \href{http://dx.doi.org/10.1103/PhysRevD.25.774}{\emph{Phys. Rev. D} {\bf 25}
  (1982) 774}.

\bibitem{Kobach:2016ami}
A.~Kobach, \emph{{Baryon Number, Lepton Number, and Operator Dimension in the
  Standard Model}},
  \href{http://dx.doi.org/10.1016/j.physletb.2016.05.050}{\emph{Phys. Lett. B}
  {\bf 758} (2016) 455--457}, [\href{http://arxiv.org/abs/1604.05726}{{\tt
  1604.05726}}].

\bibitem{Lehman:2014jma}
L.~Lehman, \emph{{Extending the Standard Model Effective Field Theory with the
  Complete Set of Dimension-7 Operators}},
  \href{http://dx.doi.org/10.1103/PhysRevD.90.125023}{\emph{Phys. Rev. D} {\bf
  90} (2014) 125023}, [\href{http://arxiv.org/abs/1410.4193}{{\tt 1410.4193}}].

\bibitem{Liao:2016hru}
Y.~Liao and X.-D. Ma, \emph{{Renormalization Group Evolution of Dimension-seven
  Baryon- and Lepton-number-violating Operators}},
  \href{http://dx.doi.org/10.1007/JHEP11(2016)043}{\emph{JHEP} {\bf 11} (2016)
  043}, [\href{http://arxiv.org/abs/1607.07309}{{\tt 1607.07309}}].

\bibitem{Cirigliano:2017djv}
V.~Cirigliano, W.~Dekens, J.~de~Vries, M.~L. Graesser and E.~Mereghetti,
  \emph{{Neutrinoless double beta decay in chiral effective field theory:
  lepton number violation at dimension seven}},
  \href{http://dx.doi.org/10.1007/JHEP12(2017)082}{\emph{JHEP} {\bf 12} (2017)
  082}, [\href{http://arxiv.org/abs/1708.09390}{{\tt 1708.09390}}].

\bibitem{Cirigliano:2018yza}
V.~Cirigliano, W.~Dekens, J.~de~Vries, M.~Graesser and E.~Mereghetti, \emph{{A
  neutrinoless double beta decay master formula from effective field theory}},
  \href{http://dx.doi.org/10.1007/JHEP12(2018)097}{\emph{JHEP} {\bf 12} (2018)
  097}, [\href{http://arxiv.org/abs/1806.02780}{{\tt 1806.02780}}].

\bibitem{Li:2019fhz}
T.~Li, X.-D. Ma and M.~A. Schmidt, \emph{{Implication of $K\to \pi \nu
  \bar{\nu}$ for generic neutrino interactions in effective field theories}},
  \href{http://dx.doi.org/10.1103/PhysRevD.101.055019}{\emph{Phys. Rev. D} {\bf
  101} (2020) 055019}, [\href{http://arxiv.org/abs/1912.10433}{{\tt
  1912.10433}}].

\bibitem{Deppisch:2020oyx}
F.~F. Deppisch, K.~Fridell and J.~Harz, \emph{{Constraining lepton number
  violating interactions in rare kaon decays}},
  \href{http://dx.doi.org/10.1007/JHEP12(2020)186}{\emph{JHEP} {\bf 12} (2020)
  186}, [\href{http://arxiv.org/abs/2009.04494}{{\tt 2009.04494}}].

\bibitem{Felkl:2021uxi}
T.~Felkl, S.~L. Li and M.~A. Schmidt, \emph{{A tale of invisibility:
  constraints on new physics in b \textrightarrow{}
  s\ensuremath{\nu}\ensuremath{\nu}}},
  \href{http://dx.doi.org/10.1007/JHEP12(2021)118}{\emph{JHEP} {\bf 12} (2021)
  118}, [\href{http://arxiv.org/abs/2111.04327}{{\tt 2111.04327}}].

\bibitem{Fuks:2020zbm}
B.~Fuks, J.~Neundorf, K.~Peters, R.~Ruiz and M.~Saimpert, \emph{{Probing the
  Weinberg operator at colliders}},
  \href{http://dx.doi.org/10.1103/PhysRevD.103.115014}{\emph{Phys. Rev. D} {\bf
  103} (2021) 115014}, [\href{http://arxiv.org/abs/2012.09882}{{\tt
  2012.09882}}].

\bibitem{Cepedello:2017lyo}
R.~Cepedello, M.~Hirsch and J.~C. Helo, \emph{{Lepton number violating
  phenomenology of d = 7 neutrino mass models}},
  \href{http://dx.doi.org/10.1007/JHEP01(2018)009}{\emph{JHEP} {\bf 01} (2018)
  009}, [\href{http://arxiv.org/abs/1709.03397}{{\tt 1709.03397}}].

\bibitem{Cepedello:2017eqf}
R.~Cepedello, M.~Hirsch and J.~C. Helo, \emph{{Loop neutrino masses from $d =
  7$ operator}}, \href{http://dx.doi.org/10.1007/JHEP07(2017)079}{\emph{JHEP}
  {\bf 07} (2017) 079}, [\href{http://arxiv.org/abs/1705.01489}{{\tt
  1705.01489}}].

\bibitem{Herrero-Garcia:2019czj}
J.~Herrero-Garc\'\i{}a and M.~A. Schmidt, \emph{{Neutrino mass models: New
  classification and model-independent upper limits on their scale}},
  \href{http://dx.doi.org/10.1140/epjc/s10052-019-7465-1}{\emph{Eur. Phys. J.
  C} {\bf 79} (2019) 938}, [\href{http://arxiv.org/abs/1903.10552}{{\tt
  1903.10552}}].

\bibitem{Liao:2010rx}
Y.~Liao, G.-Z. Ning and L.~Ren, \emph{{Flavor Violating Transitions of Charged
  Leptons from a Seesaw Mechanism of Dimension Seven}},
  \href{http://dx.doi.org/10.1103/PhysRevD.82.113003}{\emph{Phys. Rev. D} {\bf
  82} (2010) 113003}, [\href{http://arxiv.org/abs/1008.0117}{{\tt 1008.0117}}].

\bibitem{Krauss:2013gy}
M.~B. Krauss, D.~Meloni, W.~Porod and W.~Winter, \emph{{Neutrino Mass from a
  d=7 Effective Operator in an SU(5) SUSY-GUT Framework}},
  \href{http://dx.doi.org/10.1007/JHEP05(2013)121}{\emph{JHEP} {\bf 05} (2013)
  121}, [\href{http://arxiv.org/abs/1301.4221}{{\tt 1301.4221}}].

\bibitem{Gargalionis:2020xvt}
J.~Gargalionis and R.~R. Volkas, \emph{{Exploding operators for Majorana
  neutrino masses and beyond}},
  \href{http://dx.doi.org/10.1007/JHEP01(2021)074}{\emph{JHEP} {\bf 01} (2021)
  074}, [\href{http://arxiv.org/abs/2009.13537}{{\tt 2009.13537}}].

\bibitem{Bonnet:2009ej}
F.~Bonnet, D.~Hernandez, T.~Ota and W.~Winter, \emph{{Neutrino masses from
  higher than d=5 effective operators}},
  \href{http://dx.doi.org/10.1088/1126-6708/2009/10/076}{\emph{JHEP} {\bf 10}
  (2009) 076}, [\href{http://arxiv.org/abs/0907.3143}{{\tt 0907.3143}}].

\bibitem{Angel:2012ug}
P.~W. Angel, N.~L. Rodd and R.~R. Volkas, \emph{{Origin of neutrino masses at
  the LHC: $\Delta L = 2$ effective operators and their ultraviolet
  completions}},
  \href{http://dx.doi.org/10.1103/PhysRevD.87.073007}{\emph{Phys. Rev. D} {\bf
  87} (2013) 073007}, [\href{http://arxiv.org/abs/1212.6111}{{\tt 1212.6111}}].

\bibitem{Cai:2014kra}
Y.~Cai, J.~D. Clarke, M.~A. Schmidt and R.~R. Volkas, \emph{{Testing Radiative
  Neutrino Mass Models at the LHC}},
  \href{http://dx.doi.org/10.1007/JHEP02(2015)161}{\emph{JHEP} {\bf 02} (2015)
  161}, [\href{http://arxiv.org/abs/1410.0689}{{\tt 1410.0689}}].

\bibitem{Fridell:2022wbz}
K.~K. Fridell, \emph{{Phenomenology of Baryogenesis and Neutrino Physics: From
  Effective Field Theory to Simplified Models}}.
\newblock PhD thesis, Munich, Tech. U., Technical University of Munich, 8,
  2022.

\bibitem{Fridell:2023}
K.~Fridell, L.~Gr\'{a}f, J.~Harz and C.~Hati, \emph{{In Preparation}},
  \href{http://arxiv.org/abs/23xx.xxxxx}{{\tt 23xx.xxxxx}}.

\bibitem{Liao:2020zyx}
Y.~Liao, X.-D. Ma and Q.-Y. Wang, \emph{{Extending low energy effective field
  theory with a complete set of dimension-7 operators}},
  \href{http://dx.doi.org/10.1007/JHEP08(2020)162}{\emph{JHEP} {\bf 08} (2020)
  162}, [\href{http://arxiv.org/abs/2005.08013}{{\tt 2005.08013}}].

\bibitem{Jenkins:2017jig}
E.~E. Jenkins, A.~V. Manohar and P.~Stoffer, \emph{{Low-Energy Effective Field
  Theory below the Electroweak Scale: Operators and Matching}},
  \href{http://dx.doi.org/10.1007/JHEP03(2018)016}{\emph{JHEP} {\bf 03} (2018)
  016}, [\href{http://arxiv.org/abs/1709.04486}{{\tt 1709.04486}}].

\bibitem{Pilaftsis:1991ug}
A.~Pilaftsis, \emph{{Radiatively induced neutrino masses and large Higgs
  neutrino couplings in the standard model with Majorana fields}},
  \href{http://dx.doi.org/10.1007/BF01482590}{\emph{Z. Phys. C} {\bf 55} (1992)
  275--282}, [\href{http://arxiv.org/abs/hep-ph/9901206}{{\tt
  hep-ph/9901206}}].

\bibitem{Pati:1974yy}
J.~C. Pati and A.~Salam, \emph{{Lepton Number as the Fourth Color}},
  \href{http://dx.doi.org/10.1103/PhysRevD.10.275}{\emph{Phys. Rev. D} {\bf 10}
  (1974) 275--289}.

\bibitem{Mohapatra:1974gc}
R.~N. Mohapatra and J.~C. Pati, \emph{{A Natural Left-Right Symmetry}},
  \href{http://dx.doi.org/10.1103/PhysRevD.11.2558}{\emph{Phys. Rev. D} {\bf
  11} (1975) 2558}.

\bibitem{Senjanovic:1975rk}
G.~Senjanovic and R.~N. Mohapatra, \emph{{Exact Left-Right Symmetry and
  Spontaneous Violation of Parity}},
  \href{http://dx.doi.org/10.1103/PhysRevD.12.1502}{\emph{Phys. Rev. D} {\bf
  12} (1975) 1502}.

\bibitem{Hati:2018tge}
C.~Hati, S.~Patra, P.~Pritimita and U.~Sarkar, \emph{{Neutrino Masses and
  Leptogenesis in Left\textendash{}Right Symmetric Models: A Review From a
  Model Building Perspective}},
  \href{http://dx.doi.org/10.3389/fphy.2018.00019}{\emph{Front. in Phys.} {\bf
  6} (2018) 19}.

\bibitem{Keung:1983uu}
W.-Y. Keung and G.~Senjanovic, \emph{{Majorana Neutrinos and the Production of
  the Right-handed Charged Gauge Boson}},
  \href{http://dx.doi.org/10.1103/PhysRevLett.50.1427}{\emph{Phys. Rev. Lett.}
  {\bf 50} (1983) 1427}.

\bibitem{Deppisch:2015qwa}
F.~F. Deppisch, P.~S. Bhupal~Dev and A.~Pilaftsis, \emph{{Neutrinos and
  Collider Physics}},
  \href{http://dx.doi.org/10.1088/1367-2630/17/7/075019}{\emph{New J. Phys.}
  {\bf 17} (2015) 075019}, [\href{http://arxiv.org/abs/1502.06541}{{\tt
  1502.06541}}].

\bibitem{Super-Kamiokande:2020wjk}
{\scshape Super-Kamiokande} collaboration, A.~Takenaka et~al., \emph{{Search
  for proton decay via $p\to e^+\pi^0$ and $p\to \mu^+\pi^0$ with an enlarged
  fiducial volume in Super-Kamiokande I-IV}},
  \href{http://dx.doi.org/10.1103/PhysRevD.102.112011}{\emph{Phys. Rev. D} {\bf
  102} (2020) 112011}, [\href{http://arxiv.org/abs/2010.16098}{{\tt
  2010.16098}}].

\bibitem{Fornal:2017xcj}
B.~Fornal and B.~Grinstein, \emph{{SU(5) Unification without Proton Decay}},
  \href{http://dx.doi.org/10.1103/PhysRevLett.119.241801}{\emph{Phys. Rev.
  Lett.} {\bf 119} (2017) 241801}, [\href{http://arxiv.org/abs/1706.08535}{{\tt
  1706.08535}}].

\bibitem{Hambye:2017qix}
T.~Hambye and J.~Heeck, \emph{{Proton decay into charged leptons}},
  \href{http://dx.doi.org/10.1103/PhysRevLett.120.171801}{\emph{Phys. Rev.
  Lett.} {\bf 120} (2018) 171801}, [\href{http://arxiv.org/abs/1712.04871}{{\tt
  1712.04871}}].

\bibitem{Hati:2018cqp}
C.~Hati and U.~Sarkar, \emph{{$B-L$ violating nucleon decays as a probe of
  leptoquarks and implications for baryogenesis}},
  \href{http://dx.doi.org/10.1016/j.nuclphysb.2020.114985}{\emph{Nucl. Phys. B}
  {\bf 954} (2020) 114985}, [\href{http://arxiv.org/abs/1805.06081}{{\tt
  1805.06081}}].

\bibitem{Fridell:2021gag}
K.~Fridell, J.~Harz and C.~Hati, \emph{{Probing baryogenesis with
  neutron-antineutron oscillations}},
  \href{http://dx.doi.org/10.1007/JHEP11(2021)185}{\emph{JHEP} {\bf 11} (2021)
  185}, [\href{http://arxiv.org/abs/2105.06487}{{\tt 2105.06487}}].

\bibitem{delAguila:2012nu}
F.~del Aguila, A.~Aparici, S.~Bhattacharya, A.~Santamaria and J.~Wudka,
  \emph{{Effective Lagrangian approach to neutrinoless double beta decay and
  neutrino masses}},
  \href{http://dx.doi.org/10.1007/JHEP06(2012)146}{\emph{JHEP} {\bf 06} (2012)
  146}, [\href{http://arxiv.org/abs/1204.5986}{{\tt 1204.5986}}].

\bibitem{Deppisch:2013jxa}
F.~F. Deppisch, J.~Harz and M.~Hirsch, \emph{{Falsifying High-Scale
  Leptogenesis at the LHC}},
  \href{http://dx.doi.org/10.1103/PhysRevLett.112.221601}{\emph{Phys. Rev.
  Lett.} {\bf 112} (2014) 221601}, [\href{http://arxiv.org/abs/1312.4447}{{\tt
  1312.4447}}].

\bibitem{Frere:2008ct}
J.-M. Frere, T.~Hambye and G.~Vertongen, \emph{{Is leptogenesis falsifiable at
  LHC?}}, \href{http://dx.doi.org/10.1088/1126-6708/2009/01/051}{\emph{JHEP}
  {\bf 01} (2009) 051}, [\href{http://arxiv.org/abs/0806.0841}{{\tt
  0806.0841}}].

\bibitem{BhupalDev:2014hro}
P.~S. Bhupal~Dev, C.-H. Lee and R.~N. Mohapatra, \emph{{Leptogenesis
  Constraints on the Mass of Right-handed Gauge Bosons}},
  \href{http://dx.doi.org/10.1103/PhysRevD.90.095012}{\emph{Phys. Rev. D} {\bf
  90} (2014) 095012}, [\href{http://arxiv.org/abs/1408.2820}{{\tt 1408.2820}}].

\bibitem{Dhuria:2015wwa}
M.~Dhuria, C.~Hati, R.~Rangarajan and U.~Sarkar, \emph{{The $eejj$ Excess
  Signal at the LHC and Constraints on Leptogenesis}},
  \href{http://dx.doi.org/10.1088/1475-7516/2015/9/035}{\emph{JCAP} {\bf 09}
  (2015) 035}, [\href{http://arxiv.org/abs/1502.01695}{{\tt 1502.01695}}].

\bibitem{Dhuria:2015cfa}
M.~Dhuria, C.~Hati, R.~Rangarajan and U.~Sarkar, \emph{{Falsifying leptogenesis
  for a TeV scale $W^{\pm}_{R}$ at the LHC}},
  \href{http://dx.doi.org/10.1103/PhysRevD.92.031701}{\emph{Phys. Rev. D} {\bf
  92} (2015) 031701}, [\href{http://arxiv.org/abs/1503.07198}{{\tt
  1503.07198}}].

\bibitem{BhupalDev:2015khe}
P.~S. Bhupal~Dev, C.-H. Lee and R.~N. Mohapatra, \emph{{TeV Scale Lepton Number
  Violation and Baryogenesis}},
  \href{http://dx.doi.org/10.1088/1742-6596/631/1/012007}{\emph{J. Phys. Conf.
  Ser.} {\bf 631} (2015) 012007}, [\href{http://arxiv.org/abs/1503.04970}{{\tt
  1503.04970}}].

\bibitem{Aoki:2020til}
M.~Aoki, K.~Enomoto and S.~Kanemura, \emph{{Probing charged lepton number
  violation via $\ell^\pm \ell^{\prime \pm} W^\mp W^\mp$}},
  \href{http://dx.doi.org/10.1103/PhysRevD.101.115019}{\emph{Phys. Rev. D} {\bf
  101} (2020) 115019}, [\href{http://arxiv.org/abs/2002.12265}{{\tt
  2002.12265}}].

\bibitem{Fuks:2020att}
B.~Fuks, J.~Neundorf, K.~Peters, R.~Ruiz and M.~Saimpert, \emph{{Majorana
  neutrinos in same-sign $W^\pm W^\pm$ scattering at the LHC: Breaking the TeV
  barrier}}, \href{http://dx.doi.org/10.1103/PhysRevD.103.055005}{\emph{Phys.
  Rev. D} {\bf 103} (2021) 055005},
  [\href{http://arxiv.org/abs/2011.02547}{{\tt 2011.02547}}].

\bibitem{ATLAS:2023cjo}
{\scshape ATLAS} collaboration, G.~Aad et~al., \emph{{Search for heavy Majorana
  or Dirac neutrinos and right-handed W gauge bosons in final states with
  charged leptons and jets in pp collisions at $\sqrt{s}=13$~TeV with the ATLAS
  detector}},
  \href{http://dx.doi.org/10.1140/epjc/s10052-023-12021-9}{\emph{Eur. Phys. J.
  C} {\bf 83} (2023) 1164}, [\href{http://arxiv.org/abs/2304.09553}{{\tt
  2304.09553}}].

\bibitem{Alloul:2013bka}
A.~Alloul, N.~D. Christensen, C.~Degrande, C.~Duhr and B.~Fuks,
  \emph{{FeynRules 2.0 - A complete toolbox for tree-level phenomenology}},
  \href{http://dx.doi.org/10.1016/j.cpc.2014.04.012}{\emph{Comput. Phys.
  Commun.} {\bf 185} (2014) 2250--2300},
  [\href{http://arxiv.org/abs/1310.1921}{{\tt 1310.1921}}].

\bibitem{Alwall:2014hca}
J.~Alwall, R.~Frederix, S.~Frixione, V.~Hirschi, F.~Maltoni, O.~Mattelaer
  et~al., \emph{{The automated computation of tree-level and next-to-leading
  order differential cross sections, and their matching to parton shower
  simulations}}, \href{http://dx.doi.org/10.1007/JHEP07(2014)079}{\emph{JHEP}
  {\bf 07} (2014) 079}, [\href{http://arxiv.org/abs/1405.0301}{{\tt
  1405.0301}}].

\bibitem{ATLAS:2017oro}
{\scshape ATLAS} collaboration, M.~Aaboud et~al., \emph{{Study of the material
  of the ATLAS inner detector for Run 2 of the LHC}},
  \href{http://dx.doi.org/10.1088/1748-0221/12/12/P12009}{\emph{JINST} {\bf 12}
  (2017) P12009}, [\href{http://arxiv.org/abs/1707.02826}{{\tt 1707.02826}}].

\bibitem{Magliocca:2021bfg}
{\scshape ATLAS} collaboration, C.~Magliocca, \emph{{Measurement of the track
  impact parameters resolution with the ATLAS experiment at LHC using 2016-2018
  data}}, \href{http://dx.doi.org/10.1393/ncc/i2021-21055-0}{\emph{Nuovo Cim.
  C} {\bf 44} (2021) 55}.

\bibitem{Buckley:2014ana}
A.~Buckley, J.~Ferrando, S.~Lloyd, K.~Nordstr\"om, B.~Page, M.~R\"ufenacht
  et~al., \emph{{LHAPDF6: parton density access in the LHC precision era}},
  \href{http://dx.doi.org/10.1140/epjc/s10052-015-3318-8}{\emph{Eur. Phys. J.
  C} {\bf 75} (2015) 132}, [\href{http://arxiv.org/abs/1412.7420}{{\tt
  1412.7420}}].

\bibitem{Sjostrand:2014zea}
T.~Sj\"ostrand, S.~Ask, J.~R. Christiansen, R.~Corke, N.~Desai, P.~Ilten
  et~al., \emph{{An introduction to PYTHIA 8.2}},
  \href{http://dx.doi.org/10.1016/j.cpc.2015.01.024}{\emph{Comput. Phys.
  Commun.} {\bf 191} (2015) 159--177},
  [\href{http://arxiv.org/abs/1410.3012}{{\tt 1410.3012}}].

\bibitem{deFavereau:2013fsa}
{\scshape DELPHES 3} collaboration, J.~de~Favereau, C.~Delaere, P.~Demin,
  A.~Giammanco, V.~Lema\^\i{}tre, A.~Mertens et~al., \emph{{DELPHES 3, A
  modular framework for fast simulation of a generic collider experiment}},
  \href{http://dx.doi.org/10.1007/JHEP02(2014)057}{\emph{JHEP} {\bf 02} (2014)
  057}, [\href{http://arxiv.org/abs/1307.6346}{{\tt 1307.6346}}].

\bibitem{ATLAS:2018dcj}
{\scshape ATLAS} collaboration, M.~Aaboud et~al., \emph{{Search for heavy
  Majorana or Dirac neutrinos and right-handed $W$ gauge bosons in final states
  with two charged leptons and two jets at $ \sqrt{s}=13 $ TeV with the ATLAS
  detector}}, \href{http://dx.doi.org/10.1007/JHEP01(2019)016}{\emph{JHEP} {\bf
  01} (2019) 016}, [\href{http://arxiv.org/abs/1809.11105}{{\tt 1809.11105}}].

\bibitem{FCC:2018vvp}
{\scshape FCC} collaboration, A.~Abada et~al., \emph{{FCC-hh: The Hadron
  Collider}: {Future Circular Collider Conceptual Design Report Volume 3}},
  \href{http://dx.doi.org/10.1140/epjst/e2019-900087-0}{\emph{Eur. Phys. J. ST}
  {\bf 228} (2019) 755--1107}.

\bibitem{Buchmueller:2013dya}
O.~Buchmueller, M.~J. Dolan and C.~McCabe, \emph{{Beyond Effective Field Theory
  for Dark Matter Searches at the LHC}},
  \href{http://dx.doi.org/10.1007/JHEP01(2014)025}{\emph{JHEP} {\bf 01} (2014)
  025}, [\href{http://arxiv.org/abs/1308.6799}{{\tt 1308.6799}}].

\bibitem{Busoni:2013lha}
G.~Busoni, A.~De~Simone, E.~Morgante and A.~Riotto, \emph{{On the Validity of
  the Effective Field Theory for Dark Matter Searches at the LHC}},
  \href{http://dx.doi.org/10.1016/j.physletb.2013.11.069}{\emph{Phys. Lett. B}
  {\bf 728} (2014) 412--421}, [\href{http://arxiv.org/abs/1307.2253}{{\tt
  1307.2253}}].

\bibitem{Shoemaker:2011vi}
I.~M. Shoemaker and L.~Vecchi, \emph{{Unitarity and Monojet Bounds on Models
  for DAMA, CoGeNT, and CRESST-II}},
  \href{http://dx.doi.org/10.1103/PhysRevD.86.015023}{\emph{Phys. Rev. D} {\bf
  86} (2012) 015023}, [\href{http://arxiv.org/abs/1112.5457}{{\tt 1112.5457}}].

\bibitem{Pas:1999fc}
H.~Pas, M.~Hirsch, H.~Klapdor-Kleingrothaus and S.~Kovalenko, \emph{{Towards a
  superformula for neutrinoless double beta decay}},
  \href{http://dx.doi.org/10.1016/S0370-2693(99)00330-5}{\emph{Phys. Lett. B}
  {\bf 453} (1999) 194--198}.

\bibitem{Pas:2000vn}
H.~Pas, M.~Hirsch, H.~Klapdor-Kleingrothaus and S.~Kovalenko, \emph{{A
  Superformula for neutrinoless double beta decay. 2. The Short range part}},
  \href{http://dx.doi.org/10.1016/S0370-2693(00)01359-9}{\emph{Phys. Lett. B}
  {\bf 498} (2001) 35--39}, [\href{http://arxiv.org/abs/hep-ph/0008182}{{\tt
  hep-ph/0008182}}].

\bibitem{Deppisch:2012nb}
F.~F. Deppisch, M.~Hirsch and H.~Pas, \emph{{Neutrinoless Double Beta Decay and
  Physics Beyond the Standard Model}},
  \href{http://dx.doi.org/10.1088/0954-3899/39/12/124007}{\emph{J. Phys. G}
  {\bf 39} (2012) 124007}, [\href{http://arxiv.org/abs/1208.0727}{{\tt
  1208.0727}}].

\bibitem{deGouvea:2007xp}
A.~de~Gouvea and J.~Jenkins, \emph{{A Survey of Lepton Number Violation Via
  Effective Operators}},
  \href{http://dx.doi.org/10.1103/PhysRevD.77.013008}{\emph{Phys.Rev.} {\bf
  D77} (2008) 013008}, [\href{http://arxiv.org/abs/0708.1344}{{\tt
  0708.1344}}].

\bibitem{Ali:2007ec}
A.~Ali, A.~Borisov and D.~Zhuridov, \emph{{Probing new physics in the
  neutrinoless double beta decay using electron angular correlation}},
  \href{http://dx.doi.org/10.1103/PhysRevD.76.093009}{\emph{Phys. Rev. D} {\bf
  76} (2007) 093009}, [\href{http://arxiv.org/abs/0706.4165}{{\tt 0706.4165}}].

\bibitem{Graf:2018ozy}
L.~Graf, F.~F. Deppisch, F.~Iachello and J.~Kotila, \emph{{Short-Range
  Neutrinoless Double Beta Decay Mechanisms}},
  \href{http://dx.doi.org/10.1103/PhysRevD.98.095023}{\emph{Phys. Rev. D} {\bf
  98} (2018) 095023}, [\href{http://arxiv.org/abs/1806.06058}{{\tt
  1806.06058}}].

\bibitem{Deppisch:2020ztt}
F.~F. Deppisch, L.~Graf, F.~Iachello and J.~Kotila, \emph{{Analysis of Light
  Neutrino Exchange and Short-Range Mechanisms in $0\nu\beta\beta$ Decay}},
  \href{http://arxiv.org/abs/2009.10119}{{\tt 2009.10119}}.

\bibitem{Dekens:2020ttz}
W.~Dekens, J.~de~Vries, K.~Fuyuto, E.~Mereghetti and G.~Zhou, \emph{{Sterile
  neutrinos and neutrinoless double beta decay in effective field theory}},
  \href{http://dx.doi.org/10.1007/JHEP06(2020)097}{\emph{JHEP} {\bf 06} (2020)
  097}, [\href{http://arxiv.org/abs/2002.07182}{{\tt 2002.07182}}].

\bibitem{Scholer:2023bnn}
O.~Scholer, J.~de~Vries and L.~Gr\'af, \emph{{$\nu$DoBe -- A Python Tool for
  Neutrinoless Double Beta Decay}},
  \href{http://arxiv.org/abs/2304.05415}{{\tt 2304.05415}}.

\bibitem{KamLAND-Zen:2022tow}
{\scshape KamLAND-Zen} collaboration, S.~Abe et~al., \emph{{First Search for
  the Majorana Nature of Neutrinos in the Inverted Mass Ordering Region with
  KamLAND-Zen}},  \href{http://arxiv.org/abs/2203.02139}{{\tt 2203.02139}}.

\bibitem{Liao:2019tep}
Y.~Liao and X.-D. Ma, \emph{{Renormalization Group Evolution of Dimension-seven
  Operators in Standard Model Effective Field Theory and Relevant
  Phenomenology}}, \href{http://dx.doi.org/10.1007/JHEP03(2019)179}{\emph{JHEP}
  {\bf 03} (2019) 179}, [\href{http://arxiv.org/abs/1901.10302}{{\tt
  1901.10302}}].

\bibitem{KamLAND-Zen:2016pfg}
{\scshape KamLAND-Zen} collaboration, A.~Gando et~al., \emph{{Search for
  Majorana Neutrinos near the Inverted Mass Hierarchy Region with
  KamLAND-Zen}},
  \href{http://dx.doi.org/10.1103/PhysRevLett.117.082503}{\emph{Phys. Rev.
  Lett.} {\bf 117} (2016) 082503}, [\href{http://arxiv.org/abs/1605.02889}{{\tt
  1605.02889}}].

\bibitem{Graf:2022lhj}
L.~Gr\'af, M.~Lindner and O.~Scholer, \emph{{Unraveling the
  0\ensuremath{\nu}\ensuremath{\beta}\ensuremath{\beta} decay mechanisms}},
  \href{http://dx.doi.org/10.1103/PhysRevD.106.035022}{\emph{Phys. Rev. D} {\bf
  106} (2022) 035022}, [\href{http://arxiv.org/abs/2204.10845}{{\tt
  2204.10845}}].

\bibitem{Freedman:1973yd}
D.~Z. Freedman, \emph{{Coherent Neutrino Nucleus Scattering as a Probe of the
  Weak Neutral Current}},
  \href{http://dx.doi.org/10.1103/PhysRevD.9.1389}{\emph{Phys. Rev. D} {\bf 9}
  (1974) 1389--1392}.

\bibitem{Erler:2004in}
J.~Erler and M.~J. Ramsey-Musolf, \emph{{The Weak mixing angle at low
  energies}}, \href{http://dx.doi.org/10.1103/PhysRevD.72.073003}{\emph{Phys.
  Rev. D} {\bf 72} (2005) 073003},
  [\href{http://arxiv.org/abs/hep-ph/0409169}{{\tt hep-ph/0409169}}].

\bibitem{Kerman:2016jqp}
{\scshape TEXONO} collaboration, S.~Kerman, V.~Sharma, M.~Deniz, H.~T. Wong,
  J.~W. Chen, H.~B. Li et~al., \emph{{Coherency in Neutrino-Nucleus Elastic
  Scattering}}, \href{http://dx.doi.org/10.1103/PhysRevD.93.113006}{\emph{Phys.
  Rev. D} {\bf 93} (2016) 113006}, [\href{http://arxiv.org/abs/1603.08786}{{\tt
  1603.08786}}].

\bibitem{Lindner:2016wff}
M.~Lindner, W.~Rodejohann and X.-J. Xu, \emph{{Coherent Neutrino-Nucleus
  Scattering and new Neutrino Interactions}},
  \href{http://dx.doi.org/10.1007/JHEP03(2017)097}{\emph{JHEP} {\bf 03} (2017)
  097}, [\href{http://arxiv.org/abs/1612.04150}{{\tt 1612.04150}}].

\bibitem{Bolton:2021pey}
P.~D. Bolton, F.~F. Deppisch, K.~Fridell, J.~Harz, C.~Hati and S.~Kulkarni,
  \emph{{Probing Active-Sterile Neutrino Transition Magnetic Moments with
  Photon Emission from CE$\nu$NS}},
  \href{http://arxiv.org/abs/2110.02233}{{\tt 2110.02233}}.

\bibitem{Bischer:2019ttk}
I.~Bischer and W.~Rodejohann, \emph{{General neutrino interactions from an
  effective field theory perspective}},
  \href{http://dx.doi.org/10.1016/j.nuclphysb.2019.114746}{\emph{Nucl. Phys. B}
  {\bf 947} (2019) 114746}, [\href{http://arxiv.org/abs/1905.08699}{{\tt
  1905.08699}}].

\bibitem{Bolton:2020xsm}
P.~D. Bolton, F.~F. Deppisch and C.~Hati, \emph{{Probing new physics with
  long-range neutrino interactions: an effective field theory approach}},
  \href{http://dx.doi.org/10.1007/JHEP07(2020)013}{\emph{JHEP} {\bf 07} (2020)
  013}, [\href{http://arxiv.org/abs/2004.08328}{{\tt 2004.08328}}].

\bibitem{Cheng:1988im}
H.-Y. Cheng, \emph{{Low-energy Interactions of Scalar and Pseudoscalar Higgs
  Bosons With Baryons}},
  \href{http://dx.doi.org/10.1016/0370-2693(89)90402-4}{\emph{Phys. Lett. B}
  {\bf 219} (1989) 347--353}.

\bibitem{Jungman:1995df}
G.~Jungman, M.~Kamionkowski and K.~Griest, \emph{{Supersymmetric dark matter}},
  \href{http://dx.doi.org/10.1016/0370-1573(95)00058-5}{\emph{Phys. Rept.} {\bf
  267} (1996) 195--373}, [\href{http://arxiv.org/abs/hep-ph/9506380}{{\tt
  hep-ph/9506380}}].

\bibitem{Anselmino:2008jk}
M.~Anselmino, M.~Boglione, U.~D'Alesio, A.~Kotzinian, F.~Murgia, A.~Prokudin
  et~al., \emph{{Update on transversity and Collins functions from SIDIS and e+
  e- data}},
  \href{http://dx.doi.org/10.1016/j.nuclphysbps.2009.03.117}{\emph{Nucl. Phys.
  B Proc. Suppl.} {\bf 191} (2009) 98--107},
  [\href{http://arxiv.org/abs/0812.4366}{{\tt 0812.4366}}].

\bibitem{AristizabalSierra:2018eqm}
D.~Aristizabal~Sierra, V.~De~Romeri and N.~Rojas, \emph{{COHERENT analysis of
  neutrino generalized interactions}},
  \href{http://dx.doi.org/10.1103/PhysRevD.98.075018}{\emph{Phys. Rev. D} {\bf
  98} (2018) 075018}, [\href{http://arxiv.org/abs/1806.07424}{{\tt
  1806.07424}}].

\bibitem{Li:2020lba}
T.~Li, X.-D. Ma and M.~A. Schmidt, \emph{{General neutrino interactions with
  sterile neutrinos in light of coherent neutrino-nucleus scattering and meson
  invisible decays}},
  \href{http://dx.doi.org/10.1007/JHEP07(2020)152}{\emph{JHEP} {\bf 07} (2020)
  152}, [\href{http://arxiv.org/abs/2005.01543}{{\tt 2005.01543}}].

\bibitem{Bolton:2019wta}
P.~D. Bolton and F.~F. Deppisch, \emph{{Probing nonstandard lepton number
  violating interactions in neutrino oscillations}},
  \href{http://dx.doi.org/10.1103/PhysRevD.99.115011}{\emph{Phys. Rev. D} {\bf
  99} (2019) 115011}, [\href{http://arxiv.org/abs/1903.06557}{{\tt
  1903.06557}}].

\bibitem{Danko:2009qw}
{\scshape MINOS} collaboration, I.~Danko, \emph{{First Observation of
  Accelerator Muon Antineutrinos in MINOS}},  in \emph{{Meeting of the Division
  of Particles and Fields of the American Physical Society (DPF 2009)}}, 10,
  2009.
\newblock \href{http://arxiv.org/abs/0910.3439}{{\tt 0910.3439}}.

\bibitem{Formaggio:2012cpf}
J.~A. Formaggio and G.~P. Zeller, \emph{{From eV to EeV: Neutrino Cross
  Sections Across Energy Scales}},
  \href{http://dx.doi.org/10.1103/RevModPhys.84.1307}{\emph{Rev. Mod. Phys.}
  {\bf 84} (2012) 1307--1341}, [\href{http://arxiv.org/abs/1305.7513}{{\tt
  1305.7513}}].

\bibitem{Bahcall:1996qv}
J.~N. Bahcall, E.~Lisi, D.~E. Alburger, L.~De~Braeckeleer, S.~J. Freedman and
  J.~Napolitano, \emph{{Standard neutrino spectrum from B-8 decay}},
  \href{http://dx.doi.org/10.1103/PhysRevC.54.411}{\emph{Phys. Rev. C} {\bf 54}
  (1996) 411--422}, [\href{http://arxiv.org/abs/nucl-th/9601044}{{\tt
  nucl-th/9601044}}].

\bibitem{Gonzalez-Alonso:2018omy}
M.~Gonzalez-Alonso, O.~Naviliat-Cuncic and N.~Severijns, \emph{{New physics
  searches in nuclear and neutron $\beta$ decay}},
  \href{http://arxiv.org/abs/1803.08732}{{\tt 1803.08732}}.

\bibitem{Adler:1975he}
S.~L. Adler, E.~W. Colglazier, Jr., J.~B. Healy, I.~Karliner, J.~Lieberman,
  Y.~J. Ng et~al., \emph{{Renormalization Constants for Scalar, Pseudoscalar,
  and Tensor Currents}},
  \href{http://dx.doi.org/10.1103/PhysRevD.11.3309}{\emph{Phys. Rev. D} {\bf
  11} (1975) 3309}.

\bibitem{Bell:2006wi}
N.~F. Bell, M.~Gorchtein, M.~J. Ramsey-Musolf, P.~Vogel and P.~Wang,
  \emph{{Model independent bounds on magnetic moments of Majorana neutrinos}},
  \href{http://dx.doi.org/10.1016/j.physletb.2006.09.055}{\emph{Phys. Lett. B}
  {\bf 642} (2006) 377--383}, [\href{http://arxiv.org/abs/hep-ph/0606248}{{\tt
  hep-ph/0606248}}].

\bibitem{Giunti:2014ixa}
C.~Giunti and A.~Studenikin, \emph{{Neutrino electromagnetic interactions: a
  window to new physics}},
  \href{http://dx.doi.org/10.1103/RevModPhys.87.531}{\emph{Rev. Mod. Phys.}
  {\bf 87} (2015) 531}, [\href{http://arxiv.org/abs/1403.6344}{{\tt
  1403.6344}}].

\bibitem{Canas:2015yoa}
B.~C. Canas, O.~G. Miranda, A.~Parada, M.~Tortola and J.~W.~F. Valle,
  \emph{{Updating neutrino magnetic moment constraints}},
  \href{http://dx.doi.org/10.1016/j.physletb.2015.12.011}{\emph{Phys. Lett. B}
  {\bf 753} (2016) 191--198}, [\href{http://arxiv.org/abs/1510.01684}{{\tt
  1510.01684}}].

\bibitem{Raffelt:1998xu}
G.~G. Raffelt, \emph{{Comment on neutrino radiative decay limits from the
  infrared background}},
  \href{http://dx.doi.org/10.1103/PhysRevLett.81.4020}{\emph{Phys. Rev. Lett.}
  {\bf 81} (1998) 4020}, [\href{http://arxiv.org/abs/astro-ph/9808299}{{\tt
  astro-ph/9808299}}].

\bibitem{Miranda:2021kre}
O.~G. Miranda, D.~K. Papoulias, O.~Sanders, M.~T\'ortola and J.~W.~F. Valle,
  \emph{{Low-energy probes of sterile neutrino transition magnetic moments}},
  \href{http://dx.doi.org/10.1007/JHEP12(2021)191}{\emph{JHEP} {\bf 12} (2021)
  191}, [\href{http://arxiv.org/abs/2109.09545}{{\tt 2109.09545}}].

\bibitem{Tandean:1999mg}
J.~Tandean, \emph{{New physics and short distance s ---\ensuremath{>} d gamma
  transition in Omega- ---\ensuremath{>} Xi-gamma decay}},
  \href{http://dx.doi.org/10.1103/PhysRevD.61.114022}{\emph{Phys. Rev. D} {\bf
  61} (2000) 114022}, [\href{http://arxiv.org/abs/hep-ph/9912497}{{\tt
  hep-ph/9912497}}].

\bibitem{Colangelo:2003hf}
G.~Colangelo and S.~Durr, \emph{{The Pion mass in finite volume}},
  \href{http://dx.doi.org/10.1140/epjc/s2004-01593-y}{\emph{Eur. Phys. J. C}
  {\bf 33} (2004) 543--553}, [\href{http://arxiv.org/abs/hep-lat/0311023}{{\tt
  hep-lat/0311023}}].

\bibitem{ParticleDataGroup:2012pjm}
{\scshape Particle Data Group} collaboration, J.~Beringer et~al., \emph{{Review
  of Particle Physics (RPP)}},
  \href{http://dx.doi.org/10.1103/PhysRevD.86.010001}{\emph{Phys. Rev. D} {\bf
  86} (2012) 010001}.

\bibitem{Gninenko:2014sxa}
S.~N. Gninenko, \emph{{Search for invisible decays of $\pi^0, \eta, \eta', K_S$
  and $K_L$: A probe of new physics and tests using the Bell-Steinberger
  relation}}, \href{http://dx.doi.org/10.1103/PhysRevD.91.015004}{\emph{Phys.
  Rev. D} {\bf 91} (2015) 015004}, [\href{http://arxiv.org/abs/1409.2288}{{\tt
  1409.2288}}].

\bibitem{Buras:2006gb}
A.~J. Buras, M.~Gorbahn, U.~Haisch and U.~Nierste, \emph{{Charm quark
  contribution to K+ ---\ensuremath{>} pi+ nu anti-nu at
  next-to-next-to-leading order}},
  \href{http://dx.doi.org/10.1007/JHEP11(2012)167}{\emph{JHEP} {\bf 11} (2006)
  002}, [\href{http://arxiv.org/abs/hep-ph/0603079}{{\tt hep-ph/0603079}}].

\bibitem{Brod:2010hi}
J.~Brod, M.~Gorbahn and E.~Stamou, \emph{{Two-Loop Electroweak Corrections for
  the $K \to \pi \nu \bar{\nu}$ Decays}},
  \href{http://dx.doi.org/10.1103/PhysRevD.83.034030}{\emph{Phys. Rev. D} {\bf
  83} (2011) 034030}, [\href{http://arxiv.org/abs/1009.0947}{{\tt 1009.0947}}].

\bibitem{Buras:2015qea}
A.~J. Buras, D.~Buttazzo, J.~Girrbach-Noe and R.~Knegjens, \emph{{$ {K}^{+}\to
  {\pi}^{+}\nu \overline{\nu} $ and $ {K}_L\to {\pi}^0\nu \overline{\nu} $ in
  the Standard Model: status and perspectives}},
  \href{http://dx.doi.org/10.1007/JHEP11(2015)033}{\emph{JHEP} {\bf 11} (2015)
  033}, [\href{http://arxiv.org/abs/1503.02693}{{\tt 1503.02693}}].

\bibitem{Kitahara:2019lws}
T.~Kitahara, T.~Okui, G.~Perez, Y.~Soreq and K.~Tobioka, \emph{{New physics
  implications of recent search for $K_L \to \pi^0 \nu\bar{\nu}$ at KOTO}},
  \href{http://dx.doi.org/10.1103/PhysRevLett.124.071801}{\emph{Phys. Rev.
  Lett.} {\bf 124} (2020) 071801}, [\href{http://arxiv.org/abs/1909.11111}{{\tt
  1909.11111}}].

\bibitem{NA62:2020fhy}
{\scshape NA62} collaboration, E.~Cortina~Gil et~al., \emph{{An investigation
  of the very rare $ {K}^{+}\to {\pi}^{+}\nu \overline{\nu} $ decay}},
  \href{http://dx.doi.org/10.1007/JHEP11(2020)042}{\emph{JHEP} {\bf 11} (2020)
  042}, [\href{http://arxiv.org/abs/2007.08218}{{\tt 2007.08218}}].

\bibitem{Newson:2014ahc}
F.~Newson et~al., \emph{{Prospects for $K^+ \to \pi^+ \nu \bar{ \nu }$ at CERN
  in NA62}},  in \emph{{50 Years of CP Violation}}, 11, 2014.
\newblock \href{http://arxiv.org/abs/1411.0109}{{\tt 1411.0109}}.

\bibitem{Romano:2014xda}
A.~Romano, \emph{{The $K^+ \rightarrow \pi^+ \nu \bar{\nu}$ decay in the NA62
  experiment at CERN}},  \href{http://arxiv.org/abs/1411.6546}{{\tt
  1411.6546}}.

\bibitem{Grossman:1997sk}
Y.~Grossman and Y.~Nir, \emph{{K(L) ---\ensuremath{>} pi0 neutrino
  anti-neutrino beyond the standard model}},
  \href{http://dx.doi.org/10.1016/S0370-2693(97)00210-4}{\emph{Phys. Lett. B}
  {\bf 398} (1997) 163--168}, [\href{http://arxiv.org/abs/hep-ph/9701313}{{\tt
  hep-ph/9701313}}].

\bibitem{Komatsubara:2012pn}
T.~K. Komatsubara, \emph{{Experiments with K-Meson Decays}},
  \href{http://dx.doi.org/10.1016/j.ppnp.2012.04.001}{\emph{Prog. Part. Nucl.
  Phys.} {\bf 67} (2012) 995--1018},
  [\href{http://arxiv.org/abs/1203.6437}{{\tt 1203.6437}}].

\bibitem{Shiomi:2014sfa}
{\scshape KOTO} collaboration, K.~Shiomi, \emph{{$K^{0}_{L}\rightarrow \pi^{0}
  \nu \bar{\nu}$ at KOTO}},  in \emph{{8th International Workshop on the CKM
  Unitarity Triangle}}, 11, 2014.
\newblock \href{http://arxiv.org/abs/1411.4250}{{\tt 1411.4250}}.

\bibitem{KOTOWG}
\emph{{https://web2.ph.utexas.edu/~heavyquark/KOTO-intensity\_workshop.pdf}}, .

\bibitem{Liao:2019gex}
Y.~Liao, X.-D. Ma and H.-L. Wang, \emph{{Effective field theory approach to
  lepton number violating decays $K^\pm\rightarrow \pi^\mp l^{\pm}l^{\pm}$:
  short-distance contribution}},
  \href{http://dx.doi.org/10.1007/JHEP01(2020)127}{\emph{JHEP} {\bf 01} (2020)
  127}, [\href{http://arxiv.org/abs/1909.06272}{{\tt 1909.06272}}].

\bibitem{Liao:2020roy}
Y.~Liao, X.-D. Ma and H.-L. Wang, \emph{{Effective field theory approach to
  lepton number violating decays $K^\pm\rightarrow \pi^\mp l^{\pm}_\alpha
  l^{\pm}_\beta$: long-distance contribution}},
  \href{http://dx.doi.org/10.1007/JHEP03(2020)120}{\emph{JHEP} {\bf 03} (2020)
  120}, [\href{http://arxiv.org/abs/2001.07378}{{\tt 2001.07378}}].

\bibitem{Zhou:2021lnl}
G.~Zhou, \emph{{Light sterile neutrinos and lepton-number-violating kaon decays
  in effective field theory}},  \href{http://arxiv.org/abs/2112.00767}{{\tt
  2112.00767}}.

\bibitem{Aebischer:2017gaw}
J.~Aebischer, M.~Fael, C.~Greub and J.~Virto, \emph{{B physics Beyond the
  Standard Model at One Loop: Complete Renormalization Group Evolution below
  the Electroweak Scale}},
  \href{http://dx.doi.org/10.1007/JHEP09(2017)158}{\emph{JHEP} {\bf 09} (2017)
  158}, [\href{http://arxiv.org/abs/1704.06639}{{\tt 1704.06639}}].

\bibitem{Beda:2012zz}
A.~G. Beda, V.~B. Brudanin, V.~G. Egorov, D.~V. Medvedev, V.~S. Pogosov, M.~V.
  Shirchenko et~al., \emph{{The results of search for the neutrino magnetic
  moment in GEMMA experiment}},
  \href{http://dx.doi.org/10.1155/2012/350150}{\emph{Adv. High Energy Phys.}
  {\bf 2012} (2012) 350150}.

\bibitem{Borexino:2017fbd}
{\scshape Borexino} collaboration, M.~Agostini et~al., \emph{{Limiting neutrino
  magnetic moments with Borexino Phase-II solar neutrino data}},
  \href{http://dx.doi.org/10.1103/PhysRevD.96.091103}{\emph{Phys. Rev. D} {\bf
  96} (2017) 091103}, [\href{http://arxiv.org/abs/1707.09355}{{\tt
  1707.09355}}].

\bibitem{Straub:2018kue}
D.~M. Straub, \emph{{flavio: a Python package for flavour and precision
  phenomenology in the Standard Model and beyond}},
  \href{http://arxiv.org/abs/1810.08132}{{\tt 1810.08132}}.

\bibitem{Belle-II:2018jsg}
{\scshape Belle-II} collaboration, W.~Altmannshofer et~al., \emph{{The Belle II
  Physics Book}}, \href{http://dx.doi.org/10.1093/ptep/ptz106}{\emph{PTEP} {\bf
  2019} (2019) 123C01}, [\href{http://arxiv.org/abs/1808.10567}{{\tt
  1808.10567}}].

\bibitem{Belle:2017oht}
{\scshape Belle} collaboration, J.~Grygier et~al., \emph{{Search for
  $\boldsymbol{B\to h\nu\bar{\nu}}$ decays with semileptonic tagging at
  Belle}}, \href{http://dx.doi.org/10.1103/PhysRevD.96.091101}{\emph{Phys. Rev.
  D} {\bf 96} (2017) 091101}, [\href{http://arxiv.org/abs/1702.03224}{{\tt
  1702.03224}}].

\bibitem{BaBar:2013npw}
{\scshape BaBar} collaboration, J.~P. Lees et~al., \emph{{Search for $B \to
  K^{(*)} \nu \overline \nu$ and invisible quarkonium decays}},
  \href{http://dx.doi.org/10.1103/PhysRevD.87.112005}{\emph{Phys. Rev. D} {\bf
  87} (2013) 112005}, [\href{http://arxiv.org/abs/1303.7465}{{\tt 1303.7465}}].

\bibitem{Belle:2013tnz}
{\scshape Belle} collaboration, O.~Lutz et~al., \emph{{Search for $B \to
  h^{(*)} \nu \bar{\nu}$ with the full Belle $\Upsilon(4S)$ data sample}},
  \href{http://dx.doi.org/10.1103/PhysRevD.87.111103}{\emph{Phys. Rev. D} {\bf
  87} (2013) 111103}, [\href{http://arxiv.org/abs/1303.3719}{{\tt 1303.3719}}].

\bibitem{Gubernari:2018wyi}
N.~Gubernari, A.~Kokulu and D.~van Dyk, \emph{{$B\to P$ and $B\to V$ Form
  Factors from $B$-Meson Light-Cone Sum Rules beyond Leading Twist}},
  \href{http://dx.doi.org/10.1007/JHEP01(2019)150}{\emph{JHEP} {\bf 01} (2019)
  150}, [\href{http://arxiv.org/abs/1811.00983}{{\tt 1811.00983}}].

\bibitem{Bharucha:2015bzk}
A.~Bharucha, D.~M. Straub and R.~Zwicky, \emph{{$B\to V\ell^+\ell^-$ in the
  Standard Model from light-cone sum rules}},
  \href{http://dx.doi.org/10.1007/JHEP08(2016)098}{\emph{JHEP} {\bf 08} (2016)
  098}, [\href{http://arxiv.org/abs/1503.05534}{{\tt 1503.05534}}].

\bibitem{Belle-II:2023esi}
{\scshape Belle-II} collaboration, I.~Adachi et~al., \emph{{Evidence for
  $B^{+}\to K^{+}\nu\bar{\nu}$ Decays}},
  \href{http://arxiv.org/abs/2311.14647}{{\tt 2311.14647}}.

\bibitem{Fridell:2023ssf}
K.~Fridell, M.~Ghosh, T.~Okui and K.~Tobioka, \emph{{Decoding the $B \to K \nu
  \nu$ excess at Belle II: kinematics, operators, and masses}},
  \href{http://arxiv.org/abs/2312.12507}{{\tt 2312.12507}}.

\bibitem{Belle:2012unr}
{\scshape Belle} collaboration, Y.~Miyazaki et~al., \emph{{Search for
  Lepton-Flavor-Violating and Lepton-Number-Violating $\tau \to \ell h
  h^\prime$ Decay Modes}},
  \href{http://dx.doi.org/10.1016/j.physletb.2013.01.032}{\emph{Phys. Lett. B}
  {\bf 719} (2013) 346--353}, [\href{http://arxiv.org/abs/1206.5595}{{\tt
  1206.5595}}].

\bibitem{RodriguezPerez:2019nhw}
{\scshape Belle-II} collaboration, D.~Rodr\'\i{}guez~P\'erez, \emph{{Prospects
  for $\tau$ Lepton Physics at Belle II}},  in \emph{{17th Conference on Flavor
  Physics and CP Violation}}, 6, 2019.
\newblock \href{http://arxiv.org/abs/1906.08950}{{\tt 1906.08950}}.

\bibitem{Liao:2021qfj}
Y.~Liao, X.-D. Ma and H.-L. Wang, \emph{{Effective field theory approach to
  lepton number violating $\tau$ decays}},
  \href{http://arxiv.org/abs/2102.03491}{{\tt 2102.03491}}.

\bibitem{Atre:2005eb}
A.~Atre, V.~Barger and T.~Han, \emph{{Upper bounds on lepton-number violating
  processes}}, \href{http://dx.doi.org/10.1103/PhysRevD.71.113014}{\emph{Phys.
  Rev. D} {\bf 71} (2005) 113014},
  [\href{http://arxiv.org/abs/hep-ph/0502163}{{\tt hep-ph/0502163}}].

\bibitem{Atre:2009rg}
A.~Atre, T.~Han, S.~Pascoli and B.~Zhang, \emph{{The Search for Heavy Majorana
  Neutrinos}},
  \href{http://dx.doi.org/10.1088/1126-6708/2009/05/030}{\emph{JHEP} {\bf 05}
  (2009) 030}, [\href{http://arxiv.org/abs/0901.3589}{{\tt 0901.3589}}].

\bibitem{Abada:2017jjx}
A.~Abada, V.~De~Romeri, M.~Lucente, A.~M. Teixeira and T.~Toma,
  \emph{{Effective Majorana mass matrix from tau and pseudoscalar meson lepton
  number violating decays}},
  \href{http://dx.doi.org/10.1007/JHEP02(2018)169}{\emph{JHEP} {\bf 02} (2018)
  169}, [\href{http://arxiv.org/abs/1712.03984}{{\tt 1712.03984}}].

\bibitem{Abada:2019bac}
A.~Abada, C.~Hati, X.~Marcano and A.~M. Teixeira, \emph{{Interference effects
  in LNV and LFV semileptonic decays: the Majorana hypothesis}},
  \href{http://dx.doi.org/10.1007/JHEP09(2019)017}{\emph{JHEP} {\bf 09} (2019)
  017}, [\href{http://arxiv.org/abs/1904.05367}{{\tt 1904.05367}}].

\bibitem{Armbruster:2003pq}
B.~Armbruster et~al., \emph{{Improved limits on anti-nu(e) emission from mu+
  decay}}, \href{http://dx.doi.org/10.1103/PhysRevLett.90.181804}{\emph{Phys.
  Rev. Lett.} {\bf 90} (2003) 181804},
  [\href{http://arxiv.org/abs/hep-ex/0302017}{{\tt hep-ex/0302017}}].

\bibitem{Mu2e:2014fns}
{\scshape Mu2e} collaboration, L.~Bartoszek et~al., \emph{{Mu2e Technical
  Design Report}},  \href{http://arxiv.org/abs/1501.05241}{{\tt 1501.05241}}.

\bibitem{Kuno:2015tya}
Y.~Kuno, \emph{{Rare lepton decays}},
  \href{http://dx.doi.org/10.1016/j.ppnp.2015.01.003}{\emph{Prog. Part. Nucl.
  Phys.} {\bf 82} (2015) 1--20}.

\bibitem{SINDRUMII:1998mwd}
{\scshape SINDRUM II} collaboration, J.~Kaulard et~al., \emph{{Improved limit
  on the branching ratio of mu- --\ensuremath{>} e+ conversion on titanium}},
  \href{http://dx.doi.org/10.1016/S0370-2693(97)01423-8}{\emph{Phys. Lett. B}
  {\bf 422} (1998) 334--338}.

\bibitem{Natori:2014yba}
{\scshape DeeMe} collaboration, H.~Natori, \emph{{DeeMe experiment - An
  experimental search for a mu-e conversion reaction at J-PARC MLF}},
  \href{http://dx.doi.org/10.1016/j.nuclphysbps.2014.02.010}{\emph{Nucl. Phys.
  B Proc. Suppl.} {\bf 248-250} (2014) 52--57}.

\bibitem{COMET:2009qeh}
{\scshape COMET} collaboration, Y.~G. Cui et~al., \emph{{Conceptual design
  report for experimental search for lepton flavor violating mu- - e-
  conversion at sensitivity of 10**(-16) with a slow-extracted bunched proton
  beam (COMET)}}, .

\bibitem{Kuno:2012pt}
Y.~Kuno, \emph{{COMET and PRISM: Search for charged lepton flavor violation
  with muons}},
  \href{http://dx.doi.org/10.1016/j.nuclphysbps.2012.02.047}{\emph{Nucl. Phys.
  B Proc. Suppl.} {\bf 225-227} (2012) 228--231}.

\bibitem{Berryman:2016slh}
J.~M. Berryman, A.~de~Gouv\^ea, K.~J. Kelly and A.~Kobach,
  \emph{{Lepton-number-violating searches for muon to positron conversion}},
  \href{http://dx.doi.org/10.1103/PhysRevD.95.115010}{\emph{Phys. Rev. D} {\bf
  95} (2017) 115010}, [\href{http://arxiv.org/abs/1611.00032}{{\tt
  1611.00032}}].

\bibitem{Conlin:2020veq}
R.~Conlin and A.~A. Petrov, \emph{{Muonium-antimuonium oscillations in
  effective field theory}},
  \href{http://dx.doi.org/10.1103/PhysRevD.102.095001}{\emph{Phys. Rev. D} {\bf
  102} (2020) 095001}, [\href{http://arxiv.org/abs/2005.10276}{{\tt
  2005.10276}}].

\bibitem{AristizabalSierra:2021fuc}
D.~Aristizabal~Sierra, O.~G. Miranda, D.~K. Papoulias and G.~S. Garcia,
  \emph{{Neutrino magnetic and electric dipole moments: From measurements to
  parameter space}},
  \href{http://dx.doi.org/10.1103/PhysRevD.105.035027}{\emph{Phys. Rev. D} {\bf
  105} (2022) 035027}, [\href{http://arxiv.org/abs/2112.12817}{{\tt
  2112.12817}}].

\bibitem{deGouvea:2022znk}
A.~de~Gouv\^ea, G.~Jusino~S\'anchez, P.~A.~N. Machado and Z.~Tabrizi,
  \emph{{Majorana versus Dirac Constraints on the Neutrino Dipole Moments}},
  \href{http://arxiv.org/abs/2209.03373}{{\tt 2209.03373}}.

\bibitem{Magill:2018jla}
G.~Magill, R.~Plestid, M.~Pospelov and Y.-D. Tsai, \emph{{Dipole Portal to
  Heavy Neutral Leptons}},
  \href{http://dx.doi.org/10.1103/PhysRevD.98.115015}{\emph{Phys. Rev. D} {\bf
  98} (2018) 115015}, [\href{http://arxiv.org/abs/1803.03262}{{\tt
  1803.03262}}].

\bibitem{Brdar:2020quo}
V.~Brdar, A.~Greljo, J.~Kopp and T.~Opferkuch, \emph{{The Neutrino Magnetic
  Moment Portal: Cosmology, Astrophysics, and Direct Detection}},
  \href{http://dx.doi.org/10.1088/1475-7516/2021/01/039}{\emph{JCAP} {\bf 01}
  (2021) 039}, [\href{http://arxiv.org/abs/2007.15563}{{\tt 2007.15563}}].

\bibitem{Schwetz:2020xra}
T.~Schwetz, A.~Zhou and J.-Y. Zhu, \emph{{Constraining active-sterile neutrino
  transition magnetic moments at DUNE near and far detectors}},
  \href{http://dx.doi.org/10.1007/JHEP07(2021)200}{\emph{JHEP} {\bf 21} (2020)
  200}, [\href{http://arxiv.org/abs/2105.09699}{{\tt 2105.09699}}].

\end{thebibliography}\endgroup
\end{document}